\documentclass[aps,pra,twocolumn,nofootinbib]{revtex4-2}
\usepackage[colorlinks]{hyperref}
\usepackage[utf8]{inputenc}
\usepackage{graphicx}
\usepackage{multirow}
\usepackage{amssymb}
\usepackage{amsmath}
\usepackage{physics}
\usepackage{bm}

\begin{document}

\title{
Parity and Time-reversal Invariance Violation In Neutron-Nucleus Scattering
}

\author{V.~V.~Flambaum$^{1,2,3}$}
\author{A.~J.~Mansour$^1$}
\affiliation{$^1$School of Physics, University of New South Wales, Sydney 2052, Australia}
\affiliation{$^2$Helmholtz Institute Mainz, GSI Helmholtzzentrum für Schwerionenforschung, 55099 Mainz, Germany}
\affiliation{$^3$Johannes Gutenberg University Mainz, 55099 Mainz, Germany}

\begin{abstract}
Parity violating (PV) as well as parity and time-reversal invariance violating  (PTRIV) effects are enhanced a million times in neutron reactions near $p$-wave compound resonances. We present the calculation of such effects using a statistical theory based on the properties of chaotic eigenstates and discuss a possibility to extract the strength constants of PTRIV interactions from the experimental data, including nucleon-nucleon and pion-nucleon CP-violating interactions, the QCD $\theta$-term and the quark chromo-EDM.  PV effects have random sign for all target nuclei except for $^{232}\text{Th}$, where PV effects of a positive sign have been observed for ten statistically significant $p$-wave resonances, with energy smaller than 250 eV. This may be an indication of a possible regular (non-chaotic) contribution to PV effects. We link this regular effect to the doublets of opposite parity states in the rotation spectra of nuclei with an octupole deformation and suggest other target nuclei where this hypothesis may be tested. We also discuss a permanent sign contribution produced by doorway states. An estimate of the ratio of PTRIV effects to PV effects is presented. Although a polarised target is not needed for the measurement of PV effects, for the interpretation of the results, it may be convenient to do both PV and PTRIV experiments with a polarised target.

\end{abstract}

\maketitle

\section{Introduction} \label{1}
It was predicted  that the effects of parity violation are enhanced a million times in neutron reactions which occur near $p$-wave nuclear compound resonances~\cite{Sushkov80,Sushkov82,Flambaum84,Flambaum85}. Experiments performed at the Joint institute for Nuclear Research in Dubna~\cite{Alfimenkov1981,Alfimenkov1983,Alfimenkov1984} first confirmed this, and further verification was provided after an extensive experimental study was undertaken, across several locations, including the Petersburg Institute of Nuclear Physics, the Joint Institute for Nuclear Research (Dubna), KEK (Tsukuba), and in Los Alamos, see reviews~\cite{MitchellReview99,MitchellReview2001}. 

The same mechanism of enhancement can also be extended to effects which violate both parity and time-reversal invariance (PTRIV)~\cite{Bunakov1983,TRIV87,Gudkov1992,Flambaum_Gribakin95,Fadeev}. PTRIV effects in the transmission of polarized neutrons through a polarized target have been suggested in Refs.~\cite{PhysRevD.25.2013,STODOLSKY1982213}. Currently,  experiments measuring PTRIV effects are in progress, see~\cite{Snow2017,Okudaira2018,Kitaguchi2018, Beda2007,Bowman96V}.



The neutron forward scattering amplitude can schematically (think of the lowest order Born approximation) be represented in the following form: 
\begin{align}
    f(0)= a + b \{\mathbf{s} \cdot \mathbf{I}\} + c \{\mathbf{s} \cdot \mathbf{p}\} + d \{\mathbf{s} \cdot [ \mathbf{p} \times \mathbf{I}]\},
\end{align}
where  $\mathbf{s}, \mathbf{p}$ and $ \mathbf{I}$ are the operators for the neutron spin, neutron momentum and target spin correspondingly. The terms $a$ and $b\{\mathbf{s} \cdot \mathbf{I}\}$ govern the strength of the spin-independent and spin-spin strong interactions, while $c \{ \mathbf{s} \cdot \mathbf{p}\}$ and $d\{\mathbf{s} \cdot [ \mathbf{p} \times \mathbf{I}]\}$ are responsible for  PV and PTRIV effects. 
In PV experiments, the number of neutrons transmitted through the target is measured for neutrons of positive and negative helicities. 
In PTRIV experiments neutron spin $\mathbf{s}$, neutron momentum $\mathbf{p}$ and nuclear spin $\mathbf{I}$ are all perpendicular to each other. 

For PV effects, although a polarised target is not needed, for comparison with PTRIV effects it may be convenient to do both PV and PTRIV measurements with a polarised  target, by changing the orientation of neutron spin $\mathbf{s}$; for PV measurement $\mathbf{s}$ is parallel or antiparallel to the neutron momentum $\mathbf{p}$ (correlation $(\mathbf{s} \cdot \mathbf{p})$, with $\mathbf{s}$ and  $\mathbf{p}$ perpendicular to $ \mathbf{I}$). 
As we will show below, in the case where the spin of the $p$-wave compound resonance is equal to $J=I-1/2$, the ratio of PTRIV and PV effects is reduced to the ratio of the PTRIV and PV weak interaction matrix elements (in the two-resonance approximation; in the general case it is the ratio of the weighted sums of the PTRIV and PV matrix elements). This looks important since in the case of  $J=I+1/2$ and for PV effects measured with an unpolarised target, the ratio of PTRIV and PV effects contains the unknown ratio of the $p_{3/2}$ and $p_{1/2}$ neutron capture amplitudes, $M_{3/2}/M_{1/2}$. We will discuss possibilities to measure this ratio $M_{3/2}/M_{1/2}$ within the same experimental arrangement.

It is important to note that although nuclear reactions such as scattering and particle decay exhibit T-odd angular correlations, this alone is not sufficient to establish time-reversal invariance violation. These correlations may arise due to the phases from  strong, weak and electromagnetic interactions. For example, T-odd correlations in neutron beta decay are imitated by electromagnetic interaction in the final state. This is not the case in forward elastic scattering, as the initial and final state is the same. A consequence of this is that PTRIV correlation in neutron transmission cannot be imitated by scattering phases \cite{PhysRevC.90.065503,ryndin_1968,Bilenky:1964pm,Bilenky:1969pm,PhysRevD.37.1856}.
A discussion of possible systematic errors and specific schemes to eliminate  them have been presented in Refs.~\cite{PhysRevC.90.065503,PhysRevD.50.5632,masuda1992,PhysRevD.53.4070,PhysRevC.83.035501,MASUDA1998479,MASUDA2000632,KABIR198963,Baryshevsky1965,Abragam1982,MasudaBowman1996,Serebrov1996}.

In compound states with several excited particles the density of energy  levels is very high, and the residual interaction between the particles exceeds the energy intervals. As a result, excited states $|n\rangle$ in all medium and heavy nuclei near the neutron separation energy (as well as in atoms and ions with several excited electrons in an open f-shell) are chaotic superpositions of thousands or even millions of Hartree-Fock basis states $|i\rangle$, $|n\rangle = \sum_i C_i^n |i\rangle$. 
For the planning and interpretation of experiments we need reliable calculations. Although at first this seems impossible, due to the complicated nature of chaotic compound states, chaos allows us to develop a statistical theory \cite{Flam93,Flam93PRL,PhysRevE.56.5144,PhysRevA.91.052704},
similar to the Maxwell-Boltzmann theory for macroscopic systems, which allows for very accurate predictions.  
This theory allows us to calculate the root-mean-squared values of matrix elements of different operators and the transition amplitudes between chaotic states, and processes involving chaotic compound resonances in nuclei \cite{Flam93,Flam93PRL,Flambaum1994V,Flambaum1995a,Flambaum1995b,FlamGrib2000}, atoms and ions \cite{FlamGribakin1994,DzubaBerengut2017,Flambaum1998StatisticsOE,FlamGribakin2002,Harabati2017,PhysRevA.91.052704}.









Chaos implies that PTRIV and PV effects have random sign, when they are considered in a set of $p$-wave compound resonances. Such randomness has indeed been observed in many nuclei, however there exists a notable exception: PV effects in neutron capture by the target nucleus $^{232}$Th~\cite{232Th,MitchellReview2001,PhysRevC.58.1236}. These publications found that for 10 observed statistically significant PV effects in $p$-wave resonances with energies below 250 eV the sign of the PV effect was the same (positive). According to the interpretation of the experimental data in Ref.~\cite{MitchellReview2001}, PV effects in $^{233}$Th are better described as a linear combination of the constant effect and random sign effect. The constant component is slightly bigger and determines the permanent sign of the PV effects for energies below 250 eV. Above 250 eV, PV effects have both negative and positive signs. A critical review of suggested explanations of the permanent sign of PV effect in $^{233}\text{Th}$ and corresponding references may be found in~\cite{MitchellReview2001,Noguera1996}. 
	
A possible explanation for the constant sign component in PV effects may be due to the octupole deformation in the excited states of $^{233}\text{Th}$ (to avoid misunderstanding, note that there is no need for octupole deformation in the ground state of this nucleus). This is compounded by indications of octupole deformation in the ground states of $^{226}\text{Th}$ and $^{228}\text{Th}$, while $^{233} \text{Th}$ has octupole deformation in the fission channel~\cite{Blons:1984fqo,Blons1989,MitchellReview2001}, a property which is used to explain large PV effects in nuclear fission~\cite{Sushkov82,FLAMBAUM1980277,sushkovangular1981,sushkovnature1981}. A permanent sign contribution to the PV effects may be due to the mixing by PV interaction of the opposite parity doublet states in the rotational spectra, which appears in nuclei with octupole deformation and non-zero spin~\cite{Flambaum_1995,PhysRevLett.74.2638,Spevak1995}.


As known, the third well in the deformation potential energy corresponds to the octupole deformation (see e.g. Ref. \cite{Butler_2016}). Assuming resonances with energy below 250 eV in $^{233}$Th are excited compound states built on the isomer state with octupole deformation, mixing of the doublet states by the weak interaction may give a noticeable permanent sign contribution to PV effects. Note that a compound resonance wave function may contain both types of components, components with octupole deformation and components with no octupole deformation - see e.g. papers on PV in nuclear fission  ~\cite{Sushkov82,FLAMBAUM1980277,sushkovangular1981,sushkovnature1981}. The distribution function for the masses of fission fragments has maximum for significantly different fragment masses. This may be considered as evidence for the octupole deformation of components in the compound state wave function. 

Statistical theory predicts mean squared values of the weak matrix elements, requiring measurements on many compound resonances in order to extract the strength constants of the PV and PTRIV interactions with a high accuracy (PV effects with significance above 1$\sigma$ have been detected  in $\sim 150$ resonances). This may not necessarily be the case for regions of a possible constant sign effect, and is especially important for the measurements of the PTRIV effects which are expected to be very small. A possible outcome may be a limit on the strength constant rather the non-zero value of this constant. Furthermore, the need for polarised targets complicates significantly experiments for PTRIV  effects. In the case of the proposed permanent sign effect, it may be sufficient to measure one or a few resonances to extract quantitative information about PTRIV interactions. Also, if the doublet mechanism is confirmed, the measurements of PV effects may be used to search for nuclei with octupole deformation.
	
Given this, it seems that performing measurements and calculations on nuclei which may exhibit a permanent sign effect may be advantageous.
There are two main areas of proton numbers $Z$ and neutron numbers $N$ where octupole deformation is expected in the ground state, or in a low energy isomer state (below the neutron separation threshold)~\cite{Afanasjev,Butler,Nazarevich,FlambaumFeldmeier}. These areas are the lanthanides in $Z \sim 56 - 66$, $N \sim 88-97$ (including Ba, La, Ce, Eu, Gd, Dy and Sm isotopes) and the $Z \sim 88-102 $, $ N \sim 134 -194$ mass region (including  Rn, Ra, Ac, Th, Pa, U, Np, Pu and Cm isotopes). The study of PTRIV effects requires the target nuclei to be polarised, meaning isotopes with non-zero spin are necessary. Furthermore, the stability of the nuclei must be considered, as many candidate nuclei with an odd number of nucleons are unstable, meaning they are not suitable for experiments. It is also important to note that octupole deformation is required in the nucleus excited by the neutron capture. Information regarding octupole deformation can be extracted from the nuclear rotational spectra presented in the database~\cite{nudat} (see also ~\cite{FlambaumFeldmeier,FlambaumDzuba}). In nuclei with non-zero spin with octupole deformation, doublets of opposite parity states with the same spin (along the rotational bands) can be seen. Energy splitting of this doublet in the ground state typically ranges from 25 keV to 400 keV. In nuclei with zero spin and octupole deformation, the negative parity rotational band $1^{-}, 3^{-}, 5^{-},...$ is separated from the positive parity rotational band $0^{+}, 2^{+}, 4^{+},...$ by roughly 0.1 - 1 MeV. Note that this is smaller than the typical excitation energy of octupole vibration which is $\sim 2.5 -3$ MeV in most nuclei~\cite{LANE196039}.

Note the energy interval between doublet states in the case of octupole deformation is never zero, due to the Coriolis interaction and ``tunneling''. In these nuclei, the splitting of doublet states is dominated by the ``tunneling'' of the octupole bump (excess nucleons on one side) to the other side of the nucleus, removing a pre-existing degeneracy between them. As this process is similar to an octupole vibration mode, there is no sharp boundary between the static octupole deformation in the deep minimum of the potential energy and dynamical octupole deformation (low energy octupole excitation) when this minimum is shallow or flat.

In a compound state this tunneling amplitude may be  small, as tunneling of the nucleon number excess from one side of the pear shape nucleus to another side has a small probability to return the system back to exactly the same intrinsic state of all nucleons, due to a very large number of the principal components, $N \sim 10^4$, in the compound state wave function. Due to this possible suppression of the tunneling amplitude, the energy interval between the opposite parity doublet states in the compound nucleus may be significantly smaller than such an interval in the ground state.  Indeed, in the fission channel's structure, the energy interval between the opposite parity doublet states is practically invisible 
\cite{Blons:1984fqo,Blons1989}.


 Several papers~\cite{FlambaumPwave1992,Hussein1995,Feshbach1996,PhysRevC.45.R514,PhysRevC.46.2582,PhysRevC.50.1456,PhysRevLett.74.2638,PhysRevLett.68.780} described in the review \cite{MitchellReview2001} suggested mixing of the doorway states of opposite parity as a source of the permanent sign effect in $^{233}$Th. For example, in Refs.~\cite{FlambaumPwave1992,Hussein1995,Feshbach1996} the doorway states are two-particle-one-hole states excited by the first collision of neutron with nucleus (Ref. \cite{FlambaumPwave1992} also mentioned rotation and vibration excitations). 
In  section \ref{8} of  the present paper we calculate the doorway states contribution to PV and PTRIV effects. We treated doorway states as components of the nuclear compound states wave functions and found an additional term, which may have both positive and negative sign. This term may dominate if the distance between the $s$-wave compound resonance and the $p$-wave compound resonance is small, $|E_s -E_p| < d/2$, where 
 $d \approx 20$ eV  is the interval between compound resonances with the same spin and parity
 (note that such cases may produce the largest PV effects). This term may be important for the contribution of  any local doorway states (with energies close to a $p$-wave compound resonance). For distant doorway states this term is not significant, however, the total distant doorway states contribution looks too small to explain the observed constant sign effect (see discussion in review \cite{MitchellReview2001}).  

 A possible factor of enhancement in the local doorway mechanisms is the interval between the opposite parity doorway states, which accidentally happened to be very small near the neutron threshold in $^{233}$Th nucleus. Indeed, the opposite parity states do not repel each other and the probability density to have a zero energy interval $(E_+ - E_-)$ is not suppressed.  PV effects $P$ have been measured in 20 nuclei, so in one of them the interval $(E_+ - E_-)$ between the opposite parity doorway states is expected to be 20 times smaller than the average value of such an interval, and the value of the corresponding contribution to the PV effect $P$ could be 20 times bigger than a typical value of this contribution to $P$  (since $P \propto 1/(E_+ - E_-))$.  
However, in the case of the doorway-induced effect, this enhancement is limited if we take into account the finite spreading width of the doorway states, which stays in the energy denominator $E_+ -E_- +i\Gamma_{d}/2$ of the doorway state contribution to $P$. According to Ref. \cite{Kerman1963}, observations indicate $\Gamma_{d} \sim$100 keV and a distance between the doorway states of $D\sim$ 300 keV.
 Refs.~\cite{PhysRevLett.74.2638,Hussein2016} argue that  $\Gamma_{d}$ and $D$ may be smaller.
 

 In other words, the doorway state is just one of $N$ principal components of the compound state, meaning it hardly can dominate in the weak matrix element between opposite parity compound states as $N \sim 10^4$. Indeed, a sum of $N$  random sign terms increases as $N^{1/2}$, therefore one may expect that a single doorway state contribution is suppressed as $N^{-1/2} \sim \left(\frac{d}{\Gamma_{\text{spr}}}\right)^{1/2}$ relative to the statistical contribution of all $N$ terms.  To overcome this suppression,  the doorway width  should be exceptionally small, $\Gamma_{d} < $ 1 keV, for both opposite parity doorway states, and both of these states should be in close vicinity, within 1 keV, of the $p$-wave resonances of interest (see section \ref{8} of the present paper). Is is not clear if the probability of such a coincidence is significantly higher than the probability of ten random sign PV effects having the same sign.  
 



This paper is organised as follows. In section \ref{2} we present our results for the general expressions for the PV and PTRIV effects, including the angular coefficients for both polarised and unpolarised targets. In section \ref{3} we use a statistical approach to express the mean squared values of the matrix elements of the PV and PTRIV interactions by values of PV and PTRIV effects $P$ (defined as the asymmetry in the neutron transmission). In section \ref{4}  we provide a brief overview of the statistical theory of finite systems based on the properties of chaotic compound states. In section \ref{5} we use this theory to calculate the mean squared values of the matrix elements of PV and PTRIV interactions. In section \ref{6} we summarise the experimental results which indicate a possible regular component in PV effects, using the target nucleus $^{232}$Th. In section \ref{7} we present a possible octupole doublet mechanism for a regular component of PV and PTRIV effects in neutron scattering. In section \ref{8} we derive expressions for the doorway states contribution to PV and PTRIV effects. Section \ref{9} contains comments about other potential mechanisms for the permanent sign PV and PTRIV effects. In Appendix \ref{AppendixA}, we present our calculations for the angular coefficients of PV and PTRIV effects, for both a polarised and unpolarised target. Appendix \ref{AppendixB} presents a brief overview of the theory used to calculate the root-mean-squared values of the matrix elements of the PV and PTRIV interactions.


\section{Scattering amplitudes for parity Violation and time reversal violation } \label{2}

\subsection{Brief introduction to parity violation in compound nuclei}

Let us start from a simple qualitative explanation of the origin of the enhanced parity violating effect. 
For simplicity, we will initially assume that nuclear spin is zero. Resonances which are excited by $\ell=0$ neutrons are called $s$-wave resonances and have positive parity, while resonances excited by $\ell=1$ neutrons are called $p$-wave resonances and have negative parity. In such a compound nucleus close to $p$-wave resonance, the wave function can be written as

\begin{align} \label{psi}
    \psi_{p}^{\prime} =  \psi_{p} + \sum_{s} \beta_{s} \psi_{s},
\end{align}
with the mixing coefficient 

\begin{align} 
    \beta_{s} = \frac{W_{sp}}{E_{p} - E_{s}},
\end{align}
where $W_{sp} = \mel{s}{W}{p}$ is the matrix element of the weak interaction mixing $s$-wave and $p$-wave resonances. We can present the parity violating effect in terms of the widths of the $s$-wave and $p$-wave resonances. Since the width $\Gamma^{n}$ is proportional to the amplitude squared, the amplitudes of $s$ and $p$-wave captures can be represented as $\sqrt{\Gamma_{s}^{n}} \eta_{s}$  and $  i \sqrt{\Gamma_{p}^{n}} \eta_{p}$ respectively where $\eta_{s}$ and $\eta_{p}$ are the sign factors of each amplitude (equal to $\pm 1$). The cross section $\sigma$ is proportional to the square of the amplitude:

\begin{align}
    \sigma_{\pm}  \propto \left| \pm i \sqrt{\Gamma_{p}^{n}}  \eta_{p} +   \sum_{s} \frac{W_{sp}}{E_{p} - E_{s}} \sqrt{\Gamma_s^{n}} \eta_{s} \right|,
\end{align}
where $\sigma_{\pm}$ are the cross sections for the positive and negative helicities. The $\pm$ sign in the $p$-wave amplitude corresponds to a positive or negative helicity, see Section \ref{CFSA}. It follows that the longitudinal asymmetry $P$ can be expressed as~\cite{Sushkov80}
	
\begin{align}\label{P}
P = \frac{\sigma_{+} - \sigma_{-}}{\sigma_{+} + \sigma_{-}} = 2 \sum_{s} \frac{i W_{sp}}{E_s-E_p} \sqrt{\frac{\Gamma_s^n}{\Gamma_p^n}},
\end{align}
where the sign coefficients $\eta_{s}$ and $\eta_{p}$ have been moved inside the matrix element $W_{sp}$. Note, in the standard defintion of the angular wave functions, $W_{sp}$ is imaginary. A proper derivation of (\ref{P}) including the case of a non-zero nuclear spin will be presented in Section \ref{CFSA}. Here we assume that the non-resonant part of $(\sigma_{+} + \sigma_{-})$ has been subtracted, as it has been done in the PV neutron transmission experiments.

Upon analysis of (\ref{P}), two reasons for the enhancement of the PV effect can be identified. Firstly, in a nucleus excited by neutron capture, the interval $E_{s} - E_{p}$ between the chaotic compound states (resonances) of opposite parity is very small. This contribution is labelled \textit{dynamical} enhancement, and it enhances the mixing of these states (by the weak PV interaction between nucleons) by three orders of magnitude. Secondly, the admixture between the large $s$-wave amplitudes $\sqrt{\Gamma_s^n}$ and the small $p$-wave amplitudes $\sqrt{\Gamma_p^n}$ allows neutron capture in the $s$-wave channel to contribute to the $p$-wave resonance. At small neutron energies the $s$-wave amplitude is three orders of magnitude larger than the p-wave amplitude ($ \sqrt{\Gamma_{s}^{n} / \Gamma_{p}^{n}} \sim 10^{3}$). This contribution is referred to as \textit{kinematic} enhancement. The ratio of the strength of the weak interaction to that of the strong interaction is $\sim 10^{-7}$. As a result of these two $10^{3}$ factors acting together, the factor of enhancement is as large as $\sim 10^{6}$, see~\cite{Sushkov80,Sushkov82,Flambaum84,Flambaum85,Alfimenkov1983}.

The relative difference of the neutron cross sections for  positive and negative helicities has been measured in a range of nuclei. These measurements have been done in various polarised neutron transmission experiments, by flipping the neutron spin orientation $\mathbf{s}$ from parallel to anti-parallel to neutron momentum $\mathbf{p}$ by a magnetic field pulse. Once the spin has been flipped, the number of neutrons passing through a material can be counted, i.e. measuring the correlation $\mathbf{s} \cdot \mathbf{p}$, see e.g.~\cite{MitchellReview2001}. 
Further, there has also been measurements of PV correlations in neutron radiative capture, $(n,\gamma)$ reactions. The theory of these correlations is presented in~\cite{Flambaum84,Flambaum85}.

\subsection{Calculation of the forward scattering amplitude with parity violation} \label{CFSA}
In this section, we follow the derivation in Ref.~\cite{Flambaum84}, in which the authors  use a resonance diagrammatic technique to calculate the forward elastic scattering amplitude with parity violation $f_{\text{p.v}}$. Let us consider the parity violating effects in neutron optics. The angle of neutron spin rotation around the direction of motion in matter is

\begin{align}
    \varphi = \frac{2 \pi N_{0} l}{k} 2 \  \text{Re} f_{\text{p.v}},
\end{align}
where $N_{0}$ is density of the target atoms, $l$ is the neutron path length,  $k$ is the neutron momentum. The difference in total cross sections $\Delta \sigma$ for right and left handed polarised neutrons can be expressed in terms of $f_{\text{p.v}}$

\begin{align}
    \Delta \sigma = \sigma_{+} - \sigma_{-} = \frac{4 \pi}{k} 2 \ \text{Im} f_{\text{p.v}},
\end{align}
The main contribution to the forward elastic scattering amplitudes comes from mixing by the PV weak interaction of the wave functions of  compound resonances. Other contributions, such as neutron scattering on the parity violating potential of the nucleus are not enhanced by small energy denominators, and may be neglected. 

We will now present a calculation of this resonance  amplitude. In this derivation, we will skip the summation over all $s$-wave and $p$-wave resonances for brevity. Let $\mathbf{n}_{k}$ denote the neutron's initial momentum direction and  $I$ be the angular momentum of the initial nucleus.  $\mathbf{J} = \mathbf{I} + \mathbf{j}$  is the spin of the compound resonance and $\mathbf{j} = \mathbf{l} + \mathbf{s}$ denotes the momentum of the $p$-wave neutron at which the capture occurs. Let us begin by writing down the wave-function of the incident neutron
\begin{align}
    e^{ikr} |\chi_{\alpha}> & = \sum_{l,m} 4 \pi i^{l} j_{l} (kr) Y^{*}_{lm} (\mathbf{n}_{k}) Y_{lm} (\mathbf{n}_{k}) |\chi_{\alpha}>,
\end{align}
where $|\chi_{\alpha}>$ is the neutron spinor with spin projection $\alpha$ and $j_{l}(kr)$ is the spherical Bessel function. In neutron-nucleus scattering, the capture amplitude of the neutron into the $s$-resonance is
	
\begin{align} \label{swave}
C^{J J_{z}}_{I I_{z} \frac{1}{2} \alpha} \eta_{s} \sqrt{\Gamma_{s}^{n} (E) },
\end{align}
while the capture amplitude of the neutron into the $p$-resonance is 

\begin{align} \label{pwaveamplitude}
  \sum_{j j_{z} m}  C^{J J_{z}}_{I I_{z} j j_{z} } C^{j j_{z}}_{1m \frac{1}{2} \alpha } \sqrt{4 \pi} \ Y_{1m}^{*} (\mathbf{n}_{k}) i \eta_{j} \sqrt{\Gamma_{p_{j}}^{n} (E) },
\end{align}
where $ C^{J J_{z}}_{I I_{z} \frac{1}{2} \alpha}$ is the Clebsch-Gordon coefficient; $\Gamma_{pj}^{n}$ is the neutron width corresponding to the emission of a neutron with momentum $ j $ and $\eta_{s,j} = \pm 1$ is, as defined above, the sign of the amplitude~\cite{Flambaum84}. We further define the Green function of a compound nucleus

\begin{align} \label{Green}
\frac{1}{E-E_{c}+\frac{1}{2} i \Gamma_{c}}.
\end{align}
Now, in conjunction with the above rules, we may write the forward elastic scattering amplitude near the $s$-resonance~\cite{Flambaum84};

\begin{align} 
f(0)=-\frac{1}{2 k} C_{I I_{z} \frac{1}{2} \alpha}^{J J_{z}} \sqrt{\Gamma_{s}^{n}(E)} \frac{1}{E-E_{s}+\frac{1}{2} i \Gamma_{s}} C_{I I_{z} \frac{1}{2} \alpha}^{J J_{z}} \sqrt{\Gamma_{s}^{n}(E)},
\end{align}
where $-1 / 2 k$ is the common factor for the scattering amplitudes, given the neutron momentum $k$. Next, summing over $J_{z}$ (using the standard relations for the coupling of angular momenta) and averaging over $I_{z}$, we obtain the standard Breit-Wigner Formula~\cite{Flambaum84}

\begin{align}
f(0)=-\frac{1}{2 k}  \frac{g \Gamma_{s}^{n}(E)}{E-E_{s}+\frac{1}{2} i \Gamma_{s}},
\end{align}
where the factor

\begin{align} \label{gfactor}
    g = \frac{(2 J+1)}{ 2(2 I+1)},
\end{align}
appears after averaging over the initial nucleus' spin projections. Performing a similar calculation for the $p$-wave resonance yields

\begin{align}
\begin{split}
f(0)& =-\frac{1}{2 k} \sum_{j j_{z} m \atop \tilde{j} \tilde{j_{z}} \tilde{\dot{m}}} C_{I I_{z} j j_{z}}^{J J_{z}} C_{1 m_{2}^{1} \alpha}^{j j_{z}} \sqrt{4 \pi} Y_{1 m}^{*}\left(\mathbf{n}_{k}\right) \sqrt{\Gamma_{p_{j}}^{n}(E)} \\
& \times \frac{1}{E-E_{p}+\frac{1}{2} \Gamma_{p}} C_{I I_{z} \tilde{j} \tilde{j_{z}}}^{J J_{z}} C_{1 m_{2}^{1} \alpha}^{\tilde{j} \tilde{j_{z}}} \sqrt{4 \pi} Y_{1 \tilde{m}}\left(\mathbf{n}_{k}\right) \sqrt{\Gamma_{p \tilde{j}}^{n}(E)},
\end{split}
\end{align}
which, after summation over $J_{z}$ and averaging over $I_{z}$ is again equal to 

\begin{align} \label{breitwignerp}
f(0)=-\frac{1}{2 k} \frac{g \Gamma_{p}^{n}(E)}{E-E_{p}+\frac{1}{2} i \Gamma_{p}},
\end{align}
where $\Gamma_{p}^{n}=\Gamma_{p_{1/2}}^{n}+\Gamma_{p_{3/2}}^{n}$. Using the above calculations as guides, we may now write down the forward elastic scattering amplitude with parity violation 
\begin{widetext}
\begin{align}
\begin{split}
f_{\mathrm{p . v .}}(0)& =-\frac{2}{2 k} C_{I I_{z} \frac{1}{2} \alpha}^{J J_{z}} \eta_{s} \sqrt{\Gamma_{s}^{n}(E)}  \frac{1}{E-E_{s}+\frac{1}{2} i \Gamma_{s}} \times  W_{sp}  \frac{1}{E-E_{p}+\frac{1}{2} \Gamma_{p}} \\ \times & (-i) \sum_{j j_{z} m} C_{I I_{z} j j_{z}}^{J J_{z}} C_{1 m \frac{1}{2} \alpha}^{i j_{z}} \sqrt{4 \pi} Y_{1 m}\left(\mathbf{n}_{k}\right) \eta_{j} \sqrt{\Gamma_{p j}^{n}(E)},
\end{split}
\end{align}
\end{widetext}
where $W_{sp}$ is the weak interaction matrix element between compounds states. Once again, we may sum over $J_{z}$ and average over $I_{z}$ to obtain~\cite{Flambaum84}

\begin{align} \label{FSA}
f_{\mathrm{p . v .}}(0)=\pm \frac{1}{2 k} \frac{2 g \sqrt{\Gamma_{s}^{n}(E)} i W_{sp} \sqrt{\Gamma_{p \frac{1}{2}}^{n}(E)} \eta_{s} \eta_{\frac{1}{2}}}{\left(E-E_{s}+\frac{1}{2} i \Gamma_{s}\right)\left(E-E_{p}+\frac{1}{2} i \Gamma_{p}\right)}.
\end{align}
The sign of this expression corresponds to the positive or negative helicity neutron (respectively). We also note that the term with $j=3/2$ vanishes after summation over $J_{z}$. The sign factor $\eta_{s} \eta_{1/2}$ can be excluded by means of the redefinition of the states $s$ and $p$ (i.e.  we introduce it into the matrix element $W_{sp}$). 

Note that all neutron widths are energy dependent, and should be taken at neutron energy $E$. If the energy $E$ is close to $p$-wave resonance, then they must take the form $\Gamma_{s}^{n}(E_{p})$ and $\Gamma_{p}^{n}(E_{p})$.

Equation (\ref{FSA}) can now be related to the total cross section via the optical theorem

\begin{align}
    \sigma = \frac{4 \pi}{k} \text{Im} f(0).
\end{align}
Upon application of the optical theorem to the forward scattering amplitude with parity violation, we yield the predicted longitudinal asymmetry (\ref{P}).

\subsection{Forward scattering amplitude with time reversal invariance and parity violation} \label{TPFSA}
In this section, we will present our calculation of the forward scattering amplitude of neutron-nucleus scattering with time and parity violation. 
In the case of PTRIV effects, the spin of the target nucleus and neutron momentum are perpendicular. As such, it is convenient to set the nuclear target spin $I$ along the $z$-axis, thus $I = I_{z}$. Neutron spin is along $x$-axis, so the spinor has equal amplitudes 1/2 and -1/2 along $z$ -axis, $\frac{1}{\sqrt{2}} [ \ket{s_z=1/2} + \ket{s_z=-1/2}]$, and neutron momentum is along $y$ axis, therefore we substitute $\theta=\pi/2$ and $\phi=\pi/2$ as the arguments of the $ \ Y_{1m}^{*} (\theta,\phi)$ in Eq. (\ref{pwaveamplitude}). In this case, we have contributions to the capture amplitude from both the $s$-wave and $p$-wave resonances, with a total spin of $J = I + 1/2$ or $J = I - 1/2$. We will make the substitutions $M_{s}  = \sqrt{\Gamma_{s}^{n}} \eta_{s}, M_{p,j}  =  \sqrt{\Gamma_{p,j}^{n}} \eta_{j}$ for brevity. Following the method described above, we can write the time and parity violating amplitude $f_{\text{t.p.v}}$

\begin{align} \label{FSATP} 
    f_{\text{t.p.v}} & = \pm \frac{1}{k} \frac{M_s W_{sp}^{T,P}  (\alpha_{1/2,J}  M_{p,1/2} + \alpha_{3/2,J} M_{p,3/2})}
{\left(E-E_{s}+\frac{1}{2} i \Gamma_{s}\right)\left(E-E_{p}+\frac{1}{2} i \Gamma_{p}\right)},
\end{align}
where the angular coefficients $\alpha_{i.J} $ are
\begin{align} \label{alphacoeff}
\begin{split}
\alpha_{1/2,J=I+1/2}& = \frac{I}{2I+1}, \\
\alpha_{3/2,J=I+1/2}& = - \frac{\sqrt{I(2I+3)}}{2(2I+1)}, \\
\alpha_{1/2,J=I-1/2}& = \frac{I}{2I+1}, \\
\alpha_{3/2,J=I-1/2}& = -\frac{I}{2(2I+1)} \sqrt{\frac{2I-1}{I+1}}.
\end{split}
\end{align}
The calculation of this amplitude is presented in Appendix \ref{CFSAPTRIV}. In this amplitude we should sum over s-wave resonances (if the energy is close to a p-wave resonance). However, for comparison with Ref.~\cite{Gudkov2018} it is instructive to  present the ratio of the two scattering amplitudes $f_{\text{t.p.v}}$ (\ref{FSATP}) and $f_{\text{p.v}}$ (\ref{FSA}) taking into account only one s-wave resonance:

\begin{align}
  \frac{f_{\text{t.p.v}}}{f_{\text{p.v}}} & = \kappa \frac{W_{sp}^{T,P}}{W_{sp}},
\end{align}
where the angular coefficient of the ratio $\kappa$ includes amplitudes of the partial neutron widths which depend on spin channels $J = I \pm 1/2$, and can be determined via our calculations above;

\begin{align}
    \kappa(I \pm 1/2) = \frac{ \left(\alpha_{1/2,J}  M_{p,1/2} + \alpha_{3/2,J} M_{p,3/2} \right)}{g  M_{p,1/2}},
\end{align}
where $g$ is as defined by Equation (\ref{gfactor}), 

\begin{align}
    g = \frac{2J + 1}{2(2I+1)} = \begin{cases} 
      \frac{I+1}{2I+1}, & J = I + \frac{1}{2}, \\
      \frac{I}{2I+1} ,& J = I - \frac{1}{2}.
   \end{cases}
\end{align}
Hence, calculations yield

\begin{align}
\begin{split} \label{kappaplus}
   \kappa(I+1/2) & = \frac{I}{I+1} - \frac{M_{p,3/2}}{M_{p,1/2}} \frac{\sqrt{I(2I+3)}}{2(I+1)},
\end{split} \\
\begin{split} \label{kappaminus}
   \kappa(I-1/2) & = 1 - \frac{M_{p,3/2}}{M_{p,1/2}} \frac{\sqrt{2I-1}}{ 2\sqrt{I+1}}.
\end{split}
\end{align}
Our results for the factors $\kappa (I \pm 1/2)$ seem to be in agreement with the values calculated by in Ref. ~\cite{Gudkov2018}, however their calculations do not include the sign factors $\eta_{s}, \eta_{j}$ from the amplitudes $M_{s}  = \sqrt{\Gamma_{s}^{n}} \eta_{s}$ and $M_{p,j}  =  \sqrt{\Gamma_{p,j}^{n}} \eta_{j}$ (their results are expressed in terms of $\sqrt{\Gamma_{s}^{n}}$ and $\sqrt{\Gamma_{p,j}^{n}}$ assuming that both  $M_{s}$ and $M_{p,j}$ are  positive; this assumption  is not justified). 
Ref. ~\cite{Gudkov2018} limits their calculation to one s-wave resonance. This approximation is not justified in the case of the statistical theory. Finally, the experiments measure the relative difference $P$ of the number of passing neutrons for opposite orientation of neutron spin. The relative difference contains the $p$-wave amplitude in the denominator, which also should be calculated.


\subsection{Forward scattering amplitude for parity violation with a polarised target} \label{AmplitudePVPolarised}

Now, in a similar way, we may also calculate the parity violating forward scattering amplitude for a polarised target. In this case, we require neutron momentum and spin to be parallel, and both perpendicular to the nuclear target spin. Setting once again nuclear target spin along the $z$-axis, we have $I = I_{z}$. We once again have neutron spin along the $x$-axis, meaning the spinor is $\frac{1}{\sqrt{2}} [ \ket{s_{z} = 1/2} + \ket{s_{z} = -1/2}]$. However now, we require neutron momentum to also be along the $x$-axis, and thus we substitute $\theta = \pi/2$ and $\phi = 0$ as the arguments of the $Y_{1m}^{*} (\theta, \phi)$. 
Using a similar method to Section \ref{TPFSA}, we yield the following for the forward scattering amplitude with parity violation with a polarised target (see Appendix \ref{PVpolarisedAmplitude})

\begin{align} 
    f_{\text{p.v}}^{\text{P}} & = \pm \frac{1}{ k} \frac{M_s i W_{sp}  (\delta_{1/2,J}  M_{p,1/2} + \delta_{3/2,J} M_{p,3/2})}
{\left(E-E_{s}+\frac{1}{2} i \Gamma_{s}\right)\left(E-E_{p}+\frac{1}{2} i \Gamma_{p}\right)},
\end{align}
where the angular coefficients $\delta_{i.J} $ are
\begin{align} \label{delta}
\begin{split}
\delta_{1/2,J=I+1/2}& = -\frac{I+1}{2I+1}, \\
\delta_{3/2,J=I+1/2}& = -\frac{2I-1}{2(2I+1)} \sqrt{\frac{I}{2I+3}}, \\
\delta_{1/2,J=I-1/2}& = -\frac{I}{2I+1}, \\
\delta_{3/2,J=I-1/2}& = \frac{I}{2(2I+1)} \sqrt{\frac{2I-1}{I+1}}.
\end{split}
\end{align}

\subsection{Ratio of PTRIV and PV effects in experiments with a polarised target}

Now  we can present the ratio of the PTRIV and PV forward scattering amplitudes for a polarised target. We start from the results in the two-resonance approximation:

\begin{align}
    \frac{f_{\mathrm{t.p.v}}}{f_{\mathrm{p.v}}^{P}} = \frac{W_{sp}^{T,P}}{i W_{sp}} \left[   \frac{\alpha_{1/2,J}  M_{p,1/2} + \alpha_{3/2,J} M_{p,3/2}}{\delta_{1/2,J}  M_{p,1/2} + \delta_{3/2,J} M_{p,3/2}} \right].
\end{align}
Specifically, for $J=I+1/2$, we have

\begin{align}  \label{polarisedPTPVJ+}
\begin{split}
    \frac{f_{\mathrm{t.p.v}}}{f_{\mathrm{p.v}}^{P}}  = \\ 
    \frac{W_{sp}^{T,P}}{i W_{sp}} \frac{-2I \sqrt{2I+3} M_{p,1/2} + (2I+3) \sqrt{I} M_{p,3/2}}{2(I+1) \sqrt{2I+3} M_{p,1/2} + (2I-1) \sqrt{I} M_{p,3/2}},
\end{split}
\end{align}
while for $J=I-1/2$ we have a very simple result,

\begin{align} \label{polarisedPTPVJ-}
    \frac{f_{\mathrm{t.p.v}}}{f_{\mathrm{p.v}}^{P}} = -\frac{W_{sp}^{T,P}}{i W_{sp}} 
\end{align}
A more accurate treatment requires summation over $s$-wave resonances in the numerator and denominator since the PTRIV and PV matrix elements are not proportional to each other (according to our statistical theory calculation \cite{Flambaum1995b},  the relative correlator between $W_{sp}^{T,P}$ and $W_{sp}$ is 0.1). Therefore, in the above expressions, we should make the substitution
\begin{align}
\frac{W_{sp}^{T,P}}{i W_{sp}} \,\, \rightarrow  \,\,  \frac{\sum_{s}  A_{s p}W_{sp}^{T,P}}{\sum_{s} A_{s p} i W_{sp}}\,,
\end{align}
where
\begin{align} \label{Asp}
     A_{s p} = 2 \frac{1}{E_{s} - E_{p}} \sqrt{\frac{\Gamma_{s}^{n}}{\Gamma_{p}^{n}}}.
\end{align}
The width $\Gamma_{p}^{n}$ is a common factor and may be cancelled. However, after statistical averaging the ratio of PTRIV and PV effects for $J=I-1/2$  will again be given by a simple expression containing  only root-mean-squared values of the PTRIV and PV matrix elements (see below).  Eq. (\ref{polarisedPTPVJ-}) is also valid for the octupole doublet and doorway state mechanisms (the permanent sign contribution) where the two-level approximation is justified (see below).

\subsection{$p$-wave amplitude for a polarised target (PTRIV and PV configurations)} \label{pampPTRIV}

\label{pwave} 

We will now present our results for the $p$-wave amplitude in the configurations above which use a polarised target. Firstly, let us consider the case when neutron momentum, neutron spin and target spin are all perpendicular to each other, i.e. for the  configuration presented in section \ref{TPFSA}. This calculation was performed using the method presented in Section \ref{CFSA}, from which we yield (see Appendix \ref{pwavepolarised}) 

\begin{align} \label{pwaveamplitudeP}
    f_{p} = -\frac{1}{2k} \frac{\beta_{1,J} M_{p,1/2}^{2} + \beta_{13,J} M_{p,1/2} M_{p,3/2} + \beta_{3,J} M_{p,3/2}^{2} }{E - E_{p} + \frac{1}{2} i \Gamma_{p} },
\end{align}
where, in the case when $J = I+1/2$

\begin{align}
\begin{split}
    \beta_{1,J= I+1/2} & = \frac{I + 1}{2I+1} = g, \\
    \beta_{13,J= I+1/2} & = \frac{\sqrt{I}}{\sqrt{2I+3}} \frac{2I-1}{2I+1} , \\
    \beta_{3,J= I+1/2} & = \frac{2I^{2} + 5I + 9}{2(2I+3)(2I+1)},
\end{split}
\end{align}
and in the case when $J=I-1/2$

\begin{align}
\begin{split}
    \beta_{1,J = I - 1/2} & = \frac{I}{2I+1} = g, \\
    \beta_{13,J = I - 1/2} & = -\frac{I}{2I+1} \sqrt{\frac{2I-1}{I+1}}, \\
    \beta_{3,J = I - 1/2} & = \frac{I(I+4)}{(2I+1)(2I+3)}.
\end{split}
\end{align}
Next, we consider the case when neutron momentum and neutron spin are parallel to each other, and both perpendicular to the nuclear target spin (the configuration presented in section \ref{AmplitudePVPolarised}). A similar calculation verifies that the $p$-wave amplitude in this configuration coincides with Eq. (\ref{pwaveamplitudeP}) above.

\subsection{Ratio of $p_{1/2}$ and $p_{3/2}$ capture amplitudes}
The expressions presented above contain the unknown ratio of the $p_{1/2}$ and $p_{3/2}$ capture amplitudes, $T=M_{p,3/2}/M_{p,1/2}$. This ratio may be extracted via measurement of the ratio of the total $p$-wave resonance  cross sections (transmission probabilities) for the polarised and unpolarised target, i.e. from the ratio of the forward scattering amplitudes in Eqs. (\ref{pwaveamplitudeP}) and  (\ref{breitwignerp}):
\begin{equation}
\frac{\sigma(polarised)}{\sigma(unpolarised)}= \frac{1}{g}(\beta_{1,J} R_{1/2}^{2} + \beta_{13,J} R_{1/2} R_{3/2} + \beta_{3,J} R_{3/2}^{2})\,.
\end{equation}
where
\begin{align}\label{Rj}
    R_{j} = \frac{ M_{p,j}}{\sqrt{ M^{2}_{p_{1/2}} +M^{2}_{p_{3/2}}  }}.
\end{align}
Here $M_{p,1/2}^{2} + M_{p,3/2}^{2} = M_p^2=\Gamma_{p}^{n}$. Note that $R_{1/2}^{2} + R_{3/2}^{2} = 1$. This ratio $T=M_{p,3/2}/M_{p,1/2}$ may also  be extracted from the ratio of the PV forward scattering amplitudes for a polarised and unpolarised target, which is equal to
\begin{equation}
\frac{\Delta \sigma(polarised)}{\Delta \sigma(unpolarised)}= \frac{1}{g} \left[ (\delta_{1/2,J}  + \delta_{3/2,J} T) \right].
\end{equation}

\section{Statistical analysis of parity Violation and time reversal violation in compound Nuclei}\label{3}

The statistical theory predicts mean squared values of the amplitudes. However, if we treat the energy intervals between the opposite parity energy levels $E_s - E_p$ as random variables, which have a finite probability density to be zero, we obtain a meaningless infinite result for the variance of the effect $\ev{P^2}$ in Eq. (\ref{P}).
Inclusion of the widths $\Gamma_s$ into the energy denominators makes   $\ev{P^2}$ finite but so big that one would require many thousands of measurements on different compound resonances to find $\ev{P^2}$ from experiments~\cite{FlamGrib94}. The solution is to take the energy intervals from experimental data, i.e. do not treat them as random variables.

\subsection{Parity violation for zero nuclear spin}  

In this section, we follow Bowman et al.~\cite{PhysRevLett.65.1192} in their analysis using the statistical mechanism of parity violation in nuclear resonances, the effects of which are produced by the mixing of opposite parity states. As such, individual weak matrix elements between $s$- and $p$-wave resonances can be considered to be mean-zero Gaussian random variables, with variance $W^{2} = \expval{ | W_{sp} |^{2}}$. 

Let us first consider the case of targets with spin $I=0$. In such cases, we have for $s$-wave resonances, $J = 1/2^{+}$, while for $p$-wave resonances, $J = 1/2^{-}, 3/2^{-}$. 
As $p$-wave resonances with $3/2^{-}$ cannot mix with the $1/2^{+}$ $s$-wave levels via a parity violating interaction, these levels will not show parity violation.

For a given $p$-wave level, the observed asymmetry has contributions from many $s$-wave levels, as per Eq. (\ref{P}). We may rewrite this equation as
\begin{align} \label{pmu}
    P = \sum_{s} i A_{s p}  W_{s p},
\end{align}
where

\begin{align} 
     A_{s p} = 2 \frac{1}{E_{s} - E_{p}} \sqrt{\frac{\Gamma_{s}^{n}}{\Gamma_{p}^{n}}},
\end{align}
in order to separate factors  which are taken from experimental data and not affected by averaging. 
Squaring Equation (\ref{pmu}) yields

\begin{align}
\begin{split}
    P^{2}& = \left| \sum_{s}A_{s p} W_{sp} \right|^{2}\\
    & = \sum_{s} A_{sp}^{2} |W_{sp}|^{2} + \sum_{s \neq i} A_{ip} A_{sp} W_{ip} W_{sp}, \\
     \frac{P}{A_{J}^{2}} & = \frac{1}{A_{J}^{2}}\sum_{s} A_{sp}^{2} |W_{sp}|^{2} + \frac{1}{A_{J}^{2}} \sum_{s \neq i} A_{ip} A_{sp} W_{ip} W_{sp},
\end{split}
\end{align}
where we define $A_{J}^{2} \equiv \sum_{s} A_{s p}^{2}$. Now, given that each $W_{s p}$ is statistically independent, upon averaging over matrix elements the cross terms vanish. We can rewrite this expression in terms of $W$~\cite{PhysRevLett.65.1192}

\begin{align}
   W^{2} & = \expval{\frac{P^{2}}{A_{J}^{2}}} .
\end{align}
Thus, we conclude that each $P/A_{J}$ is also a Gaussian random variable, with mean zero and variance $|W|^{2}$. 
This means that given several experimental measurements for the quantity $P$, $\expval{|W_{sp}|^{2}} \equiv W^{2} $ can be extracted.

\subsection{Parity violation for an unpolarised  target with a non-zero spin} \label{PVNTS}

Let us now consider neutron scattering from an unpolarised target with nonzero spin, $I \neq 0$. Given parity $\pi$, these targets have spin and parity $( I \pm 1/2)^{\pi}$ in $s$-wave levels and $( I \pm 1/2)^{-\pi}, ( I \pm 3/2)^{-\pi}$ in $p$-wave levels. As the weak interaction is a scalar, angular momentum conservation implies only $I \pm 1/2$ has a nonzero P-odd effect. 
Only the $j=1/2$ $p$-wave capture amplitude contributes to the effects of parity violation, see Section \ref{CFSA}. The projectile $\mathbf{j}$ is further coupled to the spin of the target nucleus $\mathbf{I}$ to form the total spin $\mathbf{J} = \mathbf{j} + \mathbf{I}$. 
Once again defining $  M_{s} \equiv \eta_{s} \sqrt{\Gamma_{s}^{n}}$ and $ M_{p} \equiv  \eta_{p}\sqrt{\Gamma_{p}^{n}}$ as the neutron decay amplitudes of the levels $s$ and $p$ respectively, we can rewrite Eq. (\ref{P}) as (see Section \ref{CFSA})

\begin{align} \label{PIneq0}
    P =  R_{1/2}\sum_{s} \frac{2 iW_{s p}}{E_{s} - E_{p}} \sqrt{\frac{\Gamma_{s}^{n}}{\Gamma_{p}^{n}}},
\end{align}
where $R_{1/2}$ is given by Eq. (\ref{Rj}).
Let us rewrite Equation (\ref{PIneq0}) in the form
\begin{align}
    P & = \sum_{s} i \tilde{W}_{sp} A_{sp}, 
\end{align}
where $\tilde{W}_{sp} \equiv W_{sp} R_{1/2}$ and coefficients   $A_{s p}$ are given by Eq. (\ref{Asp}).
%
Thus, in a similar method to the case when $I=0$, we obtain
\begin{align}
    \expval{|\tilde{W}_{sp}|^{2}} & = \expval{ \frac{P^{2}}{A_{J}^{2}}}, 
\end{align}
where $A_{J}^{2} \equiv \sum_{s} A_{s p}^{2}$. Assuming that $W_{sp}$ and $R_{1/2}$ are statistically independent, their product can be averaged separately. Then, the average squared matrix elements between compound states are

\begin{align}
    W^{2} & =  \frac{1}{\expval{R_{1/2}^{2}}} \expval{ \frac{P^{2}}{A_{J}^{2}}}.
\end{align}
$\expval{R_{1/2}^{2}}$ can be extracted experimentally, and for masses near $A = 110$, it typically lies between 0.6 and 0.8~\cite{MitchellReview2001}. 

\subsection{Time reversal and parity violation for a polarised target} \label{TPVNTS}

In this section, we consider the PTRIV configuration, and perform a similar calculation to that above. Firstly, we note that the $p$-wave forward scattering amplitude for neutron momentum, neutron spin and target spin perpendicular to each other, does not coincide with Eq. (\ref{breitwignerp}). As per the result presented in section \ref{pwave}, the $p$-wave forward scattering amplitude in this case is given by Equation (\ref{pwaveamplitudeP}). Using the calculation of the PTRIV amplitude in section \ref{TPFSA}, we can now write PTRIV effect in a similar form to that of the unpolarised P-odd, T-even effect (\ref{P})
\begin{widetext}
\begin{align} \label{PTexpression}
\begin{split}
    P_{T,P} & =  \left[ \frac{(\alpha_{1/2,J} M_{p,1/2} + \alpha_{3/2,J} M_{p,3/2}) M_{p}}{\beta_{1,J} M_{p,1/2}^{2} + \beta_{13,J} M_{p,1/2} M_{p,3/2} + \beta_{3,J} M_{p,3/2}^{2} }  \right] \sum_{s}  \frac{W^{T,P}_{sp}}{E_{s} - E_{p}}  \sqrt{\frac{\Gamma_{s}^{n}}{\Gamma_{p}^{n}}}, \\
    \end{split}
\end{align}
\end{widetext}
where $\alpha_{p,J}$ are the angular factors for the PTRIV amplitude in Eq. (\ref{alphacoeff})  and $W_{sp}^{T,P}$ is the matrix element of the PTRIV interaction. We can further rewrite this equation in the familiar form


\begin{align} \label{PTPProduct}
     P_{T,P} = \sum_{s} \tilde{W}_{sp}^{T,P} A_{sp},
\end{align}
where $A_{sp}$ is as defined in (\ref{Asp}), and  

\begin{align}
\nonumber
    \tilde{W}_{sp}^{T,P} = \left[ \frac{\alpha_{1/2,J} R_{1/2} + \alpha_{3/2,J} R_{3/2}}{\beta_{1,J} R_{1/2}^{2} + \beta_{13,J} R_{1/2} R_{3/2} + \beta_{3,J} R_{3/2}^{2} }  \right]  W_{sp}^{T,P},
\end{align}
where $R_{j}$ is as defined in Eq. (\ref{Rj}). Once again, we aim to determine the average of this relation. In the same way as above, we yield

\begin{align}
    \expval{(\tilde{W}_{sp}^{T,P})^{2}} = \expval{ \frac{P_{T,P}^{2}}{A_{J}^{2}}}.
\end{align}
Assuming that $R_{1/2}, R_{3/2}$ and $W_{sp}^{T,P}$ are statistically independent, the products $R_{1/2}^{2} W_{sp}^{T,P}$ and $R_{3/2}^{2} W_{sp}^{T,P}$  can be averaged separately, and using the same method as Section \ref{PVNTS}, we may determine the average squared matrix elements of the T,P-odd interaction. Neglecting the terms linear in $R_{1/2}$ or $R_{3/2}$ (as they have a random sign), assuming that $\expval{R_{j}^{4}} \approx \expval{R_{j}^{2}}^{2}$ and averaging the numerator and denominator separately gives
\begin{widetext}
\begin{align}
\begin{split}
    \expval{(\tilde{W}_{sp}^{T,P})^{2}} & \approx \left[ \frac{\alpha_{1/2,J}^{2} \expval{R_{1/2}^{2}} + \alpha_{3/2,J}^{2} \expval{R_{3/2}^{2}}}{\beta_{1,J}^{2} \expval{R_{1/2}^{2}}^{2} + \beta_{13,J}^{2} \expval{R_{1/2}^{2}} \expval{R_{3/2}^{2}} + \beta_{3,J}^{2} \expval{R_{3/2}^{2}}^{2} }  \right]  \expval{(W_{sp}^{T,P})^{2}}.
\end{split}
\end{align}
Thus, we conclude that the average squared matrix elements of time and parity violating effects are

\begin{align} \label{experimentPT}
    (W^{T,P})^{2} & \approx \left[ \frac{\beta_{1,J}^{2} \expval{R_{1/2}^{2}}^{2} + \beta_{13,J}^{2} \expval{R_{1/2}^{2}} \expval{R_{3/2}^{2}} + \beta_{3,J}^{2} \expval{R_{3/2}^{2}}^{2}}{\alpha_{1/2,J}^{2} \expval{R_{1/2}^{2}} + \alpha_{3/2,J}^{2} \expval{R_{3/2}^{2}}} \right] \expval{ \frac{P_{T,P}^{2} }{A_{J}^{2}}}.
\end{align}
Note that $R_{1/2}^{2} + R_{3/2}^{2} = 1$. According to \cite{MitchellReview2001}, $\ev{R_{1/2}^{2}}\simeq 0.7$ and $\expval{R_{3/2}^{2}} \simeq 0.3 $ for masses near A=110. 
\end{widetext}

\subsection{Parity violation for a polarised target}
We also may perform a similar calculation for the parity violating interaction with a polarised target. Here, we once again note that the $p$-wave forward scattering amplitude in this configuration coincides with Eq. (\ref{pwaveamplitudeP}). Thus, we may write the PV effect, and its corresponding mean-squared matrix elements for a polarised target (replacing the coefficients $\alpha_{j,J}$ with $\delta_{j,J}$ in Equations (\ref{PTexpression}) and (\ref{experimentPT}))

\begin{widetext}
\begin{align}
    P_{\text{polarised}} & =  \left[ \frac{(\delta_{1/2,J} M_{p,1/2} + \delta_{3/2,J} M_{p,3/2}) M_{p}}{\beta_{1,J} M_{p,1/2}^{2} + \beta_{13,J} M_{p,1/2} M_{p,3/2} + \beta_{3,J} M_{p,3/2}^{2} }  \right] \sum_{s}  \frac{iW_{sp}}{E_{s} - E_{p}}  \sqrt{\frac{\Gamma_{s}^{n}}{\Gamma_{p}^{n}}}, \\
 |W|^{2} & \approx \left[ \frac{\beta_{1,J}^{2} \expval{R_{1/2}^{2}}^{2} + \beta_{13,J}^{2} \expval{R_{1/2}^{2}} \expval{R_{3/2}^{2}} + \beta_{3,J}^{2} \expval{R_{3/2}^{2}}^{2}}{\delta_{1/2,J}^{2} \expval{R_{1/2}^{2}} + \delta_{3/2,J}^{2} \expval{R_{3/2}^{2}}} \right] \expval{ \frac{P_{\text{polarised}}^{2} }{A_{J}^{2}}}.
\end{align}
\end{widetext}

\section{ Statistical theory of finite systems based on the properties of chaotic eigenstates.}\label{4}

The number of combinations for the distribution of $n$ particles over $m$ orbitals, $m!/[n!(m-n)!]$, increases exponentially with the number of particles. Therefore, in compound  states with several excited particles, the density of energy levels is exponentially high, and the residual interaction between the particles exceeds the energy intervals. As a result, the excited states $|n\rangle$ in all medium and heavy nuclei near the neutron separation energy (as well as in atoms and ions with several excited electrons in an open f-shell) are chaotic superpositions of thousands or even millions of Hartree-Fock basis states $|i\rangle$, $|n\rangle = \sum_i C_i^n |i\rangle$. 
Chaos allows us to  develop a statistical theory, including a method  to calculate the matrix elements between chaotic states in finite systems (in excited nuclei, atoms and molecules) \cite{Flam93,Flam93PRL,PhysRevE.56.5144,PhysRevA.91.052704}.

Following Ref.~\cite{NuclearStructure}, we treat the expansion coefficients $C_i^n$  as Gaussian random variables, with average values $\overline {C_i^n}=0$ and variance 
\begin{align} \label{TheAboveFunction1}
\nonumber 
    \overline{C^{2}(E_{\alpha})} & = \frac{1}{\overline{N}} \Delta (\Gamma_{\text{spr}}, E - E_{\alpha}), \\ 
    \Delta (\Gamma_{\text{spr}}, E - E_{\alpha}) & = \frac{\Gamma_{\text{spr}}^{2}/4}{(E - E_{\alpha} )^{2} + \Gamma_{\text{spr}}^{2}/4}. 
\end{align}
where  $\overline{N}= \frac{\pi \Gamma_{\text{spr}}}{2d}$ 
is the number of the principal components in the compound state found from the normalization condition $\sum_i (C_i^n)^2 =1$,  $\Gamma_{spr}$ is the spreading width calculated using Fermi's golden rule and $d$ is the average energy interval between compound states with the same angular spin and parity (see details in Appendix \ref{AppendixB}). 

The function $\overline{(C_i^n)^2}\equiv f(E^n - E_i)$ gives the probability to find the basis component $\ket{i}$ in the compound state $\ket{n}$,  i.e. it plays the role of the statistical partition function. The difference from conventional statistical theory is that the partition function depends on the total energy of the isolated system $E^n$ instead of temperature for a system in a thermostat (recall the Boltzmann factor $\exp(-E_i/T)$).  One may compare this with the microcanonical distribution where equipartition is assumed within the shell of the states with fixed energy $E_i$. Expectation values of matrix elements of any operator $\hat O$ in a chaotic compound state are found as 
\begin{equation}
\overline{|\langle n|\hat O|n \rangle|}=\sum_{i} \overline{(C_i^n)^2} |\langle i|\hat O|i \rangle|^2.
\end{equation}
For example, substituting the occupation number operator $\hat \nu=a^{\dagger}_k a_k$ into this expression gives the distribution of the orbital occupation numbers $\nu$ in finite chaotic systems which replaces the Fermi-Dirac (or Bose-Einstein) distribution. The average values of the non-diagonal matrix elements of any perturbation operator $W$ are equal to zero, $\overline{\langle n|W|m \rangle}=0$, while the average values of the squared matrix elements 
\begin{equation}\label{singleW}
\overline{|\langle n|W|m \rangle |^2}=\sum_{i,j} \overline{(C_i^n)^2} \overline{(C_j^m)^2}\, |\langle i|W|j \rangle |^2
\end{equation}
are reduced to the sum over simple matrix elements between the Hartree-Fock states $|\langle i|W|j \rangle|^2$. A convenient formula for the  root-mean-squared values of the  matrix elements has been derived in Ref. ~\cite{Flam93PRL} (we also present the derivation in Appendix \ref{AppendixB}):
\begin{widetext}
\begin{align} \label{PMSME1}
\begin{split}
   W\equiv \sqrt{\overline{|W_{sp}|^{2}}}=\sqrt{\frac{2 d}{\pi \Gamma_{\mathrm{spr}}}}\bigg\{\sum_{a b c d} \nu_{a}\left(1-\nu_{b}\right) \nu_{c}\left(1-\nu_{d}\right) \frac{1}{4}\left|\tilde{W}_{a b, c d}-\tilde{W}_{a d, c b}\right|^{2}  \times \Delta\left(\Gamma_{\mathrm{spr}}, \varepsilon_{a}-\varepsilon_{b}+\varepsilon_{c}-\varepsilon_{d}\right)\bigg\}^{\frac{1}{2}},
\end{split}
\end{align}
\end{widetext}
where the summation goes over orbitals $a,b,c,d$;  $E-E_{\alpha} = \varepsilon_{a}-\varepsilon_{b}+\varepsilon_{c}-\varepsilon_{d}$ is the change in energy. The function $\Delta\left(\Gamma_{\mathrm{spr}}, \varepsilon_{a}-\varepsilon_{b}+\varepsilon_{c}-\varepsilon_{d}\right)$, defined in Eq. (\ref{TheAboveFunction1}),
can be viewed as an approximate energy conservation law, with accuracy up to the spreading width of the basis states~\cite{Flam93PRL}. Indeed, 
$ \Delta(\Gamma_{\text{spr}}, E - E_{\alpha}) \rightarrow \frac{\pi \Gamma_{\text{spr}}}{2} \delta (E - E_{\alpha})$,
when $\Gamma_{\text{spr}} \rightarrow 0$. 

In Refs. \cite{Flam93,Flam93PRL,Flambaum1994V} we calculated the PV matrix elements. The matrix elements of the  P,T- violating interactions have been obtained in our papers \cite{Flambaum1995a,Flambaum1995b,Fadeev}. 

Let us now consider the correlator between two different operators (e.g. P-violating and T,P-violating). In general, we obtain \cite{Flambaum1995b}  
\begin{align}
\nonumber
\overline{\mel{n}{W_{P}}{m}  \mel{m}{W_{T,P}}{n}  } = \\ 
\sum_{i,j} \overline{(C_i^n)^2} \overline{(C_j^m)^2}  \times \mel{i}{W_{P}}{j}   \mel{j}{W_{T,P}}{i}.
\end{align}
Note that our theory predicts the results averaged over several compound resonances.

We have done many tests comparing the statistical theory results with both experimental data and with numerical simulations for electromagnetic amplitudes, electron recombination rates, and parity violation effects in nuclei - see e.g. Refs. \cite{Flam93,Flam93PRL,Flambaum1994V,FlamGrib2000,FlamGribakin1994,DzubaBerengut2017,Flambaum1998StatisticsOE,FlamGribakin2002,Harabati2017,PhysRevA.91.052704}. For example, we obtained an enhancement (of the order of $10^{3}$) of the electron recombination rate with highly charged tungsten ions (charge $q=18$--25) due to the very dense spectrum of chaotic compound resonances \cite{DzubaBerengut2017,FlamGribakin2002,Harabati2017,PhysRevA.91.052704}. The results agree with experimental data that is only available for lower charge $q=18$--21 ions. These results are important for the thermonuclear reactors in which the diverters are made from tungsten. The tungsten ions contaminate the plasma and significantly affect the energy output.

\section{Matrix elements of PV and PTRIV interactions between nuclear compound states} \label{5}

In this section we present a brief summary of the calculations of the PV and PTRIV matrix elements between chaotic compound states performed in Refs. \cite{Flam93,Flam93PRL,Flambaum1994V,Flambaum1995a,Flambaum1995b,Fadeev}. The details are presented in Appendix \ref{AppendixB}. The parity violating weak potential of nucleons in a nucleus may be presented as
\begin{align} \label{WeakInteraction}
    \hat W= \frac{Gg_{p,n}}{2 \sqrt{2}m} \{ (\sigma \mathbf{p}), \rho   \},
\end{align}
where $G$ is the Fermi constant, $m$ is the mass of the nucleon, $\sigma$ and $\mathbf{p}$ are the neutron's sigma matrix (doubled spin operator)  and momentum respectively, $\rho$ is the  nuclear number density and $g_{p,n}$ are the nucleon dimensionless constants which are of the order of unity. 
The calculation described in Appendix \ref{AppendixB} gives the following result for the root-mean-squared value of the matrix element between compound states:
\begin{align}\label{WP}
    W & = 0.57\ \text{meV}\sqrt{g_n^2 + 0.76 g_p^2}.
\end{align}
The values of $W$ are actually proportional to  $(\overline{N})^{-1/2} \propto d^{1/2}$, where $d$ is the average interval between resonances with the same spin and parity, which determines the number of principal components in the compound state, $\overline{N} = \frac{\pi \Gamma_{\text{spr}}}{2d}$ - see details in Appendix \ref{AppendixB}. This specific number for $W$ has  been calculated  for $d=17$ eV in $^{233}$Th. A more universal parameter is the weak spreading width $\Gamma_W= 2 \pi W^2/d$, where the dependence on $d$ cancels out.

The constants  $g_{p}$ and $g_{n}$ may be expressed in terms of the weak nucleon-meson interaction constants $h$ and $f$ ~\cite{Flam93,FLAMBAUM1984367,DESPLANQUES1980449,PhysRevC.56.1641}

\begin{align}
\begin{split}
g_{p}& = 2 \times 10^{5} v_{\rho}\biggl[176 \frac{v_{\pi}}{v_{\rho}} f_{\pi}-19.5 h_{\rho}^{0}-4.7 h_{\rho}^{1} \\ &  +1.3 h_{\rho}^{2}-11.3\left(h_{\omega}^{0}+h_{\omega}^{1}\right)\biggr], \\
\end{split} \\
\begin{split}
g_{n} & = 2 \times 10^{5} v_{\rho}\biggl[-118 \frac{v_{\pi}}{v_{\rho}} f_{\pi}-18.9 h_{\rho}^{0}+8.4 h_{\rho}^{1} \\ &  -1.3 h_{\rho}^{2}-12.8\left(h_{\omega}^{0}+h_{\omega}^{1}\right) \biggr],
\end{split}
\end{align}
where $h$ and $f$ are the weak $NN$-meson couplings, and $v_{\pi}$ and $v_{\rho}$ are constants which account for the repulsion between nucleons at small distances and for a finite range of the interaction potential. These quantities were found to be $v_{\rho} = 0.4$ and $v_{\pi} = 0.16$, as in~\cite{FLAMBAUM1984367,PhysRevC.56.1641}.


\begin{table} 
\centering
\begin{tabular}{ccc}
\hline
\hline
Reference & ~$g_{p}$~   & ~$g_{n}$~  \\
\hline
DDH (1980)~\cite{DESPLANQUES1980449,PhysRevC.56.1641}   & 4.5     & 0.2     \\
ND (1986)~\cite{NOGUERA1986189}    & 4       & 1       \\ 
DZ (1986)\cite{DUBOVIK1986100}   & 2.4     & 1.1     \\ 
FCDH (1991)~\cite{PhysRevC.43.863}   & 2.7     & -0.1    \\ 
Wasem (2012)~\cite{Wasem2012} & 2.6     & 1.5     \\
NPDGamma (2018)~\cite{PhysRevLett.121.242002} & 3.4 & 0.9 \\
\hline
\hline
\end{tabular}
\caption{Values of $g_p$ and $g_n$ based on the meson exchange constants
from different publications: 
Desplanques,
Donoghue, and Holstein (DDH) \cite{DESPLANQUES1980449,PhysRevC.56.1641}; Noguera and Desplanques
(ND) \cite{NOGUERA1986189}; Dubovik and Zenkin (DZ) \cite{DUBOVIK1986100}; Feldman, Crawford,
Dubach, and Holstein (FCDH) \cite{PhysRevC.43.863}.
In the line of Wasem \cite{Wasem2012} , the best DDH values for all the values of $h$ were used, except $f_\pi = h^1_\pi$, which was recently derived by the lattice QCD methods  \cite{Wasem2012,Haxton2013} to be $h^1_\pi = 1.1 \cdot 10^{-7}$ (as presented in~\cite{Fadeev}). A recent experiment measuring P-violation in the neutron radiative capture by proton \cite{PhysRevLett.121.242002}
gave $h_{\pi}^{1}=[2.6 \pm 1.2 (stat.) \pm 0.2 (sys.)] \times 10^{-7}$ which is larger than the theoretical estimate $h^1_\pi = 1.1 \cdot 10^{-7}$. Using this experimental value, and the rest from DDH, gives slightly larger $g_p=3.4 \pm 0.8$ and smaller $g_n=0.9 \pm 0.6$ which are close to the values $g_p=4$ and $g_n=1$ used in the numerical calculation of PV matrix elements in Ref. \cite{Flam93PRL}.} 
\label{tabg}
\end{table}
Now, using the updated values of the meson-nucleus interaction constants published by the NPDGamma collaboration~\cite{PhysRevLett.121.242002}, we obtain  $g_{p}=3.4 $ and $g_{n} = 0.9$ - see the last line in Table \ref{tabg}. This gives $W=1.78$ meV which corresponds to $d$=17 eV, the interval between $s$ (or $p_{1/2}$) resonances in $^{233}$Th. It is in agreement with the experimental value of $1.39_{-0.38}^{+0.55}$ meV in  in $^{233}$Th ~\cite{PhysRevLett.65.1192,232Th}. The scaling of $W$ with $d^{1/2}$ is also in agreement with the measured parity violating effects in other nuclei presented in the review \cite{MitchellReview2001}.

In the short-range approximation, the PTRIV potential of nucleons in a nucleus may be presented as~\cite{Flambaum:154087,FLAMBAUM1986750}
\begin{align}\label{WTP}
\hat W_{T,P} = \frac{G}{2\sqrt{2}m} \eta_{p,n} (\sigma \cdot \nabla ) \rho(r),
\end{align}
where $\eta_{p}, \eta_{n}$ are dimensionless constants which characterise the strength of the interaction for protons and neutrons respectively. The matrix elements of the operator  $\hat W_{T,P}$ between discrete spectrum states in the standard definition of the angular wave functions are real. According to the calculation presented in Appendix \ref{AppendixB}, the root-mean-squared matrix elements $W^{T,P}$ between nuclear compound states is equal to
\begin{align}
W_{T,P}  = 0.15 \ \text{meV} \sqrt{\eta_{n}^{2} + 0.76 \eta_{p}^{2}}\, .
\end{align}
The value of $W_{T,P}$ is also proportional to  $\overline{N}^{-1/2} \propto d^{1/2}$. Specific values for $W$ and $W_{T,P}$ have been calculated for  $d=17$ eV in $^{233}$Th. However, in the ratio of  $W_{T,P}/W$ the number of principal components $\overline{N}$ cancels out, and the result may be extended to all compound nuclei: 
\begin{align} \label{ratio}
    \frac{W_{T,P}}{W} = 0.09 \sqrt{\eta_{n}^{2} + 0.76 \eta_{p}^{2}} .
\end{align}
If, following Refs.~\cite{Flambaum1995b,PhysRevLett.113.103003}, we take $|\eta_p| = |\eta_n|$, this ratio becomes   
       \begin{align} \label{ratio2}
         \frac{w}{v} &=
 0.12  |\eta_n|  \, . 
     \end{align}

The relative correlator between PV and TRIV matrix elements is ~\cite{Flambaum1995b}  
 \begin{align}
    C = \frac{ \left| \overline{ \mel{p}{W}{s} \mel{s}{W_{T,P}}{p} }  \right|}{W W_{T,P}} \approx 0.1.
\end{align}

PTRIV nuclear forces are dominated by  $\pi_0$ meson exchange. Such an exchange is described by the interaction \cite{Haxton83,Khriplovich2000,Dmitriev03}

\begin{align} \label{piTP}
\begin{split}
\mathcal{W}\left(\boldsymbol{r}_{1}-\boldsymbol{r}_{2}\right) &=-\frac{\bar{g}}{8 \pi m_{N}}\left[\nabla_{1}\left(\frac{e^{-m_{\pi} r_{12}}}{r_{12}}\right)\right] \cdot\left\{\left(\boldsymbol{\sigma}_{1}-\boldsymbol{\sigma}_{2}\right)\right.\\
& \times\left[\bar{g}_{0} \boldsymbol{\tau}_{1} \cdot \boldsymbol{\tau}_{2}+\bar{g}_{2}\left(\boldsymbol{\tau}_{1} \cdot \boldsymbol{\tau}_{2}-3 \tau_{1 z} \tau_{2 z}\right)\right] \\
&\left.+\bar{g}_{1}\left(\tau_{1 z} \boldsymbol{\sigma}_{1}-\tau_{2 z} \boldsymbol{\sigma}_{2}\right)\right\}
\end{split}
\end{align}
where $\bar{g}=13.6$ is the strong-force T,P-conserving $\pi NN$ coupling constant, $\bar{g}_0, \bar{g}_1$, and $\bar{g}_2$, are the strengths of the isoscalar, isovector, and isotensor T,P-violating couplings, respectively, $m_N$ is the nucleon mass, $m_{\pi}$ is the pion mass, $\boldsymbol{\sigma}$ is the nucleon spin, $\boldsymbol{\tau}$ is the nucleon Pauli isospin matrix in vector form, and $r_{12}$ is the separation between nucleons.

The strength constants $\eta_{n,p}$ can be expressed in terms of different fundamental interactions. The PTRIV interaction between nucleons is dominated by the pion exchange. Ref.~\cite{PhysRevLett.113.103003} gives the following values:

\begin{align}
\eta_n=-\eta_p =   (-  g_s \bar{g}_{0} + 5 g_s\bar{g}_{1} + 2 g_s \bar{g}_{2} ) 10^{6}.
\end{align} 
Thus, the ratio (\ref{ratio}) becomes
\begin{align} \label{estimateratioTP}
\frac{W_{TP}}{W} = 0.12 |\eta_{n}| = | (-1.2 g_s\bar{g}_{0} + 6.0 g_s \bar{g}_{1} + 2.4 g_s \bar{g}_{2} )10^{5}|.
\end{align}
Similarly, we can express $\eta$ in terms of the QCD $\theta$-term constant. Using 
results presented in Refs ~\cite{Yamanaka2017,Vries2015}
\begin{align}
g_s \bar{g}_{0} &=-0.21 \ \theta, \\
g_s \bar{g}_{1} &= 0.046 \ \theta , 
\end{align}
 we have 
\begin{align} \label{etawrttheta}
\eta_n=-\eta_p = 4.4 \times 10^{5} \ \theta,\\ 
        \frac{W_{TP}}{W} = 5.3 \times 10^{4} |\theta|.
\end{align}
Expressing $\eta$ via the quark chromo-EDMs ${\tilde d_u}$ and  ${\tilde d_d}$: $g_s {\bar g}_0 = 0.8 \times 10^{15}({\tilde d_u} +{\tilde d_d})$/cm,  $g_s {\bar g}_1 = 4 \times 10^{15}({\tilde d_u} - {\tilde d_d})$/cm~\cite{POSPELOV2005119} gives:
\begin{align}
 \eta_n=-\eta_p =  (-0.8 (\tilde{d_{u}} + \tilde{d_{d}}) + 20  (\tilde{d_{u}} - \tilde{d_{d}})) 10^{21}/\text{cm}.
\end{align}
\begin{align}
    \frac{W_{TP}}{W} = |  (-1.0 ({\tilde d_u} + {\tilde d_d}) +24 ({\tilde d_u} - {\tilde d_d}))  10^{20} |  / \text{cm}
\end{align}
Note that the current limits on the CP-violation parameters presented above correspond to $\frac{W_{TP}}{W} < 10^{-5}$ \cite{Fadeev}. The expected experimental sensitivity is an order of magnitude better, $10^{-6}$  ~\cite{PhysRevC.90.065503,Palos2018}.





Finally, a PTRIV interaction, similar to the pion-exchange-induced Eq. (\ref{piTP}), may be due to exchange by any
scalar particle which has both scalar (with the interaction constant $g^s$) and pseudoscalar  (with the interaction constant $g^p$) couplings to nucleons. The most popular examples are the dark-matter candidates axion \cite{Moody,Marsh} and relaxion \cite{Graham,Gupta,Flacke}, which have  very small masses.\footnote{The limits on the T,P-violating electron-nucleon interactions mediated by the axion or relaxion exchange from EDM measurements
were obtained in Ref. \cite{Stadnik2017}, where more references may be found.} A numerical estimate shows that due to the long range of the interaction the matrix elements in the small-mass case ($e^{- mr} \approx 1)$ are $\sim 1.5$ times larger than the pion exchange matrix elements; i.e., we have instead of Eq. (\ref{estimateratioTP}) the following estimate:
\begin{align}\label{wva}
        \frac{W_{T,P}}{W} & \sim  |1 \times 10^6 g^s g^p|   \, . 
     \end{align}
The limit on $g^sg^p$ may be obtained from the proton EDM calculation,\footnote{The calculation is similar to that for electron EDM \cite{Stadnik2017}.}
\begin{equation} \label{dp}
d_p=\frac{g^sg^pe}{8 \pi^2 m_p}\,,
    \end{equation}
and measurement \cite{Graner2016}, $|d_p|<2\times 10^{-25}e$ cm, $|g^s g^p| < 1 \times 10^{-9}$. Using limits from the proton EDM and the $^{199}$Hg nuclear-Schiff-moment measurements in Ref. \cite{Swallows2013}, the authors of Ref. \cite{musolf} concluded that the limit on $|g^s g^p|$ is between $10^{-9}$ and $10^{-11}$. This gives a rather weak limit on $W_{T,P}/W$ induced by axion exchange:    
\begin{equation} \label{limita}
    \frac{W_{T,P}}{W} < 10^{-3} - 10^{-5} \, .
\end{equation}
With the expected experimental sensitivity  $10^{-6}$  ~\cite{PhysRevC.90.065503,Palos2018}, limits on the axion interaction constants may be significantly improved.

\section{A possible regular component of parity Violation in neutron scattering: experimental results} \label{6}

As aforementioned, there were a large number of experiments performed which confirmed the existence of the enhanced longitudinal asymmetry (\ref{P}). The first to do so directly was performed at the Joint Institute for Nuclear Research (JINR), Dubna.
Firstly parity violation was measured for the
$p$-wave resonance at 1.3 eV in $^{117}\text{Sn}$~\cite{Alfimenkov1981}. Then the $p$-wave resonances in $^{139}\text{La}$, $^{111}\text{Cd}$ and $^{81}\text{Br}$ were probed, where the weak matrix element was found to be $\sim 1$ meV, while the mixing coefficients were inferred to be of the order $10^{-4}$~\cite{Alfimenkov1983,Alfimenkov1984}. These findings were in agreement with the predictions made in Ref. \cite{Sushkov80}. 

These initial observations of parity violation in $p$-wave resonances were limited to one or two $p$-wave resonances per target in the neutron energy region up to $\sim 10$ eV. The subsequent breakthrough came from the formation of the Time Reversal Invariance and Parity at Low Energies (TRIPLE) collaboration, who were able to optimise the experiment to be able to probe a larger number of resonances per nucleus. After initial success in the measurements on $^{238}\text{U}$~\cite{PhysRevLett.65.1192} and $^{139}\text{La}$~\cite{PhysRevC.44.2187}, measurements on $^{232}\text{Th}$~\cite{232Th} were completed, with a number of statistically significant PV effects observed, which all had the same sign, seemingly contradicting the statistical nature of the reaction mechanism. Since these initial experiments, there have been numerous measurements on the nuclear resonances of a vast range of nuclei. The review~\cite{MitchellReview2001} contains data for 20 nuclei and several hundred resonances.

\subsection{Parity violation for neutron resonances in $^{232}\text{Th}$}
The unexpected outcome of early measurements conducted by the TRIPLE collaboration was by Frankle et al.~\cite{232Th} who reported on measurements of 23 $p$-wave resonances in $^{232}\text{Th}$, with energies ranging from $E_{n} = 8$ to $E_{n} = 392$ eV. Among these resonances, seven had PV effects which all contained a relative significance of $2.4 \sigma$ or higher. However, contrary to the expectations, the longitudinal asymmetry of these resonances had a constant sign.

The outcomes of this experiment seemed to develop more questions than answers. It became clear that the current experimental set-up was not sufficient to extensively study parity violating effects in neutron resonances. Furthermore, the two-level approximation used in the analysis was not adequate, implying the need for the development of a statistical approach which includes the contributions of many $s$-wave resonances, and has the ability to more accurately describe complex resonance structures. However, it is important to note that the collaboration's observations were consistent with the notion that sufficiently precise experimentation can detect parity violation in every $p$-wave resonance with the spin $J=1/2$ (which may be mixed by the weak interaction with $s$-wave resonances with the spin $J=1/2$).

In order to confirm or reject the non-random nature of PV effects  in $^{232}\text{Th}$, a new experimental system was developed, with an upgraded polarizer, spin flipper and neutron detector. Using the new system, transmission experiments were once again performed on $^{238}\text{U}$ and $^{232}\text{Th}$. These experiments refined previous results with much better statistics, and were able to detect more longitudinal asymmetries. Specifically in $^{232}\text{Th}$, the permanent  sign observation in measured parity violating effects was confirmed, and extended to 10 in a row~\cite{PhysRevC.58.1236}. The probability of obtaining ten out of ten randomly distributed quantities with the same sign is $\approx 0.2 $\%. The magnitude of the PV effect was found to be in agreement with the statistical theory calculations and consistent with experimental results in other nuclei where PV effects have a random sign. Moreover, upon probing neutron energies higher that 250 eV, the PV effect in $^{232}\text{Th}$ has 4 resonances with a negative sign, and 2 with a positive sign, see Refs.~\cite{PhysRevCLastExperiment,MitchellReview2001}.

Upgraded results for the resonances below 285 eV are shown in Table \ref{Th232Table} (reproduced from~\cite{PhysRevC.58.1236} with added data for the weak matrix elements defined as $|W|=P/A_J$ ). Measurements were taken with no initial knowledge of the $J$ value, and as such, $J= 1/2$ was assigned to resonances with a large parity violating asymmetry ($P \geq 3 \sigma$). These resonances are denoted with parentheses. 




\section{A possible octupole doublet mechanism for a regular component of PV and PTRIV effects in neutron scattering}\label{7}

	
The target of $^{232}$Th (and consequently, the compound nucleus $^{233}$Th) may be a special case due to some peculiarities of its structure. Several  thorium  isotopes display strong octupole correlations. Octupole deformation may be present in the ground state  or in excited states  near the neutron threshold.

The wave function of a deformed nucleus is the product of the internal nuclear wave function and rotational wave function. For a given internal function with a non-zero projection of nuclear spin $I$ on symmetry axis $n$, $K=(nI) \neq 0$, presence of octupole deformation (or of any axially symmetric shape that has no symmetry with respect to reflection in the equatorial plane) leads to rotational doublets with definite parity $P=\pm 1$~\cite{NuclearStructure}, 
	
\begin{align}\label{doublet}
\begin{split}
\left| \Psi_{MK;P}^{I} \right\rangle = \sqrt{\frac{2 I+1}{8 \pi}} \{ D_{MK}^{I} ( \varphi, \theta, 0) \lvert a ; K \rangle 
+ P  (-1)^{I+K} \\ D_{M-K}^{I} (\varphi, \theta, 0) \lvert a ; -K \rangle \}.
\end{split}
\end{align}

Here Wigner functions $D_{M \pm K}^{I} ( \varphi, \theta, 0)$ describe nuclear rotation and states $\lvert a ; \pm K \rangle$ are internal states of the rotating nucleus. 

\begin{widetext}
\begin{center}
\begin{table}[ht]
\centering
\begin{tabular}{ccccc}
\hline
\hline
~$E(\mathrm{eV})$~ & ~$P \ (\%)$~\cite{PhysRevC.58.1236} & ~$ P/ \Delta P$~\cite{PhysRevC.58.1236} & ~$|A_{J}|$ (1/eV)~\cite{PhysRevC.58.1236}  \\
\hline
$8.36032$ & ($1.78 \pm 0.09$) & $19.8$ & 25.0    \\
13.1377 & $0.16 \pm 0.14$ & 1.1 & 38.5   \\
36.982 & $-0.01 \pm 0.17$ & -0.1 & 20.5   \\
38.232 & ($6.41 \pm 0.32$) & 20.0 & 27.1   \\
41.066 & $-0.09 \pm 0.27$ & -0.3 & 27.0  \\
47.068 & ($2.52 \pm 0.13$) & 19.4 & 17.3   \\
49.941 & $-0.24 \pm 0.39$ & -0.6 & 40.0   \\
58.786 & $0.02 \pm 0.03$ & 0.7 & 58.3   \\
64.575 & ($14.16 \pm 0.41$) & 34.5 & 103.0   \\
90.139 & $0.21 \pm 0.19$ & 1.1 & 11.6   \\
98.057 & ($0.70 \pm 0.22$) & 3.2 & 12.9   \\
103.63 & $0.22 \pm 0.16$ & 1.4 & 13.4   \\
128.17 & ($2.31 \pm 0.12$) & 19.2 & 13.6   \\
145.83 & $0.00 \pm 0.10$ & 0.0 & 2.89   \\
148.06 & $-0.11 \pm 0.34$ & -0.3 & 12.4   \\
167.11 & ($3.21 \pm 0.10$) & 32.1 & 33.8   \\
178.86 & $0.19 \pm 0.28$ & 0.7 & 15.5   \\
196.20 & ($0.90 \pm 0.18$) & 5.0 & 11.4   \\
202.58 & ($1.10 \pm 0.25$) & 4.4 & 11.2   \\
210.91 & $-0.23 \pm 0.32$ & -0.7 & 10.5   \\
231.95 & ($4.77 \pm 0.68$) & 7.0 & 12.6  \\
234.07 & $-0.16 \pm 0.45$ & -0.4 & 10.1   \\
242.25 & $0.18 \pm 0.17$ & 1.0 & 7.04  \\
276.45 & $0.46 \pm 0.76$ & 0.6 & 17.1   \\
\hline
\hline
\end{tabular}
\begin{tabular}{r}
\hline
\hline
$|W| \ ( \mathrm{  meV})$\\
\hline
($0.712 \pm 0.036$) \\
$0.0416 \pm 0.036 $\\
$0.00488 \pm 0.083 $ \\
($2.37 \pm 0.12) $\\ 
$0.0333 \pm 0.10$ \\ 
($1.46 \pm 0.075$) \\
$0.0600 \pm 0.098$ \\
$0.00343 \pm 0.0052$ \\
($1.37 \pm 0.040$) \\
$0.181 \pm 0.16$ \\
($0.543 \pm 0.17$) \\ 
$0.164 \pm 0.12$ \\
($1.70 \pm 0.088$) \\ 
$0$ \\
$0.0887 \pm 0.27$ \\
($0.950 \pm 0.030$) \\ 
$0.123 \pm 0.18$ \\ 
($0.789 \pm 0.16$) \\
($0.982 \pm 0.22$) \\
$0.219 \pm 0.30$ \\
($3.79 \pm 0.54$) \\
$0.158 \pm 0.45$ \\
$0.256 \pm 0.24$ \\
$0.269 \pm 0.44$ \\
\hline
\hline
\end{tabular}
\caption{P-odd asymmetries in $^{232}\mathrm{Th}$.}
\label{Th232Table}
\end{table}
\end{center}
\end{widetext}

Let us start with the calculation of the PTRIV effect. In other relevant calculations of static PTRIV effects, e.g. in the calculations of the nuclear Schiff moments \cite{PhysRevLett.76.4316,PhysRevC.56.1357,FlambaumFeldmeier}, the lower energy component of the doublet is the ground state of the nuclei, whereas in our case of PTRIV in neutron scattering the doublet states are excited compound states of a nucleus with octupole deformation. These states are superpositions of simple quasiparticle configurations $\lvert \Phi_{i} ; \pm K \rangle$:
	
\begin{align}
\lvert a ; \pm K \rangle  = \sum_{i} C_{i}^{a} \lvert \Phi_{\alpha} ; \pm K \rangle,
\end{align}
where the expansion coefficients $C_{i}^{a}$ are not dependent on the sign of $K$, given the fact that strong and electromagnetic interactions preserve party and time reversal invariance.

Both the T-P,odd interaction $W_{T,P}$ (\ref{WTP}) and $K = (nI)$ are pseudoscalars, meaning the following relationship holds:
	
\begin{align}
\ev**{W_{T,P}}{a;K} = - \mel**{a;-K}{W_{T,P}}{a; -K}.
\end{align}
This results in the matrix element of $W_{T,P}$ between the doublet states of opposite parity being reduced to the expectation value over the internal nuclear state:

\begin{align}
\mel**{\Psi^{I}_{MK;+1}}{W_{T,P}}{\Psi^{I}_{MK;-1}} = \mel**{a;K}{W_{T,P}}{a; K},
\end{align}
where
	
\begin{align}
\mel**{a;K}{W_{T,P}}{a; K} = \sum_{i} \left( C_{i}^{a} \right)^{2} \ev**{W_{T,P}}{\Phi_{i} ; K}.
\end{align}
The matrix elements between simple basis states has been estimated in Ref. ~\cite{PhysRevC.56.1357}:

\begin{align}
\ev**{W_{T,P}}{\Phi_{i} ; K} \approx \frac {\beta_{3} \eta}{  A^{1/3}} \ \text{eV},
\end{align}
where   $\beta_{3}$ is the  octupole deformation parameter and  $\eta$ is the dimensionless strength constant of the nuclear PTRIV potential $\hat W_{T,P}$ in Eq. (\ref{WTP}). Implementing the normalisation condition $\sum_{i}\left (C_{i}^{a}\right)^{2}=1 $, the PTRIV matrix element is approximately equal to~\cite{PhysRevC.56.1357}
	
\begin{align} \label{doubletTP}
W_{T,P}^{\pm} \equiv \mel**{\Psi^{I}_{MK;+1}}{W_{T,P}}{\Psi^{I}_{MK;-1}} \simeq \frac{\beta_{3} \eta}{ A^{1/3}} \ \text{eV},
\end{align}
Hence, the expression for the proposed constant sign component of the T,P-odd effect will bear a similar form to (\ref{P}), and can be written as 
\begin{widetext}	
\begin{align} \label{PToddeffect}
\begin{split}
& P_{T,P} \simeq\left[ \frac{(\alpha_{1/2,J} M_{p,1/2} + \alpha_{3/2,J} M_{p,3/2}) M_{p}}{\beta_{1,J} M_{p,1/2}^{2} + \beta_{13,J} M_{p,1/2} M_{p,3/2} + \beta_{3,J} M_{p,3/2}^{2} }  \right] \frac{\beta_{3} \eta}{ A^{1/3}} \frac{[\text{eV}]}{E_{s,\text{doublet}} - E_{p}}  \sqrt{\frac{\Gamma_{s}^{n}}{\Gamma_{p}^{n}}} .
\end{split}
\end{align}
\end{widetext}
Here, the sign factors cannot be absorbed inside the matrix element $W^{T,P}_{\pm}$, as was the case for the parity violating, time-conserving effect, due to contributions from both the $p_{1/2}$ and $p_{3/2}$ amplitudes.  The parameters of $E_{s,\text{doublet}} - E_{p}$ and the kinematic factor $\sqrt{\Gamma_s^n/ \Gamma_p^n}$ are unknown at the present time. However, they also appear in P-odd (but T conserving) effects, meaning upon calculating the ratio of the T,P-odd effect  to the P-odd effect, these parameters cancel. 

In the derivation of Eq. (\ref{PToddeffect}) we assumed that the  opposite parity doublet component to the $p$-wave compound state is a stationary state of a definite energy. In fact, this state may have a small spreading width. Indeed, we assume that the opposite parity component of the doublet has the same internal state $\lvert a ; \pm K \rangle$ in Eq. (\ref{doublet}). However, the  ``exact'' copy of the  internal  state $\lvert a ; \pm K \rangle$ of the  $p$-wave resonance in the doublet component of opposite  parity may be mixed with nearby $s$-wave resonances by the Coriolis interaction. The Coriolis force is relatively weak, therefore, the corresponding spreading width should be small. This means that every $s$-wave resonance within this small spreading width may have this doublet state as a component. In this case, we should replace $1/(E_{s,\text{doublet}} - E_{p})$ with $\mathfrak{Re}(1/[E_{s,\text{doublet}} - E_{p} - i \Gamma _{spr}/2 ])$.  There is also an additional contribution coming from close $s$ wave resonance, similar to that calculated in the Section \ref{8}. 
Since this situation happens in both PV and PTRIV effects, such spreading does not significantly affect the ratio of PTRIV and PV  effects. 






	

With regards to the P-odd weak matrix element, the expectation value of the P-odd weak interaction matrix element in the body frame of the nucleus with octupole deformation vanishes, $\left\langle\Phi_{i} ;K\left|W\right| \Phi_{i} ; K\right\rangle =0 $. This is a consequence of time-reversal invariance~\cite{Sushkov82,Flambaum_1995}. Therefore, the direct matrix element of $W$ between the opposite parity components of the same doublet  vanishes. As noted in Ref.~\cite{Flambaum_1995}, the mixing with the opposite parity component of another doublet state is allowed. 

Let us explain how the P-odd effect appears. We need an additional interaction $H'$ which conserves P and T but can mix components of different doublets, $\ket{\Psi_{M K;P }^{a I}}$ and $\ket{\Psi_{M K;P }^{b I}}$, with the same parity P.  This may be the same interaction which produces mixing of the doublet components, $\ket{a;K}$ and $\ket{a;-K}$, and results in the splitting of the opposite parity states of the same doublet, $  \ket{\Psi_{M K;+ }^{a I}}$ and $\ket{\Psi_{M K;- }^{a I}} $. For example, $H'$ may be  tunneling of an excess cluster of nucleons or the Coriolis force ~\cite{NuclearStructure}.
As a result of the two interactions $W$ and $H'$ combining, the total rotational function acquires an admixture of a component of the same doublet with opposite parity~\cite{Flambaum_1995}

\begin{align} \label{P1}
\left|\Psi_{M K;P }^{a I}\right\rangle \rightarrow\left|\tilde{\Psi}_{M K;P }^{a l}\right\rangle=\left|\Psi_{M K ;P}^{a I}\right\rangle+\beta_{P}\left|\Psi_{M K;P}^{a I}\right\rangle,
\end{align}
where the mixing amplitude $\beta_{P}$ is
\begin{align} \label{P2}
\beta_{P} & = 
-2 \frac{{1}}{ E_{p} - E_{s, \text{doublet}} } \\ & \times  \sum_{b} \frac{ \mel**{a;-K}{H'}{b;K} \mel**{b;K}{W_{P}}{a;K}}{E_{p} - E_{b}},
\end{align}
where $E_{p}$ is the energy of the $p$-wave resonance. Since the energy spectrum is very dense, the additional order of the perturbation theory involving $H'$ does not produce any significant suppression of the PV effect. This conclusion is supported  by the non-conservation of the quantum number $K$ due to the enhancement of the Coriolis force effect in compound states, see~\cite{Kadmenskii1982,Kadmenskii1984}. Moreover, the energy denominator $E_p -E_b$ is also due to $H'$, so the effective strength constant of interaction $H'$, even if it  is small, cancels out. Therefore, the ratio of the PTRIV effect to the PV effect in the doublet mechanism is approximately the same or only slightly bigger than the corresponding ratio in the statistical mechanism considered in the previous sections. 

\subsection{Specific target nuclei where octupole doublet mechanism may produce regular PV and PTRIV effects} \label{candidatenuclei}

	In this section, we present a few candidate nuclei which may be used to search for the permanent sign  PV and PTRIV effects due to the octupole doublet mechanism. We start from Lanthanide nuclei with zero spin, which would only be suitable for PV measurements. Firstly, we have $^{148}_{60}$Nd$_{88}$, which according to the theoretical results has octupole deformation in the ground state~\cite{Afanasjev}. This nucleus is stable, with a natural abundance of 5.8\%. The rotational spectra look consistent with octupole deformation, and the negative and positive parity bands are separated by $\simeq 1000$ keV. Ref. ~\cite{Afanasjev} also identifies $^{150}_{62}$Sm$_{88}$ to have octupole deformation.  This nucleus is stable, with a natural abundance of 7.4\%. However, the spectra do not exhibit the expected band structure starting from the ground state (octupole deformation may still exist in excited states). Other stable isotopes of Nd and Sm, as well as other nuclei with a comparable number of protons and neutrons are also potential candidates for a permanent sign PV effect. 
	
	
	Next, we consider nuclei with non-zero spin, which would be candidates for both PV and PTRIV measurements. $^{139}_{57}$La$_{82}$ may be suitable ($Z=56$), however the neutron number $N=82 +1=83$ is outside the desired interval  $N=88-92$. After neutron capture, it is possible that $^{140}_{57}$La$_{83}$ has an excited isomeric state with octupole deformation. Further, the spectra of the stable $^{153}_{63}$Eu$_{90}$ isotope ($I^P=5/2^+$) indicates octupole deformation in the ground state~\cite{FlambaumFeldmeier,FlambaumDzuba}, with an energy band gap of 97 keV. This is a good candidate to search for PV and PTRIV effects. Finally, theoretical data again suggests the existence of octupole deformation in $^{149}_{62}$Sm$_{87}$, $I^P=7/2^-$, which after neutron capture becomes $^{150}_{62}$Sm$_{88}$~\cite{Afanasjev}. Other candidates in the Lanthanides region include $^{141}_{59}$Pr$_{82}$, $I^P=5/2^+$, $^{143}_{60}$Nd$_{83}$, $I^P=7/2^-$, $^{145}_{60}$Nd$_{85}$, $I^P=7/2^-$, $^{147}_{62}$Sm$_{85}$, $I^P=7/2^-$, $^{151}_{63}$Eu$_{88}$, $I^P=5/2^+$, $^{155}_{64}$Gd$_{91}$, $I^P=3/2^-$, $^{157}_{64}$Gd$_{93}$, $I^P=3/2^-$, $^{159}_{65}$Tb$_{91}$, $I^P=3/2^+$, $^{161}_{66}$Dy$_{91}$, $I^P=5/2^+$, and $^{163}_{66}$Dy$_{93}$, $I^P=5/2^-$, which all have non-zero spin. 
	
	In the actinide region, Ref.~\cite{Afanasjev} suggests the existence of octupole deformation in many unstable even-even nuclei, which are hardly suitable for neutron experiments. However, more stable nuclei may have octupole deformation in low energy excited states, the most obvious one being the entire reason for research into this area, $^{232}_{90}$Th$_{142}$, as per discoveries made by~\cite{MitchellReview2001,PhysRevC.58.1236}. This isotope has 4 extra neutrons as compared to $^{228}_{90}$Th$_{138}$ ($^{228}_{90}$Th$_{138}$  has a lifetime of 1.9 years). There is evidence of octupole deformation in $^{228}_{90}$Th$_{138}$~\cite{Thorium228}. 
	
\begin{figure}[!h] 
\centering
    \includegraphics[scale = 0.3]{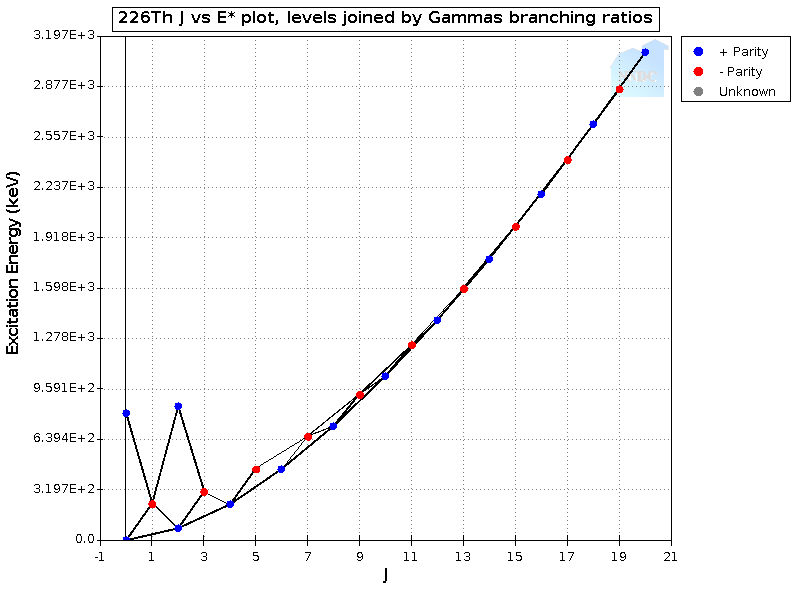}
    \caption{Plot of the Excitation Energy $E$ (keV) vs Angular Momantum $J$ in $^{226}\text{Th}$. At higher values of $J$, the negative parity rotational band ($11^{-}, 13^{-}, 15^{-},\dots$) and the positive parity rotational band ($12^{+}, 14^{+}, 16^{+}, \dots$) coincide. This is a clear indication of octupole deformation. The maximum distance between these bands occurs at $1^{-}$ (a), and is $\simeq 200$ keV, i.e. it is small on the nuclear scale.
    This graph was obtained using the database~\cite{nudat}.}
    \label{226Th}

\end{figure}

	As an example, let us analyse the rotational spectrum of $^{226}\text{Th}$, using the database~\cite{nudat}. Upon inspection of the spectra (Figure \ref{226Th}), we see that there is a common rotational band, containing both the positive ($0^{+}, 2^{+}, 4^{+}, \dots$) and negative parity states ($1^{-}, 3^{-}, 5^{-}, \dots$). This property is a clear indicator of the presence of octupole deformation in this isotope, as such nuclei permit both odd and even values of $J$ on the rotational band. This is in contrast to nuclei with only quadrupole deformation, as in this case only even levels are permitted on the rotational band ($0^{+}, 2^{+}, \dots$). Thus, we see that this isotope $^{226}\text{Th}$ appears to have a static octupole deformation. The energy interval between the $0^{+}$ state and $1^{-}$ state is $230$ keV. If we subtract an ordinary rotational energy difference between the $J=1$ and $J=0$ states, the energy gap between the odd and even parts of the rotational bands does not exceed 200 keV. This is much smaller than the typical energy gap for the dipole and octupole excitations, which is a few MeV. For higher $J$ there is practically no gap.

To conclude this section we should stress that the existence of octupole deformation in  stable nuclei is not certain, so the absence of the permanent sign PV effect in some of these nuclei can not prove that the doublet mechanism is not efficient. However, if the permanent sign effect is observed, this would provide important evidence in favour of the existence of octupole deformation in stable nuclei, which is a hotly debated topic.      

\section{Doorway states contribution to PV effects} \label{8}

Let us assume for simplicity a nuclear target spin $I=0$, as in $^{232}$Th.
To begin, we will first separate the contribution of the two-particle-one-hole  doorway components $\psi_{1s}$ and $\psi_2$ in the wave functions of positive and negative parity compound states corresponding to $s$-wave and $p$-wave resonances:
\begin{align} \label{psi2}
\psi_s& =C_{1s} \psi_{1s} + \text{other components},\\
\psi_p& =C_2 \psi_2 + \text{other components}. 
\end{align}
The amplitudes of neutron capture may be expressed in terms of the amplitudes of capture to $s$-wave and $p$-wave doorway states $M_1$ and $M_2$:  $M_s=C_{1s}M_1$ and $M_p=C_{2}M_2$, and the weak matrix element between compound states may be presented as 
\begin{align} \label{psi3}
W_{sp}=C_{1s} C_2 W_{1,2} + \text{other contributions}.
\end{align}
Thus, the contribution of the doorway states to the PV forward scattering amplitude Eq. (\ref{FSA}) may be presented as 
\begin{align} \label{FSAd}
f^{(d)}_{\mathrm{p . v .}}(0)=\pm \frac{1}{2 k} \frac{2 g M_1 M_2  i W_{1,2} |C_2|^2}{ \left(E-E_{p}+\frac{1}{2} i \Gamma_{p}\right)} S\,,
\end{align}
where for $E \approx E_p$
\begin{align} \label{S}
S=\sum_s \frac{|C_{1s}|^2}{\left(E_p-E_{s}+\frac{1}{2} i \Gamma_{s}\right)}\,.
\end{align}
This expression for $S$ has two maximums. The first is for $E_s\approx E_{p}$ ($s$-wave resonances close to $p$-wave resonance)  and the second is for $E_{s}\approx E_1$ ($s$-wave resonances close to the energy of the doorway states $E_1$), where 
\begin{align} \label{C1s}
\overline{|C_{1s}|^2}= \frac{1}{N} \frac{\Gamma_{\text{d}}^{2}/4}{(E_s- E_1 )^{2} + \Gamma_{\text{d}}^{2}/4}
\end{align}
has maximum. Here  $N = \frac{\pi \Gamma_{\text{d}}}{2d}$ is the normalization constant. Note that the doorway state spreading width   $\Gamma_{\text{d}}$ is the sum of two widths, the decay width to the continuum $\Gamma_d^{\uparrow} \equiv \Gamma_d^{out}$ (decay out) and the ``decay'' width to the other compound state components $\Gamma_d^{\downarrow} \equiv \Gamma_d^{in}$ (decay in, mixing of the doorway state with compound states). According to Ref.~\cite{Hussein2016}, the width to decay out $\Gamma_d^{out}$ =0.18 keV, is much smaller than the ``decay in'' width, which according to Ref.~\cite{Hussein1995} is $\Gamma_d^{in} \sim$ 30 keV. This allows us to treat the doorway state as a component of the nuclear compound state wave function and apply perturbation theory for stationary states.

 
{\bf First case: close $s$-wave resonances}.
Using  $\Gamma_s^{n}=|C_{1s}|^2 |M_1|^2$, the  sum $S$ may be presented as 
 \begin{align} \label{SB}
S_1 & = \frac{B}{|M_1|^2}\,,\\
B &  \equiv \sum_s \frac{\Gamma_s^{n}}{E_p -E_s},
\end{align}
where the sum $B$ may be found using experimental data. For a qualitative comparison with other contributions it is sufficient to keep one $s$-wave resonance (close to $p$-wave resonance) in the sum $S$ in Eq. (\ref{S}).
 
{\bf Second case: distant  $s$-wave resonances close to the doorway state energy}. In this case the distance $d$ between compound states is much smaller than $E_1 -E_p$, meaning a large number of $s$-wave resonances contribute. To evaluate $S$ we may replace the summation over $s$ by the integral $dE_s/d$:
 
 \begin{align} 
 \nonumber
\frac{\Gamma_{\text{d}}^2}{4 \overline{N}d}  \mathfrak{Re} \int _{-\infty}^{\infty} \frac{dE_s}{[(E_s- E_1 )^{2} + \Gamma_{\text{d}}^{2}/4]\left(E_p-E_{s}+\frac{1}{2} i \Gamma_{s}\right)}\,.
\end{align}
%

Note that this integral is similar to the normalisation integral for $\overline{|C_{1s}|^2}$, with an extra factor in the denominator,  $(E_p-E_{s}+\frac{1}{2} i \Gamma_{s})$ - see Eq.(\ref{S}). Performing the integration by closing the loop at infinity in the complex plane and using  Cauchy's residue theorem, we obtain 
\begin{align} \label{S2}
S_2=\frac{E_p - E_1}{(E_p- E_1 )^{2} + \Gamma_{\text{d}}^{2}/4}\,.
\end{align}
Note that $S_2$  vanishes if the $s$ doorway energy $E_1$ coincides with the position of the $p$-wave compound resonance. This is a result of the cancellation between the contributions of the $s$-wave resonances with $E_s < E_p$ and $E_s > E_p$.  

The ratio of the contributions of the distant $s$-wave resonances and close $s$-wave resonances  may be presented as  
\begin{align} \label{S1S2}
\frac{S_2}{S_1}=\frac{\pi}{2} x z
\end{align}
where $x=2(E_1-E_p)/\Gamma_{\text{d}}$, $z=2(E_s-E_p)/d$. The maximum of $S_2$ is for $x=1$ (the distance of the $s$ wave doorway to the $p$ compound resonance is $|E_p -E_1|=\Gamma_{\text{d}}/2$),  $z\lesssim 1$  (the $p$ resonance is between $s$  resonances, $|E_p -E_s| \lesssim  d/2$). In this case typically $S_2 \sim S_1$. However, for the biggest PV effects we may have $(E_p -E_s) \ll  d$, and in this case $S_1$ may dominate. Only $S_2$ and its contribution to the PV effect has a permanent sign for all $p$-wave resonances on one side of $E_1$. The sign of $S_1$ fluctuates as $E_s-E_p$ may be of any sign.

Now we may compare the permanent sign contribution ($S_2$) of a doorway state to the statistical contribution of all compound state components:
\begin{align} \label{dPVrefined}
\frac{f^{(d)}_{\mathrm{p . v .}}}{f_{\mathrm{p . v .}}} \sim \frac{xz}{\sqrt{(x^2+1)(y^2+1)}} \frac{\left(d\overline{\Gamma}_{\text{spr}}\right)^{1/2}}{\Gamma_{\text{d}}}\,, 
\end{align}
where  $y=2(E_2-E_p)/\Gamma_{\text{d}}$.
We have taken into account that the weak matrix element between compound states is suppressed by $1/\overline{N}^{1/2}$ in comparison with the  matrix element between simple states - see Eq. (\ref{PMSME}) in Appendix \ref{AppendixB}:
\begin{align} \label{ratioPV}
\frac{W_{sp}}{W_{1,2}}  \sim \frac{1}{\overline{N}^{1/2}} \sim \left(\frac{d}{\overline{\Gamma}_{\text{spr}}}\right)^{1/2}\,.
\end{align}
The maximum of the first factor in Eq. (\ref{dPVrefined}) is for $x=1$ and $y=0$. From its definition, $z \lesssim 1$. Therefore,
\begin{align} \label{dPV}
\frac{f^{(d)}_{\mathrm{p . v .}}}{f_{\mathrm{p . v .}}} 
\lesssim \frac{\left(d\overline{\Gamma}_{\text{spr}}\right)^{1/2}}{\Gamma_{\text{d}}} \,. 
\end{align}
In $^{233}$Th the average interval between compound states is $d=17$ eV. To have a noticeable permanent sign contribution of the doorway states we require the energies of the $s$ and $p$ doorway states to be close to $p$-wave compound resonances ($|E_1-E_p| \approx  \Gamma_{\text{d}}/2$, $E_2 \approx  E_p$) as well as a very small doorway spreading width $\Gamma_{\text{d}} < $ 1 keV.
However, at the moment we do not see any reason for the inequality $\Gamma_{\text{d}} \ll \overline{\Gamma}_{\text{spr}} $ to hold (i.e. the spreading width of the doorway states to be much smaller than the average spreading width of the compound state components).
Numerical simulations \cite{FlamGribakin1994,Izrailev1996,Horoi1995,Zelevinsky1995} have  shown that spreading widths for different components of compound states are approximately the same.
Fluctuations of the spreading widths are small due to a large number  $k$ of ``decay'' channels of each component, $\delta \Gamma_{spr}/\Gamma_{spr} \sim k^{-1/2}$ \cite{Flambaum_Gribakin95}. 

\section{Comments about other mechanisms of the permanent sign PV and PTRIV effects} \label{9}

There are also other explanations of the permanent sign effect in $^{232}$Th. The contributions from Distant doorway states~\cite{PhysRevC.45.R514,PhysRevC.46.2582,PhysRevC.50.1456,PhysRevLett.74.2638,PhysRevLett.68.780} appear to be too small. The neutron PV potential scattering contribution~\cite{Zaratsky,FlambaumPwave1992}, as well the contributions calculated using optical potentials~\cite{Koonin1992,Carlson1993,Hussein1995}, also seem to be too small. For a discussion of these  mechanisms, see the review~\cite{MitchellReview2001}.

Ref.~\cite{FlambaumPwave1992}, in the section "quasi-elastic mechanism", attempted to find a coherent contribution to PV and PTRIV effects related to the components of the compound state wave function, which may be presented as  $\Psi_n \Psi'_{\text{target}}$, where $\Psi_n$ is the wave function of the projectile neutron (which obtains an additional  resonance term localised mainly inside the nucleus  when neutron energy is close to that of the compound resonance\cite{Zaratsky,FlambaumPwave1992};  this term may be treated as a component of the compound state wave function, a doorway state)  and $\Psi'_{\text{target}}$ is the low-energy excitation of the target nucleus, e.g. the nuclear rotation, vibration or particle excitation. However, the result is inconclusive. Ref.~\cite{FlambaumPwave1992} also derived an analytical estimate for the ratio of PTRIV effect to PV effect for the potential scattering  and `"quasi-elastic" mechanisms:


\begin{align}
    \frac{f_{T,P}}{f_{P}} = \frac{\eta}{2g} \frac{(1 + f_{0}/R)}{(1+2f_{0}/R)} C(J,I),
\end{align}
where $C(J,I)$ is the angular coefficient (similar to that considered in  section \ref{2}), $R$ is the nuclear radius  and  

\begin{align}
    f_{0} = -a - \frac{1}{2k} \frac{\Gamma_{s}^{n}}{E_p - E_{s} + \frac{1}{2} i \Gamma_{s}}.
\end{align}
Here $a$ is the off-resonance neutron scattering length. 

Mixing of the projectile neutron $p$ wave state with $p$ wave states above the centrifugal barrier as a mechanism of enhancement, considered in Ref.~\cite{Lewenkopf1992}, raises a purely theoretical question. There are always states above the barrier $\psi^{\text{above}}$, and we may arrange mixing with them by some small perturbation. Consider the simplest example of a small correction to the barrier potential $\delta V$. If we use first order perturbation theory to include mixing with a state above the barrier as a correction to the wave function, $\delta \psi= C_i \psi^{\text{above}}_i$,  this may seem like an enhancement of the tunneling probability since $\psi^{\text{above}}_i$ does not decrease under the barrier. However, a  small correction to the particle interaction can not remove the suppression of the tunneling amplitude when the energy of the particle is deep below the barrier energy (for example, a classical particle still can not travel through a wall).  This example shows that using  perturbation theory may be insufficient when we consider the under-barrier effect. Anyway, the contribution considered in Ref.~\cite{Lewenkopf1992} seems to be too small (see discussion in review \cite{MitchellReview2001}).

\section{Conclusion} \label{10}

In this work we considered the parity violating (PV) effect and the parity and time reversal invariance violating (PTRIV) effect in elastic neutron transmission. While for PV effects a polarised target is not needed, for comparison with the PTRIV effect it may be convenient to do both PV and PTRIV measurements with a polarised target. In the case where the spin of $p$-wave compound resonances is equal to $J=I-1/2$, the ratio of the PTRIV and PV effects is reduced to the ratio of the PTRIV and PV weak interaction matrix elements, i.e. the $p_{1/2}$ and $p_{3/2}$ capture amplitudes and all angular factors cancel out. This greatly simplifies the interpretation of the results. In the  case of  $J=I+1/2$ for polarised target and for PV effects measured with an unpolarised target, the ratio of PTRIV and PV effects contains the unknown ratio of the $p_{3/2}$ and $p_{1/2}$ neutron capture amplitudes, $M_{3/2}/M_{1/2}$. We presented two possibilities to measure this ratio $M_{3/2}/M_{1/2}$ within the same experimental arrangement.

Furthermore, statistical theory formulas linking root-mean-squared values of the weak matrix elements $W$ and $W_{PT}$ for PV and PTRIV interactions with experimental observables were presented, in both the polarised and unpolarised (for PV) target cases. We also presented numerical values of these matrix elements for different models of PV and CP-violating interactions. 


Possible explanations for the constant sign PV effects in neutron capture by $^{232}$Th  nucleus, observed in Refs.  ~\cite{232Th,MitchellReview2001,PhysRevC.58.1236}, have been discussed and the corresponding expressions for the PV and PTRIV effects have been presented. Our first preference is mixing by the weak interaction of the opposite parity doublet states in the nuclei with octupole deformation, in the excited state produced by the neutron capture. This mechanism was suggested in Refs.~\cite{Flambaum_1995,PhysRevLett.74.2638,Spevak1995}. There are experimental and theoretical indications that  $^{233}$Th may have a significant component with octupole deformation in the compound states formed after the neutron capture. We suggested a number of other target nuclei where such a mechanism may manifest itself, in order to test this hypothesis via PV experiments. If true, such nuclei with the constant sign effect may be convenient for the interpretation of PTRIV measurements since the ratio of the PTRIV and PV effects does not fluctuate as a random variable (within statistical theory this ratio is a random variable and requires proper statistical treatment). Also, if the doublet mechanism is confirmed, measurements of PV effects may be used to search for nuclei with octupole deformation.
  
Other possible explanations~\cite{FlambaumPwave1992,Hussein1995,Feshbach1996,PhysRevLett.74.2638} of the permanent sign PV effect in $^{232}$Th are  based on the contribution of local doorway states (with energies close to the $p$-wave compound resonance).  For example, the two-particle-one-hole doorway state mechanism was suggested in Refs.~\cite{Hussein1995,Feshbach1996}. We considered the contribution of local doorway states to PV and PTRIV effects. We found an additional contribution, which has an arbitrary sign and may be bigger than the permanent sign contribution considered in Refs.~\cite{Hussein1995,Feshbach1996,PhysRevLett.74.2638}, if there is an $s$-wave resonance very close to the $p$-wave resonance where PV is measured (note that in this case we expect the largest PV effects). As for the permanent sign contributions~\cite{Hussein1995,Feshbach1996,PhysRevLett.74.2638}, in order for the local doorway state mechanism to dominate, we require both doorway states of positive and negative parity to be close to the $p$-wave compound resonance and to have exceptionally small spreading widths. Therefore, these mechanisms do not seem to be the most plausible explanations.

  
We  would like to comment about a misconception  present in the literature. It is usually assumed that in the case of a random sign of the PV and PTRIV interaction matrix elements, the average value of the observed effects $\ev{P}$ must be zero. This is incorrect due to PV and PTRIV effects having singular dependence on the energy interval between the opposite parity states,  $P \propto 1/(E_s - E_p)$. If we treat the energy interval $E_s - E_p$ as a random variable, which has a finite probability density to be zero, we obtain an infinite variance of the effect $\ev{P^2}$. As a result,  $P$ is an example of a random-sign observable non-vanishing upon averaging~\cite{FlamGrib94}. The average value $\ev{P}$ does not decrease when we increase the number of measurements $n$, it tends to a constant. There is a simple qualitative explanation for this fact. For random terms with a finite variance $\sigma$, the average of $n$ random terms decreases with $n$ as $\sigma^{1/2} n^{-1/2}$. However, if one does $n$ measurements of $P$, in one of the measurements the energy interval $(E_s - E_p)$ will be $n$ times smaller than the average interval and $P$ will be $n$ times bigger than a typical value, $P_{\text{big}}  \sim n P_{\text{typ}}$. This single term in the sum for the average value of all $P$ has contribution $P_{\text{big}}/n \sim n P_{\text{typ}}/n = P_{\text{typ}}$, which does not decrease with $n$. Inclusion of the widths $\Gamma_s$ into the energy denominators makes  $\ev{P^2}$ finite, but so big that one would require many thousands of measurements on different compound resonances to achieve $\ev{P} \approx 0$~\cite{FlamGrib94}.

  
Finally, we conclude that with the expected experimental sensitivity  $10^{-6}$  ~\cite{PhysRevC.90.065503,Palos2018}, limits on the axion PTRIV interaction constants and other mechanisms of PTRIV may be significantly improved.

\section{Acknowledgements}

This work is supported by the Australian Research Council  Grants No. DP190100974 and DP200100150 and the JGU Gutenberg Research Fellowship.

\appendix

\section{Calculations of angular coefficients for PV and PTRIV effects} \label{AppendixA}

\subsection{Calculation of the Forward Scattering Amplitude with Time and Parity Violation} \label{CFSAPTRIV}
In this section, we will present our calculation of the forward scattering amplitude with time and parity violation, i.e. in the case when neutron momentum, neutron spin and target spin are all perpendicular to each other. Let us first consider the capture amplitude into $s$-wave resonance. For $J = I + 1/2$ we have the amplitude
 
\begin{align} \label{As+}
\begin{split}
\frac{M_{s}}{\sqrt{2}} C_{I \ I \ 1/2 \ 1/2}^{I + 1/2 \ I +1/2} \ket{I + 1/2, I + 1/2} \\  + \frac{M_{s}}{\sqrt{2}} C_{I \ I \ 1/2 \ -1/2}^{I + 1/2 \ I -1/2} \ket{I + 1/2, I - 1/2} ,
\end{split}
\end{align}

while for $J = I - 1/2$, we have
\begin{align} \label{As-}
\frac{M_{s}}{\sqrt{2}} C_{I \ I \ 1/2 \ -1/2}^{I - 1/2 \ I -1/2} \ket{I - 1/2, I - 1/2} 
\end{align}
Now, considering the capture amplitude into $p$-wave resonance, for $J = I + 1/2$, we have
\begin{widetext}
\begin{align}
\begin{split}
    - \frac{ M_{p,3/2}}{2} \left[  C_{I \ I \ 3/2 \ 1/2}^{I + 1/2 \ I + 1/2} \ket{I + 1/2, I + 1/2} 
    + C_{I \ I \ 3/2 \ -1/2}^{I + 1/2 \ I - 1/2} \ket{I + 1/2, I - 1/2} \right] \\  - \frac{ M_{p,1/2}}{\sqrt{2}} \left[  C_{I \ I \ 1/2 \ 1/2}^{I + 1/2 \ I + 1/2} \ket{I + 1/2, I + 1/2} 
    - C_{I \ I \ 1/2 \ -1/2}^{I + 1/2 \ I - 1/2} \ket{I + 1/2, I - 1/2} \right] ,
    \end{split}
\end{align}
while for $J = I - 1/2$

\begin{align}
\begin{split}
     - \frac{  M_{p,3/2}}{2} C_{I \ I \ 3/2 \ -1/2}^{I - 1/2 \ I - 1/2} \ket{I - 1/2, I - 1/2}  + \frac{  M_{p,1/2}}{\sqrt{2}}  C_{I \ I \ 1/2 \ -1/2}^{I - 1/2 \ I - 1/2} \ket{I - 1/2, I - 1/2}.
\end{split}
\end{align}

The Clebsch-Gordan coefficients of these states constitute the angular factors $\zeta$ of these capture amplitudes. Combining the contributions from both the $s$-wave and $p$-wave resonances, we can calculate $\zeta$ for $J = I \pm 1/2$
\begin{align}
\begin{split}
    \zeta(I+1/2) & = \frac{M_{s}}{\sqrt{2}} \left\{  - \frac{ M_{p,3/2}}{2}  \left( C_{I \ I \ 1/2 \ 1/2}^{I + 1/2 \ I +1/2} C_{I \ I \ 3/2 \ 1/2}^{I + 1/2 \ I +1/2} 
     + C_{I \ I \ 1/2 \ -1/2}^{I + 1/2 \ I -1/2} C_{I \ I \ 3/2 \ -1/2}^{I + 1/2 \ I -1/2}     \right) \right.  \\ 
    & \left. - \frac{ M_{p,1/2}}{\sqrt{2}} \biggl[ \left(C_{I \ I \ 1/2 \ 1/2}^{I + 1/2 \ I +1/2} \right)^{2} - \left(C_{I \ I \ 1/2 \ -1/2}^{I + 1/2 \ I -1/2} \right)^{2}   \biggr] \right\}, \\
\zeta(I-1/2) & = \frac{M_{s}}{\sqrt{2}} \left\{ - \frac{ M_{p,3/2}}{2} C_{I \ I \ 1/2 \ -1/2}^{I -1/2 \ I -1/2} C_{I \ I \ 3/2 \ -1/2}^{I -1/2 \ I -1/2} 
     + \frac{M_{p,1/2}}{\sqrt{2}} ( C_{I \ I \ 1/2 \ -1/2}^{I -1/2 \ I -1/2})^{2} \right\}.
\end{split}
\end{align}
\end{widetext}
The values of the Clebsh-Gordon coefficients are
\begin{align*} 
\begin{split}
    C_{I \ I \ 1/2 \ 1/2}^{I +1/2 \ I +1/2} & = 1, \\
    C_{I \ I \ 1/2 \ -1/2}^{I +1/2 \ I -1/2} & = \frac{1}{\sqrt{2 I + 1}} ,\\
    C_{I \ I \ 3/2 \ 1/2}^{I +1/2 \ I +1/2} & =  \sqrt{\frac{I}{I + 3/2}},\\
    C_{I \ I \ 3/2 \ -1/2}^{I +1/2 \ I -1/2} & = \sqrt{\frac{8I}{(2I+3)(2I+1)}}, \\
    C_{I \ I \ 1/2 \ -1/2}^{I -1/2 \ I -1/2} & = \sqrt{\frac{2I}{2I+1}}, \\
    C_{I \ I \ 3/2 \ -1/2}^{I -1/2 \ I -1/2} & = \sqrt{\frac{I(2I-1)}{(I + 1)(2I+1)}}.
\end{split}
\end{align*}
Substituting these values for the $J = I+1/2$ case yields


\begin{align}
\begin{split} \label{zetaplus}
\zeta(I+1/2) & =  \frac{ M_{s}}{2I + 1} \left\{ M_{p,1/2}I  - \frac{M_{p,3/2}}{2} \sqrt{I(2I + 3)}     \right\}.
\end{split}
\end{align}
Similarly, for the $J= I - 1/2$ case

\begin{align}
\begin{split} \label{zetaminus}
    \zeta (I-1/2) & = \frac{ M_{s}I}{(2I+1)} \left\{  M_{p,1/2} - \frac{M_{p,3/2}}{2} \sqrt{\frac{2I -1}{I + 1}}  \right\}. 
\end{split}
\end{align}
Therefore following the method of~\cite{Flambaum84}, we can now write the time and parity violating amplitude $f_{\text{t.p.v}}$

\begin{align} 
    f_{\text{t.p.v}} & = \pm \frac{1}{ k} \frac{M_s W_{sp}^{T,P}  (\alpha_{1/2,J}  M_{p,1/2} + \alpha_{3/2,J} M_{p,3/2})}
{\left(E-E_{s}+\frac{1}{2} i \Gamma_{s}\right)\left(E-E_{p}+\frac{1}{2} i \Gamma_{p}\right)},
\end{align}
where the angular coefficients $\alpha_{i.J} $ are
\begin{align} 
\begin{split}
\alpha_{1/2,J=I+1/2}& = \frac{I}{2I+1}, \\
\alpha_{3/2,J=I+1/2}& = - \frac{\sqrt{I(2I+3)}}{2(2I+1)}, \\
\alpha_{1/2,J=I-1/2}& = \frac{I}{2I+1}, \\
\alpha_{3/2,J=I-1/2}& = -\frac{I}{2(2I+1)} \sqrt{\frac{2I-1}{I+1}}.
\end{split}
\end{align}

\subsection{Calculation of the forward scattering amplitude for parity violation with a polarised target} \label{PVpolarisedAmplitude}
Now, we will present our calculation for the forward scattering amplitude for parity violation with a polarised target. In this case, we require neutron momentum and spin to be parallel, and both perpendicular to the nuclear target spin. We begin by noting that the $s$-wave capture amplitudes coincide with that of the T,P-odd case, (Equations (\ref{As+}, \ref{As-})). For a total spin of $J = I+1/2$, the $p$-wave capture amplitude is

\begin{widetext}
\begin{align} \label{PVp1}
\begin{split}
    - i \frac{ M_{p,3/2}}{2} \left[  C_{I \ I \ 3/2 \ 1/2}^{I + 1/2 \ I + 1/2} \ket{I + 1/2, I + 1/2} 
    - C_{I \ I \ 3/2 \ -1/2}^{I + 1/2 \ I - 1/2} \ket{I + 1/2, I - 1/2} \right] \\  - i\frac{ M_{p,1/2}}{\sqrt{2}} \left[  C_{I \ I \ 1/2 \ 1/2}^{I + 1/2 \ I + 1/2} \ket{I + 1/2, I + 1/2} 
    + C_{I \ I \ 1/2 \ -1/2}^{I + 1/2 \ I - 1/2} \ket{I + 1/2, I - 1/2} \right] ,
    \end{split}
\end{align}
while for $J = I - 1/2$

\begin{align}
\begin{split} \label{PVp2}
     i \frac{  M_{p,3/2}}{2} C_{I \ I \ 3/2 \ -1/2}^{I - 1/2 \ I - 1/2} \ket{I - 1/2, I - 1/2}  - i\frac{  M_{p,1/2}}{\sqrt{2}}  C_{I \ I \ 1/2 \ -1/2}^{I - 1/2 \ I - 1/2} \ket{I - 1/2, I - 1/2}.
\end{split}
\end{align}
\end{widetext}
Performing a similar calculation to Appendix \ref{CFSAPTRIV}, we yield the following for the forward scattering amplitude with parity violation (with a polarised target):

\begin{align} 
    f_{\text{p.v}}^{\text{P}} & = \pm \frac{1}{ k} \frac{M_s i W_{sp}  (\delta_{1/2,J}  M_{p,1/2} + \delta_{3/2,J} M_{p,3/2})}
{\left(E-E_{s}+\frac{1}{2} i \Gamma_{s}\right)\left(E-E_{p}+\frac{1}{2} i \Gamma_{p}\right)},
\end{align}
where the angular coefficients $\delta_{i.J} $ are
\begin{align} 
\begin{split}
\delta_{1/2,J=I+1/2}& = -\frac{I+1}{2I+1}, \\
\delta_{3/2,J=I+1/2}& = -\frac{2I-1}{2(2I+1)} \sqrt{\frac{I}{2I+3}}, \\
\delta_{1/2,J=I-1/2}& = -\frac{I}{2I+1}, \\
\delta_{3/2,J=I-1/2}& = \frac{I}{2(2I+1)} \sqrt{\frac{2I-1}{I+1}}.
\end{split}
\end{align}

\subsection{Calculation of the $p$-wave amplitude in the PTIRV experiment configuration} \label{pwavepolarised}

In this section, will present our calculation for the $p$-wave amplitude in the PTRIV configuration. In this case, neutron momentum, neutron spin and target spin are all perpendicular to each other. Let us first consider the $p$-wave capture amplitude for the $J = I + 1/2$ case. Noting that in calculations of the $p$-wave amplitude, in addition to $j_{z}=\pm 1/2$, we  have contributions from $ j=3/2, j_{z}=-3/2$, the capture amplitude becomes

\begin{widetext}
\begin{align}
\begin{split}
    A_{p}(I+1/2)&  = - \frac{ M_{p,3/2}}{2} \left[ \sqrt{3} \ C_{I \ I \ 3/2 \ -3/2}^{I + 1/2 \ I - 3/2} \ket{I + 1/2, I - 3/2} +  C_{I \ I \ 3/2 \ 1/2}^{I + 1/2 \ I + 1/2} \ket{I + 1/2, I + 1/2} \right. \\
    & \ \left.  + C_{I \ I \ 3/2 \ -1/2}^{I + 1/2 \ I - 1/2} \ket{I + 1/2, I - 1/2} \right]   - \frac{ M_{p,1/2}}{\sqrt{2}} \left[  C_{I \ I \ 1/2 \ 1/2}^{I + 1/2 \ I + 1/2} \ket{I + 1/2, I + 1/2} \right.
   \\ & \ \left.  - C_{I \ I \ 1/2 \ -1/2}^{I + 1/2 \ I - 1/2} \ket{I + 1/2, I - 1/2} \right].
    \end{split}
\end{align}
where
\end{widetext}
\begin{align}
    C_{I \ I \ 3/2 \ -3/2}^{I + 1/2 \ I - 3/2} = \sqrt{ \frac{6}{(2I+1)(2I+3)}}.
\end{align}
Using the same method presented in Appendix \ref{CFSAPTRIV}, the $p$-wave amplitude is equal to 

\begin{align}
    f_{p} = -\frac{1}{2k} \frac{(A_{p})^{2}}{E - E_{p} + \frac{1}{2} i \Gamma_{p}}. 
\end{align}
Thus, we must now calculate explicitly the square of the capture amplitude. We note that states of differing projections do not interact, meaning their cross term is equal to zero. Substitution of the Clebsh-Gordan coefficients gives

\begin{align}
\begin{split}
    (A_{p})^{2} & = \beta_{1,J = I+1/2} M_{p,1/2}^{2} + \beta_{13,J = I+1/2} M_{p,1/2} M_{p,3/2} \\ & + \beta_{3,J = I+1/2} M_{p,3/2}^{2},
    \end{split}
\end{align}
where

\begin{align} \label{betaJplus}
\begin{split}
    \beta_{1,J = I+1/2} & = \frac{I + 1}{2I+1} = g, \\
    \beta_{13,J = I+1/2} & = \frac{\sqrt{I}}{\sqrt{2I+3}} \frac{2I-1}{2I+1} , \\
    \beta_{3,J = I+1/2} & = \frac{2I^{2} + 5I + 9}{2(2I+3)(2I+1)}.
\end{split}
\end{align}
Now, let us consider the $J = I - 1/2$ case. In this case, noting that there are contributions from $j = 3/2, j_{z} = -3/2$, the $p$-wave capture amplitude is
\begin{widetext}

\begin{align}
\begin{split}
    A_{p} (I - 1/2) & = \frac{ M_{p,3/2}}{2} \left[ \sqrt{3} \ C_{I \ I \ 3/2 \ -3/2}^{I - 1/2 \ I - 3/2} \ket{I - 1/2, I - 3/2} +  C_{I \ I \ 3/2 \ -1/2}^{I - 1/2 \ I - 1/2} \ket{I - 1/2, I - 1/2}  \right]  \\ & \ +
    \frac{M_{p,1/2}}{\sqrt{2}} C_{I \ I \ 1/2 \ -1/2}^{I - 1/2 \ I - 1/2} \ket{I-1/2,I-1/2} ,
\end{split}
\end{align}
\end{widetext}

where

\begin{align}
     C_{I \ I \ 3/2 \ -3/2}^{I - 1/2 \ I - 3/2} = \sqrt{\frac{3I}{(1+I)(1+2I)}}.
\end{align}

Thus, in a similar way to above, we yield

\begin{align}
\begin{split}
(A_{p})^{2} & = \beta_{1,J = I-1/2} M_{p,1/2}^{2} + \beta_{13,J = I-1/2} M_{p,1/2} M_{p,3/2} \\ & + \beta_{3,J = I-1/2} M_{p,3/2}^{2},
\end{split}
\end{align}

where

\begin{align} \label{betaJminus}
\begin{split}
    \beta_{1,J = I-1/2} & = \frac{I}{2I+1} = g, \\
    \beta_{13,J = I-1/2} & = -\frac{I}{2I+1} \sqrt{\frac{2I-1}{I+1}}, \\
    \beta_{3,J = I-1/2} & = \frac{I(I+4)}{(2I+1)(2I+3)}.
\end{split}
\end{align}
Therefore, the $p$-wave forward scattering amplitude can be written as

\begin{align} 
    f_{p} = - \frac{1}{2k} \frac{\beta_{1,J} M_{p,1/2}^{2} + \beta_{13,J} M_{p,1/2} M_{p,3/2} + \beta_{3,J} M_{p,3/2}^{2} }{E - E_{p} + \frac{1}{2} i \Gamma_{p} },
\end{align}
where the coefficients for each individual case are given by Equations (\ref{betaJplus}, \ref{betaJminus}). A similar calculation using the capture amplitudes (\ref{PVp1}, \ref{PVp2}) shows that the $p$-wave amplitude for parity violation with a polarised target (which requires neutron momentum and spin to be parallel, and both perpendicular to the nuclear target spin) coincides with this expression.

\section{Calculation of the matrix elements of PV and PTRIV interaction between chaotic compound states.} \label{AppendixB}

In this section we present the calculation of the weak matrix elements between nuclear compound states, following the calculation performed by Ref.~\cite{Flam93PRL}. The short-range weak interaction nuclear PV potential acting on a nucleon may be presented in the following form:


\begin{align} 
    \hat W= \frac{Gg_{p,n}}{2 \sqrt{2}m} \{ (\sigma \mathbf{p}), \rho   \},
\end{align}
where $G$ is the Fermi constant, $m$ is the mass of the nucleon, $\sigma$ and $\mathbf{p}$ are the nucleon sigma matrix and momentum respectively, $\rho$ is the core nuclear number density and $g_{p,n}$ are dimensionless constants which are of the order of unity (for example, Ref.~\cite{FLAMBAUM1984367} obtained the proton constant to be $g_{p} = 4.6$ and the neutron constant to be $g_{n} \lesssim 1$). Since $\mathbf{p} = - i \nabla$, the matrix elements of $\hat W$ between discrete states are imaginary for a standard definition of angular wave functions. For a given compound state with angular momentum $j$ and parity $\pi$, the wave function may be expressed as

\begin{align} \label{CompoundState}
    \ket{j^{\pi}} = \sum_{\alpha} C_{\alpha} \ket{\alpha}, \ \ \ket{\alpha} = (a^{\dagger} bc^{\dagger} de^{\dagger} \hdots)_{j^{\pi}} \ket{0}, 
\end{align}
where here, the states $\ket{\alpha}$ are many particle excitations over the ground state $\ket{0}$.  Thus, the eigenstates $\ket{j^{\pi}}$ are a chaotic superposition of a large number of Hartree-Fock basis states $\ket{\alpha}$. We note here that the normalisation sum of the compound state (\ref{CompoundState}), $\sum_{\alpha} C_{\alpha}^{2} = 1$, is dominated by it's ``principal components''. Defining the energy of the compound state to be $E$, the energies of the principal components are within the interval $[E - \Gamma_{\text{spr}}/2, E + \Gamma_{\text{spr}}/2]$, where $\Gamma_{\text{spr}}$ is the component's spreading width. These components are produced by an excitation of nucleons inside nuclear valence shells. As per the book \textit{Nuclear Structure}~\cite{NuclearStructure}, the expansion coefficients $C_{\alpha}$ can be treated as Gaussian random variables (with $\expval{C_{\alpha}} = 0$), and can be written as

\begin{align} \label{TheAboveFunction}
\begin{split}
    \overline{C^{2}(E_{\alpha})} & = \frac{1}{\overline{N}} \Delta (\Gamma_{\text{spr}}, E - E_{\alpha}), , \ \ \ \overline{N} = \frac{\pi \Gamma_{\text{spr}}}{2d} ,\\
    \Delta (\Gamma_{\text{spr}}, E - E_{\alpha}) & = \frac{\Gamma_{\text{spr}}^{2}/4}{(E - E_{\alpha} )^{2} + \Gamma_{\text{spr}}^{2}/4}. 
\end{split}
\end{align}
Here, $E_{\alpha}$ is the energy of an arbitrary many-particle configuration and $\overline{N}$ is the number of principle components, which is expressed in terms of the average energy distance between nuclear compound resonances (with identical parities  and angular momenta) $d$. The factor $\Delta$ is a \textit{Breit-Wigner-type factor}, which governs the energy distance  $\left| E - E_{\alpha}\right| \leq \Gamma_{\text{spr}}/2$ at which states may be called principal components with the weight $\sim 1/\overline{N}$. 
 Thus, we can see that (\ref{TheAboveFunction}) calculates the probability to find the basis component $\ket{\alpha}$ in the compound state $\ket{j^{\pi}}$, and hence acts as a microcanonical partition function, which depends on the energy  of the isolated  system $E$. The canonical  statistical partition function for a system in a thermostat with  temperature $T$ gives the probability 
$\propto \exp(-E_{\alpha}/T)$. 

Now, we know that the weak interaction (\ref{WeakInteraction}) only mixes single-particle states with the same angular momentum, and opposite  parity. No such states are present in the valence shell, meaning it follows that the weak matrix element between two compound states of close energy is dominated by weak transitions between the ``principal'' components, $\ket{j^{\pi})}$ of one resonance, and the ``small'' components of the other ~\cite{Flam93PRL,Zaratsky}.
 
This means that the excitation of particles from the valence shell requires an energy as large as $ \sim 8$ MeV (which is much larger than the matrix elements due to the residual strong interaction $V$) leading out a configuration from the partition function of ``principal'' components according to (\ref{TheAboveFunction})~\cite{Flam93PRL}. Therefore, via the use of first order perturbation theory in the residual strong interaction $V$

\begin{align} \label{StrongInteraction}
    V = \frac{1}{2} \sum_{ab,cd} a^{\dagger} b V_{ab,cd} c^{\dagger} d,
\end{align}
an appropriate set of ``small'' configurations can be generated, meaning we can write the matrix elements of the weak interaction between compound states as

\begin{align} \label{Mel}
    \mel{s}{W}{p} = \sum_{\alpha} \frac{\mel{(s}{V}{\alpha} \mel{\alpha}{W}{p)}}{E - E_{\alpha}} +  \sum_{\beta} \frac{\mel{(s}{W}{\beta} \mel{\beta}{V}{p)}}{E - E_{\beta}},
\end{align}
where here, $\ket{\alpha}$ and $\ket{\beta}$ are small components, and $\ket{s)}$ and $\ket{p)}$ are the principal components of the compound states. As Eq. (\ref{WeakInteraction}) is a single particle operator, it can be included in the mean nuclear field, and thus transfer the perturbation theory expansion in the single-particle orbitals:


\begin{align}
    \tilde{\psi}_{a} = \psi_{a} + \sum_{A} \frac{\mel{\psi_{A}}{W}{\psi_{a}}}{\varepsilon_{a} - \varepsilon_{A}} \psi_{A},
\end{align}
where here, $\varepsilon_{a}$ and $\varepsilon_{A}$ are the energies of the orbitals $\psi_{a} $ and $\psi_{A}$ respectively, which have opposing parity. Therefore, the two-body  weak interaction may be renormalized by the strong interaction $V$ ~\cite{Flam93PRL,Flambaum1994V,Flambaum1995a}, and defining $\tilde{W}_{ab,cd} \equiv V(\tilde{a} \tilde{b},\tilde{c} \tilde{d})$, we can express it as~\cite{Flam93PRL}

\begin{align}\label{2bodyW}
\begin{split}
    \tilde{W}_{abcd} =  \sum_{A} V_{Ab,cd} \frac{\mel{\psi_{A}}{W}{\psi_{a}}}{\varepsilon_{a} - \varepsilon_{A}} + \sum_{B} V_{aB,cd} \frac{\mel{\psi_{B}}{W}{\psi_{b}}}{\varepsilon_{b} - \varepsilon_{B}} +  \\  \sum_{C} V_{ab,Cd}  \frac{\mel{\psi_{C}}{W}{\psi_{c}}}{\varepsilon_{c} - \varepsilon_{C}} + \sum_{d} V_{ab,cD} \frac{\mel{\psi_{D}}{W}{\psi_{d}}}{\varepsilon_{d} -
    \varepsilon_{d}}.
    \end{split}
\end{align}
Thus, rewriting (\ref{Mel}) as $\mel{s}{W}{p} = \mel{(s}{\tilde{W}}{p)}$, we see that the matrix elements between compound states can be expressed in terms of the matrix elements between single-shell particle states, meaning we can avoid considering explicitly the ``small'' components of each compound state. This is favourable, as it is not clear whether these components can be described by the same spreading widths as the principal components (Equation (\ref{TheAboveFunction})). The mean squared of this matrix element is

\begin{align}
    \overline{W_{sp}^{2}} \equiv \overline{\mel{p}{W}{s} \mel{s}{W}{p}} = 
    \overline{\mel{(p}{\tilde{W}}{s)} \mel{(s}{\tilde{W}}{p)}}  .
\end{align}
Expanding out the compound state $\ket{s}$ in terms of it's components using (\ref{CompoundState}) gives
\begin{align} \label{ExpectationValues}
   \overline{W_{sp}^{2}} & = \sum_{\alpha \beta} \overline{C_{\alpha} C_{\beta} \mel{(p}{\tilde{W}}{\alpha} \mel{\beta}{\tilde{W}}{p)}   }. 
\end{align}
Now, given the fact that the coefficients  $C_{\alpha}$ and $ C_{\beta}$ are statistically independent, we can rewrite their product as (see Ref.~\cite{NuclearStructure})

\begin{align}
\begin{split}
    \overline{C_{\alpha} C_{\beta}} & = \overline{C_{\alpha}^{2}} \delta_{\alpha \beta}, \\
    & = \delta_{\alpha \beta} \frac{1}{\overline{N}} \Delta (\Gamma_{\text{spr}}, E - E_{\alpha}),
\end{split}
\end{align}
using (\ref{TheAboveFunction}). Thus, combining these expressions, the mean squared of the matrix element can be written as 

\begin{align} \label{summationoveralpha}
   \overline{W_{sp}^{2}}   & = \sum_{\alpha} \frac{1}{\overline{N}} \Delta (\Gamma_{\text{spr}}, E - E_{\alpha}) \overline{ \mel{(p}{\tilde{W}}{\alpha } \mel{\alpha}{\tilde{W}}{p)}   }.
\end{align}
From here, we can use the fact that in the second quantisation, the summation over $\alpha$ in (\ref{summationoveralpha}) is equivalent to summation over the different components of the interaction (\ref{2bodyW}), meaning we are left with calculating $\mel{(p}{\tilde{W} \tilde{W}}{p)}$.  Applying this method  to (\ref{summationoveralpha}), we obtain~\cite{Flam93PRL}
\begin{widetext}
\begin{align} \label{PMSME}
\begin{split}
   \sqrt{\overline{W_{sp}^{2}}}=\sqrt{\frac{2 d}{\pi \Gamma_{\mathrm{spr}}}}\bigg\{\sum_{a b c d} \nu_{a}\left(1-\nu_{b}\right) \nu_{c}\left(1-\nu_{d}\right) \frac{1}{4}\left|\tilde{W}_{a b, c d}-\tilde{W}_{a d, c b}\right|^{2}  \times \Delta\left(\Gamma_{\mathrm{spr}}, \varepsilon_{a}-\varepsilon_{b}+\varepsilon_{c}-\varepsilon_{d}\right)\bigg\}^{\frac{1}{2}},
\end{split}
\end{align}
\end{widetext}
where $E - E_{\alpha} = \varepsilon_{a}-\varepsilon_{b}+\varepsilon_{c}-\varepsilon_{d}$ is the change in energy. The function 

\begin{align}
\Delta\left(\Gamma_{\mathrm{spr}}, \varepsilon_{a}-\varepsilon_{b}+\varepsilon_{c}-\varepsilon_{d}\right), 
\end{align}
can be viewed as an approximate energy conservation law, with accuracy up to the width of the states~\cite{Flam93PRL}. Indeed, 
\begin{align}
 \Delta(\Gamma_{\text{spr}}, E - E_{\alpha}) \rightarrow \frac{\pi \Gamma_{\text{spr}}}{2} \delta (E - E_{\alpha}),
\end{align}
when $\Gamma_{\text{spr}} \rightarrow 0$~\cite{Flam93PRL}.

For the calculation of the nucleon  orbital occupation numbers, defined as $\mel{(p}{a^{\dagger}b}{p)} = \delta_{ab} \nu_{a}$, we can replace the current microcanonical ensemble with the equivalent canonical ensemble, as per~\cite{Flam93PRL}. In general, the canonical ensemble may be chosen for a system with a large number of degrees of freedom via the introduction of the chemical potentials $\lambda_{n}, \lambda_{p}$ and the effective nuclear temperature $T$. Thus, we have that the expectation value in (\ref{ExpectationValues}) can be reduced to a standard canonical ensemble average. In doing so, we have that $\mel{(p}{a^{\dagger}b}{p)} = \delta_{ab} \nu_{a}^{T}$, where $\nu_{a}^{T}$ is the finite temperature Fermi occupation probability;

\begin{align}\label{nuT}
    \nu_{a}^{T} = \frac{1}{\exp[(\varepsilon_{a} - \lambda)/T] + 1}.
\end{align}
 Numerical simulations \cite{FlamGribakin1994,Izrailev1997} have shown that the orbital occupation numbers $\nu_{a}$ are indeed very close to the Fermi-Dirac distribution $\nu_{a}^{T}$. These formulas (\ref{PMSME}, \ref{nuT}) were used in~\cite{Flam93PRL} to perform numerical calculations of the root-mean-square weak matrix element between compound states, $\sqrt{\overline{|W_{sp}^{2}}|}\equiv W$. Furthermore, a calculation of this form can also be performed to determine the time and parity violating matrix elements between compound states, see Ref. ~\cite{Flambaum1995b}.

\begin{figure}[!h] 
    \centering
    \includegraphics[scale = 0.1]{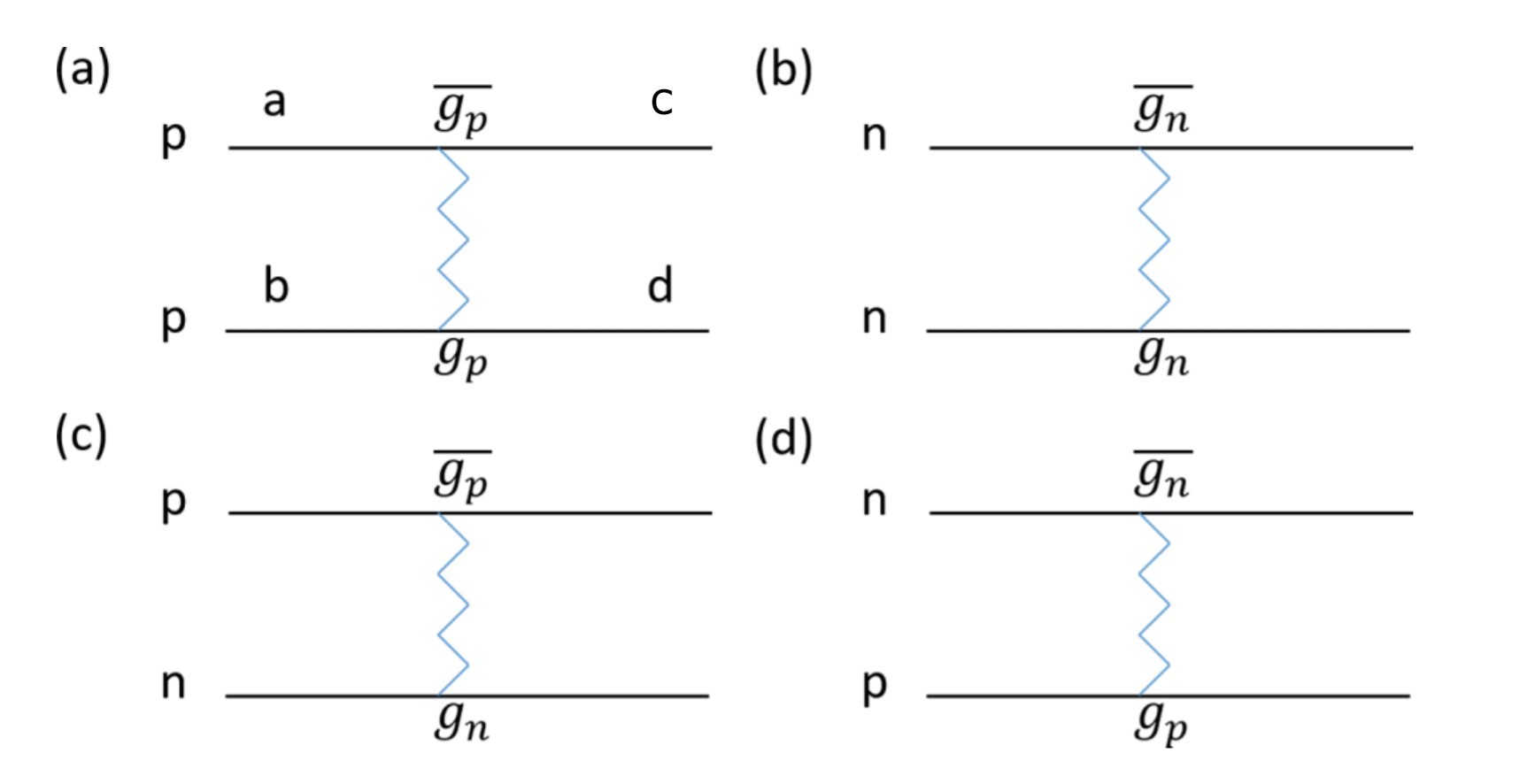}
    \caption{Possible configurations of weak interactions $W_{ab,cd}$~\cite{Flam93PRL} between protons and neutrons in the nucleus. (a) Interactions between two protons; (b) interactions between two neutrons; (c) and (d) interactions between a proton and a neutron. These interactions contribute to the PV matrix elements in Eq. (\ref{PMSME}) by $(W_{np} + W_{pn})^{2} = W_{np}^{2} + W_{pn}^{2} + 2 W_{np} W_{pn} $ (as presented in~\cite{Fadeev}). }
    \label{Interactions}
\end{figure}

 Numerical calculations of $W$ and $W_{P,T}$ were done in Refs.~\cite{Flam93PRL,Flambaum1995b} for specific values of $g_{p}, g_{n}$ in the P-odd interaction and $\eta_{p}, \eta_{n}$ in T,P odd interaction (which appear in the PV and TRIV operators respectively). As the values of these constants have been refined over time, we must first take this into account. Let us rewrite $W_{P,T}$ and $W$ in terms of the nucleon interaction constants $\eta_{p}, \eta_{n}$ and $g_{p},g_{n}$. Using~\cite{Flam93PRL,Flambaum1995b}

\begin{align}
\begin{split} \label{w1}
    W_{P,T} = \\ \sqrt{\frac{2 d}{\pi \Gamma_{\mathrm{spr}}}} \sqrt{ \left( \Sigma_{pp}^{(T,P)} \eta_{p}  \right)^{2} + \left( \Sigma_{nn}^{(T,P)} \eta_{n}  \right)^{2} + \left( \Sigma_{pn}^{(T,P)} \eta_{p} \eta_{n}  \right)   }, 
    \end{split} \\
    \begin{split} \label{v1}
    W  = \\\sqrt{\frac{2 d}{\pi \Gamma_{\mathrm{spr}}}} \sqrt{ \left( \Sigma_{pp}^{(P)} g_{p}  \right)^{2} + \left( \Sigma_{nn}^{(P)} g_{n}  \right)^{2} + \left( \Sigma_{pn}^{(P)} g_{p}g_{n}  \right)   }.
    \end{split}
\end{align}
Here, $\Sigma$ represents the sums of the weighted square matrix elements of the weak interaction between nucleon orbitals defined in Equation (\ref{PMSME})~\cite{Fadeev}. The cross terms $\Sigma_{pn}^{(P)} g_{p}g_{n} $ and $\Sigma_{pn}^{(T,P)} \eta_{p} \eta_{n} $ provide small contributions in comparison to the other terms, as they contain products of different matrix elements, with random sign. This is not the case for the terms which have squared interaction constants. As such, Equations (\ref{w1}, \ref{v1}) can be further simplified

\begin{align}
\begin{split} \label{wK}
W_{P,T} & = K_{T,P} \sqrt{\eta^{2}_{n} + k \eta^{2}_{p}}, 
\end{split} \\ 
\begin{split} \label{vK}
W & = K_{P} \sqrt{g^{2}_{n} + k g^{2}_{p}},
\end{split}
\end{align}
where the value of the constant $k$ should be slightly smaller than 1~\cite{Fadeev}, as in heavy nuclei, the number of neutrons $N = 1.5Z$, where $Z$ is the number of protons in the nucleus. $K_{T,P}$ and $K_{P}$ are constants which follow from (\ref{w1}, \ref{v1}). To estimate the behaviour of these mean squared matrix elements under changes to the strength constants, we can assume that the   $\Sigma$ in (\ref{w1}, \ref{v1}) are proportional to the number of interaction terms in the nucleus. These interactions are depicted in Figure \ref{Interactions}. Thus, there are $Z^{2}/2$ terms for interaction between protons, $N^{2}/2$ terms for interaction between neutrons, and $ZN$ for interactions between a proton and a neutron. Therefore, one can write~\cite{Fadeev}

\begin{align}
\nonumber
    k  = \frac{Z^{2} + 2 ZN}{N^{2} + 2ZN}\simeq 0.76.
\end{align}

Ref.~\cite{Flam93PRL} completed calculations of the parity violating mean squared matrix element $  W = 2.08$ meV for $g_{p} = 4, g_{n} = 1$, yielding

\begin{align} \label{vold}
    W & = K_{P} \sqrt{1 +  16k } \simeq 2.08 \ \text{meV},\\
 \implies     W & = 0.57\ \text{meV}\sqrt{g_n^2 + 0.76 g_p^2}\,.
\end{align}
 For the PTRIV mean squared matrix elements calculations in \cite{Flambaum1995b} yielded $W_{P,T} = 0.2 |\eta_{n}|$ meV, assuming $\eta_{p} = \eta_{n}$. Thus, we can apply this result here to yield

\begin{align}
    W_{P,T} & = K_{p} \sqrt{\eta_{n}^{2} + 0.76 \eta_{n}^{2}} = 0.2 |\eta| \text{meV} ,\\
 \implies    W_{P,T} & = 0.15 \ \text{meV} \sqrt{\eta_{n}^{2} + 0.76 \eta_{p}^{2}}.
\end{align}

\section{Possible evidence for static or dynamical octupole deformation in rotational spectra of various nuclei}

In this Appendix, we will discuss the rotational spectra for a few of the proposed nuclei in Section \ref{candidatenuclei}. For a target nucleus to be suitable for scattering experiments, we require a lifetime which exceeds $\sim 10^{6}$ years. The following list of candidates are proposed based on their rotational spectra, which may be obtained using the database~\cite{nudat}. Here we discuss static or dynamical octupole deformation (a soft octupole vibration mode) in the ground state. The soft octupole vibration mode indicates that the minimum of the deformation potential, corresponding to the octupole deformation,  may be  below neutron threshold. Some potential candidate nuclei include:

\begin{itemize}


    
    \item $^{153}_{63}$Eu$_{90}$ (stable): Rotational band including $E_{0}$ $(I^{P} = 5/2^{+})$ and $E \approx 83$ keV $(I^{P} = 7/2^{+})$. Opposite positive parity band is evident, with the relatively small energy intervals, which implies the existence of doublets of opposite parity states in this candidate nucleus.  Interval between the doublet   $(I^{P} = 5/2^{+})$ and  $(I^{P} = 5/2^{-})$ is $ 97$ keV. This may be an indication of static or dynamical octupole deformation.  
    
    
    
    \item $^{155}_{64}$Gd$_{91}$ (stable): Rotational band including  $E_{0}$ $(I^{P} = 3/2^{-})$ and $E \approx 86$ keV $(I^{P} = 5/2^{-})$. Opposite parity band is evident, but it possibly corresponds to a different internal nuclear state. Interval between the states  $(I^{P} = 3/2^{-})$ and  $(I^{P} = 3/2^{+})$ is $105$ keV. Octupole deformation is not excluded. 
    
    \item $^{157}_{64}$Gd$_{93}$ (stable): Rotational band including  $E_{0}$ $(I^{P} = 3/2^{-})$ and $E \approx 64$ keV $(I^{P} = 5/2^{-})$. Opposite parity band appears with the relatively large energy interval  between  $(I^{P} = 3/2^{-})$ and  $(I^{P} = 3/2^{+})$ equal to  $ 474$ keV. Possible soft octupole mode. Rotational opposite parity band with the small interval 10 keV appears at $I=9/2$. 
    
    \item $^{159}_{65}$Tb$_{94}$ (stable): Rotational band including  $E_{0}$ $(I^{P} = 3/2^{+})$ and $E \approx 58$ keV $(I^{P} = 5/2^{+})$. Opposite parity band appears starting from $I=5/2$ with the energy interval 305 keV. There may be soft octupole mode. 
    
    \item $^{161}_{66}$Dy$_{95}$ (stable): Rotational band including $E_{0}$ $(I^{P} = 5/2^{-})$ and $E \approx 103$ keV $(I^{P} = 7/2^{-})$. Opposite parity band is evident, ground state doublet splitting 25 keV. However, this interval increases with $I$. Possible octupole deformation.  
    
    \item $^{163}_{66}$Dy$_{97}$ (stable): Rotational band including $E_{0}$ $(I^{P} = 5/2^{-})$ and $E \approx 73$ keV $(I^{P} = 7/2^{-})$. Opposite positive parity band is evident, interval with the ground state 251 keV. However,  positive  parity band appears to cross the negative parity band. This implies that these bands may have differing moments of inertia. Thus, we conclude that there is weak evidence suggesting the existence of octupole deformation in this nuclide.

    \item $^{233}_{92}$U$_{141}$ (half-life $ = 1.6 \times 10^{6}$ years): Rotational band including  $E_{0}$ $(I^{P} = 5/2^{+})$ and $E \approx 40$ keV $(I^{P} = 7/2^{+})$. Opposite parity band is evident, ground state doublet splitting 300 keV. Possible soft octupole vibration mode. 
    
     \item $^{237}_{93}$Np$_{144}$ (half-life $ = 2.1 \times 10^{6}$ years): Rotational band including  $E_{0}$ $(I^{P} = 5/2^{+})$ and $E \approx 33$ keV $(I^{P} = 7/2^{+})$. Opposite parity band is evident. Interval between the doublet   $(I^{P} = 5/2^{+})$ and  $(I^{P} = 5/2^{-})$ is $ 59.5$ keV. Possible octupole deformation.

\end{itemize}

\bibliographystyle{apsrev4-2}
\bibliography{References1.bib}

\begin{thebibliography}{125}%
\makeatletter
\providecommand \@ifxundefined [1]{%
 \@ifx{#1\undefined}
}%
\providecommand \@ifnum [1]{%
 \ifnum #1\expandafter \@firstoftwo
 \else \expandafter \@secondoftwo
 \fi
}%
\providecommand \@ifx [1]{%
 \ifx #1\expandafter \@firstoftwo
 \else \expandafter \@secondoftwo
 \fi
}%
\providecommand \natexlab [1]{#1}%
\providecommand \enquote  [1]{``#1''}%
\providecommand \bibnamefont  [1]{#1}%
\providecommand \bibfnamefont [1]{#1}%
\providecommand \citenamefont [1]{#1}%
\providecommand \href@noop [0]{\@secondoftwo}%
\providecommand \href [0]{\begingroup \@sanitize@url \@href}%
\providecommand \@href[1]{\@@startlink{#1}\@@href}%
\providecommand \@@href[1]{\endgroup#1\@@endlink}%
\providecommand \@sanitize@url [0]{\catcode `\\12\catcode `\$12\catcode
  `\&12\catcode `\#12\catcode `\^12\catcode `\_12\catcode `\%12\relax}%
\providecommand \@@startlink[1]{}%
\providecommand \@@endlink[0]{}%
\providecommand \url  [0]{\begingroup\@sanitize@url \@url }%
\providecommand \@url [1]{\endgroup\@href {#1}{\urlprefix }}%
\providecommand \urlprefix  [0]{URL }%
\providecommand \Eprint [0]{\href }%
\providecommand \doibase [0]{https://doi.org/}%
\providecommand \selectlanguage [0]{\@gobble}%
\providecommand \bibinfo  [0]{\@secondoftwo}%
\providecommand \bibfield  [0]{\@secondoftwo}%
\providecommand \translation [1]{[#1]}%
\providecommand \BibitemOpen [0]{}%
\providecommand \bibitemStop [0]{}%
\providecommand \bibitemNoStop [0]{.\EOS\space}%
\providecommand \EOS [0]{\spacefactor3000\relax}%
\providecommand \BibitemShut  [1]{\csname bibitem#1\endcsname}%
\let\auto@bib@innerbib\@empty
\bibitem [{\citenamefont {Sushkov}\ and\ \citenamefont
  {Flambaum}(1980)}]{Sushkov80}%
  \BibitemOpen
  \bibfield  {author} {\bibinfo {author} {\bibfnamefont {O.}~\bibnamefont
  {Sushkov}}\ and\ \bibinfo {author} {\bibfnamefont {V.}~\bibnamefont
  {Flambaum}},\ }\href@noop {} {\bibfield  {journal} {\bibinfo  {journal} {JETP
  Lett.}\ }\textbf {\bibinfo {volume} {32}},\ \bibinfo {pages} {353} (\bibinfo
  {year} {1980})}\BibitemShut {NoStop}%
\bibitem [{\citenamefont {Sushkov}\ and\ \citenamefont
  {Flambaum}(1982)}]{Sushkov82}%
  \BibitemOpen
  \bibfield  {author} {\bibinfo {author} {\bibfnamefont {O.}~\bibnamefont
  {Sushkov}}\ and\ \bibinfo {author} {\bibfnamefont {V.}~\bibnamefont
  {Flambaum}},\ }\href@noop {} {\bibfield  {journal} {\bibinfo  {journal} {Sov.
  Phys. Usp.}\ }\textbf {\bibinfo {volume} {25}},\ \bibinfo {pages} {1}
  (\bibinfo {year} {1982})}\BibitemShut {NoStop}%
\bibitem [{\citenamefont {Flambaum}\ and\ \citenamefont
  {Sushkov}(1984)}]{Flambaum84}%
  \BibitemOpen
  \bibfield  {author} {\bibinfo {author} {\bibfnamefont {V.}~\bibnamefont
  {Flambaum}}\ and\ \bibinfo {author} {\bibfnamefont {O.}~\bibnamefont
  {Sushkov}},\ }\href
  {https://doi.org/https://doi.org/10.1016/0375-9474(84)90383-X} {\bibfield
  {journal} {\bibinfo  {journal} {Nuclear Physics A}\ }\textbf {\bibinfo
  {volume} {412}},\ \bibinfo {pages} {13 } (\bibinfo {year}
  {1984})}\BibitemShut {NoStop}%
\bibitem [{\citenamefont {Flambaum}\ and\ \citenamefont
  {Sushkov}(1985)}]{Flambaum85}%
  \BibitemOpen
  \bibfield  {author} {\bibinfo {author} {\bibfnamefont {V.}~\bibnamefont
  {Flambaum}}\ and\ \bibinfo {author} {\bibfnamefont {O.}~\bibnamefont
  {Sushkov}},\ }\href
  {https://doi.org/https://doi.org/10.1016/0375-9474(85)90469-5} {\bibfield
  {journal} {\bibinfo  {journal} {Nuclear Physics A}\ }\textbf {\bibinfo
  {volume} {435}},\ \bibinfo {pages} {352 } (\bibinfo {year}
  {1985})}\BibitemShut {NoStop}%
\bibitem [{\citenamefont {Alfimenkov}\ \emph {et~al.}(1981)\citenamefont
  {Alfimenkov}, \citenamefont {Borzakov}, \citenamefont {Tkhuan}, \citenamefont
  {Mareev}, \citenamefont {Pikelner}, \citenamefont {Ruben}, \citenamefont
  {Khrykin},\ and\ \citenamefont {Sharapov}}]{Alfimenkov1981}%
  \BibitemOpen
  \bibfield  {author} {\bibinfo {author} {\bibfnamefont {V.}~\bibnamefont
  {Alfimenkov}}, \bibinfo {author} {\bibfnamefont {S.}~\bibnamefont
  {Borzakov}}, \bibinfo {author} {\bibfnamefont {V.}~\bibnamefont {Tkhuan}},
  \bibinfo {author} {\bibfnamefont {Y.}~\bibnamefont {Mareev}}, \bibinfo
  {author} {\bibfnamefont {L.}~\bibnamefont {Pikelner}}, \bibinfo {author}
  {\bibfnamefont {D.}~\bibnamefont {Ruben}}, \bibinfo {author} {\bibfnamefont
  {A.}~\bibnamefont {Khrykin}},\ and\ \bibinfo {author} {\bibfnamefont
  {E.}~\bibnamefont {Sharapov}},\ }\href@noop {} {\bibfield  {journal}
  {\bibinfo  {journal} {JETP Lett.}\ }\textbf {\bibinfo {volume} {34}},\
  \bibinfo {pages} {295} (\bibinfo {year} {1981})}\BibitemShut {NoStop}%
\bibitem [{\citenamefont {Alfimenkov}\ \emph {et~al.}(1983)\citenamefont
  {Alfimenkov}, \citenamefont {Borzakov}, \citenamefont {{Van Thuan}},
  \citenamefont {Mareev}, \citenamefont {Pikelner}, \citenamefont {Khrykin},\
  and\ \citenamefont {Sharapov}}]{Alfimenkov1983}%
  \BibitemOpen
  \bibfield  {author} {\bibinfo {author} {\bibfnamefont {V.}~\bibnamefont
  {Alfimenkov}}, \bibinfo {author} {\bibfnamefont {S.}~\bibnamefont
  {Borzakov}}, \bibinfo {author} {\bibfnamefont {V.}~\bibnamefont {{Van
  Thuan}}}, \bibinfo {author} {\bibfnamefont {Y.}~\bibnamefont {Mareev}},
  \bibinfo {author} {\bibfnamefont {L.}~\bibnamefont {Pikelner}}, \bibinfo
  {author} {\bibfnamefont {.~A.}\ \bibnamefont {Khrykin}},\ and\ \bibinfo
  {author} {\bibfnamefont {E.}~\bibnamefont {Sharapov}},\ }\href
  {https://doi.org/https://doi.org/10.1016/0375-9474(83)90649-8} {\bibfield
  {journal} {\bibinfo  {journal} {Nuclear \\ Physics A}\ }\textbf {\bibinfo
  {volume} {398}},\ \bibinfo {pages} {93 } (\bibinfo {year}
  {1983})}\BibitemShut {NoStop}%
\bibitem [{\citenamefont {Alfimenkov}\ \emph {et~al.}(1984)\citenamefont
  {Alfimenkov}, \citenamefont {Borzakov}, \citenamefont {Tkhuan}, \citenamefont
  {Mareev}, \citenamefont {Pikelner}, \citenamefont {Frank}, \citenamefont
  {Khrykin},\ and\ \citenamefont {Sharapov}}]{Alfimenkov1984}%
  \BibitemOpen
  \bibfield  {author} {\bibinfo {author} {\bibfnamefont {V.}~\bibnamefont
  {Alfimenkov}}, \bibinfo {author} {\bibfnamefont {S.}~\bibnamefont
  {Borzakov}}, \bibinfo {author} {\bibfnamefont {V.}~\bibnamefont {Tkhuan}},
  \bibinfo {author} {\bibfnamefont {Y.}~\bibnamefont {Mareev}}, \bibinfo
  {author} {\bibfnamefont {L.}~\bibnamefont {Pikelner}}, \bibinfo {author}
  {\bibfnamefont {I.}~\bibnamefont {Frank}}, \bibinfo {author} {\bibfnamefont
  {A.}~\bibnamefont {Khrykin}},\ and\ \bibinfo {author} {\bibfnamefont
  {E.}~\bibnamefont {Sharapov}},\ }\href@noop {} {\bibfield  {journal}
  {\bibinfo  {journal} {JETP Lett.}\ }\textbf {\bibinfo {volume} {39:8}}
  (\bibinfo {year} {1984})}\BibitemShut {NoStop}%
\bibitem [{\citenamefont {Mitchell}\ \emph {et~al.}(1999)\citenamefont
  {Mitchell}, \citenamefont {Bowman},\ and\ \citenamefont
  {Weidenm\"uller}}]{MitchellReview99}%
  \BibitemOpen
  \bibfield  {author} {\bibinfo {author} {\bibfnamefont {G.~E.}\ \bibnamefont
  {Mitchell}}, \bibinfo {author} {\bibfnamefont {J.~D.}\ \bibnamefont
  {Bowman}},\ and\ \bibinfo {author} {\bibfnamefont {H.~A.}\ \bibnamefont
  {Weidenm\"uller}},\ }\href {https://doi.org/10.1103/RevModPhys.71.445}
  {\bibfield  {journal} {\bibinfo  {journal} {Rev. Mod. Phys.}\ }\textbf
  {\bibinfo {volume} {71}},\ \bibinfo {pages} {445} (\bibinfo {year}
  {1999})}\BibitemShut {NoStop}%
\bibitem [{\citenamefont {Mitchell}\ \emph {et~al.}(2001)\citenamefont
  {Mitchell}, \citenamefont {Bowman}, \citenamefont {Penttilä},\ and\
  \citenamefont {Sharapov}}]{MitchellReview2001}%
  \BibitemOpen
  \bibfield  {author} {\bibinfo {author} {\bibfnamefont {G.}~\bibnamefont
  {Mitchell}}, \bibinfo {author} {\bibfnamefont {J.}~\bibnamefont {Bowman}},
  \bibinfo {author} {\bibfnamefont {S.}~\bibnamefont {Penttilä}},\ and\
  \bibinfo {author} {\bibfnamefont {E.}~\bibnamefont {Sharapov}},\ }\href
  {https://doi.org/https://doi.org/10.1016/S0370-1573(01)00016-3} {\bibfield
  {journal} {\bibinfo  {journal} {Physics Reports}\ }\textbf {\bibinfo {volume}
  {354}},\ \bibinfo {pages} {157 } (\bibinfo {year} {2001})}\BibitemShut
  {NoStop}%
\bibitem [{\citenamefont {Bunakov}\ and\ \citenamefont
  {Gudkov}(1983)}]{Bunakov1983}%
  \BibitemOpen
  \bibfield  {author} {\bibinfo {author} {\bibfnamefont {V.}~\bibnamefont
  {Bunakov}}\ and\ \bibinfo {author} {\bibfnamefont {V.}~\bibnamefont
  {Gudkov}},\ }\href
  {https://doi.org/https://doi.org/10.1016/0375-9474(83)90338-X} {\bibfield
  {journal} {\bibinfo  {journal} {Nuclear Physics A}\ }\textbf {\bibinfo
  {volume} {401}},\ \bibinfo {pages} {93 } (\bibinfo {year}
  {1983})}\BibitemShut {NoStop}%
\bibitem [{\citenamefont {Gould}\ and\ \citenamefont {Bowman}(1987)}]{TRIV87}%
  \BibitemOpen
  \bibfield  {author} {\bibinfo {author} {\bibfnamefont {C.~R.}\ \bibnamefont
  {Gould}}\ and\ \bibinfo {author} {\bibfnamefont {N.~R.}\ \bibnamefont
  {Bowman}, \bibfnamefont {J.~D.~Robertson}},\ }\href@noop {} {\emph {\bibinfo
  {title} {Tests of time reversal invariance in neutron physics}}}\ (\bibinfo
  {publisher} {Singapore : World Scientific},\ \bibinfo {year}
  {1987})\BibitemShut {NoStop}%
\bibitem [{\citenamefont {Gudkov}(1992)}]{Gudkov1992}%
  \BibitemOpen
  \bibfield  {author} {\bibinfo {author} {\bibfnamefont {V.}~\bibnamefont
  {Gudkov}},\ }\href
  {https://doi.org/https://doi.org/10.1016/0370-1573(92)90121-F} {\bibfield
  {journal} {\bibinfo  {journal} {Physics Reports}\ }\textbf {\bibinfo {volume}
  {212}},\ \bibinfo {pages} {77 } (\bibinfo {year} {1992})}\BibitemShut
  {NoStop}%
\bibitem [{\citenamefont {Flambaum}\ and\ \citenamefont
  {Gribakin}(1995)}]{Flambaum_Gribakin95}%
  \BibitemOpen
  \bibfield  {author} {\bibinfo {author} {\bibfnamefont {V.}~\bibnamefont
  {Flambaum}}\ and\ \bibinfo {author} {\bibfnamefont {G.}~\bibnamefont
  {Gribakin}},\ }\href
  {https://doi.org/https://doi.org/10.1016/0146-6410(95)00045-K} {\bibfield
  {journal} {\bibinfo  {journal} {Progress in Particle and Nuclear Physics}\
  }\textbf {\bibinfo {volume} {35}},\ \bibinfo {pages} {423 } (\bibinfo {year}
  {1995})}\BibitemShut {NoStop}%
\bibitem [{\citenamefont {Fadeev}\ and\ \citenamefont
  {Flambaum}(2019)}]{Fadeev}%
  \BibitemOpen
  \bibfield  {author} {\bibinfo {author} {\bibfnamefont {P.}~\bibnamefont
  {Fadeev}}\ and\ \bibinfo {author} {\bibfnamefont {V.~V.}\ \bibnamefont
  {Flambaum}},\ }\href {https://doi.org/10.1103/physrevc.100.015504} {\bibfield
   {journal} {\bibinfo  {journal} {Phys. Rev. C}\ }\textbf {\bibinfo {volume}
  {100}} (\bibinfo {year} {2019})}\BibitemShut {NoStop}%
\bibitem [{\citenamefont {Kabir}(1982)}]{PhysRevD.25.2013}%
  \BibitemOpen
  \bibfield  {author} {\bibinfo {author} {\bibfnamefont {P.~K.}\ \bibnamefont
  {Kabir}},\ }\href {https://doi.org/10.1103/PhysRevD.25.2013} {\bibfield
  {journal} {\bibinfo  {journal} {Phys. Rev. D}\ }\textbf {\bibinfo {volume}
  {25}},\ \bibinfo {pages} {2013} (\bibinfo {year} {1982})}\BibitemShut
  {NoStop}%
\bibitem [{\citenamefont {Stodolsky}(1982)}]{STODOLSKY1982213}%
  \BibitemOpen
  \bibfield  {author} {\bibinfo {author} {\bibfnamefont {L.}~\bibnamefont
  {Stodolsky}},\ }\href
  {https://doi.org/https://doi.org/10.1016/0550-3213(82)90287-5} {\bibfield
  {journal} {\bibinfo  {journal} {Nuclear Physics B}\ }\textbf {\bibinfo
  {volume} {197}},\ \bibinfo {pages} {213} (\bibinfo {year}
  {1982})}\BibitemShut {NoStop}%
\bibitem [{\citenamefont {Snow}(2017)}]{Snow2017}%
  \BibitemOpen
  \bibfield  {author} {\bibinfo {author} {\bibfnamefont {W.~M.}\ \bibnamefont
  {Snow}},\ }\href {https://doi.org/10.22323/1.294.0007} {\bibfield  {journal}
  {\bibinfo  {journal} {PoS}\ }\textbf {\bibinfo {volume} {KMI2017}},\ \bibinfo
  {pages} {007} (\bibinfo {year} {2017})}\BibitemShut {NoStop}%
\bibitem [{\citenamefont {Okudaira}\ \emph {et~al.}(2018)\citenamefont
  {Okudaira}, \citenamefont {Takada}, \citenamefont {Hirota}, \citenamefont
  {Kimura}, \citenamefont {Kitaguchi}, \citenamefont {Koga}, \citenamefont
  {Nagamoto}, \citenamefont {Nakao}, \citenamefont {Okada}, \citenamefont
  {Sakai}, \citenamefont {Shimizu}, \citenamefont {Yamamoto},\ and\
  \citenamefont {Yoshioka}}]{Okudaira2018}%
  \BibitemOpen
  \bibfield  {author} {\bibinfo {author} {\bibfnamefont {T.}~\bibnamefont
  {Okudaira}}, \bibinfo {author} {\bibfnamefont {S.}~\bibnamefont {Takada}},
  \bibinfo {author} {\bibfnamefont {K.}~\bibnamefont {Hirota}}, \bibinfo
  {author} {\bibfnamefont {A.}~\bibnamefont {Kimura}}, \bibinfo {author}
  {\bibfnamefont {M.}~\bibnamefont {Kitaguchi}}, \bibinfo {author}
  {\bibfnamefont {J.}~\bibnamefont {Koga}}, \bibinfo {author} {\bibfnamefont
  {K.}~\bibnamefont {Nagamoto}}, \bibinfo {author} {\bibfnamefont
  {T.}~\bibnamefont {Nakao}}, \bibinfo {author} {\bibfnamefont
  {A.}~\bibnamefont {Okada}}, \bibinfo {author} {\bibfnamefont
  {K.}~\bibnamefont {Sakai}}, \bibinfo {author} {\bibfnamefont {H.~M.}\
  \bibnamefont {Shimizu}}, \bibinfo {author} {\bibfnamefont {T.}~\bibnamefont
  {Yamamoto}},\ and\ \bibinfo {author} {\bibfnamefont {T.}~\bibnamefont
  {Yoshioka}},\ }\href {https://doi.org/10.1103/PhysRevC.97.034622} {\bibfield
  {journal} {\bibinfo  {journal} {Phys. Rev. C}\ }\textbf {\bibinfo {volume}
  {97}},\ \bibinfo {pages} {034622} (\bibinfo {year} {2018})}\BibitemShut
  {NoStop}%
\bibitem [{\citenamefont {Yamamoto}\ \emph {et~al.}(2017)\citenamefont
  {Yamamoto}, \citenamefont {Shimizu}, \citenamefont {Kitaguchi}, \citenamefont
  {Hirota}, \citenamefont {Okudaira}, \citenamefont {Haddock}, \citenamefont
  {Oi}, \citenamefont {Ito}, \citenamefont {Endo}, \citenamefont {Takada},
  \citenamefont {Koga}, \citenamefont {Yoshioka}, \citenamefont {Ino},
  \citenamefont {Asahi}, \citenamefont {Momose}, \citenamefont {Iwata},
  \citenamefont {Sakai}, \citenamefont {Oku}, \citenamefont {Kimura},
  \citenamefont {Hino}, \citenamefont {Shima},\ and\ \citenamefont
  {Yamagata}}]{Kitaguchi2018}%
  \BibitemOpen
  \bibfield  {author} {\bibinfo {author} {\bibfnamefont {T.}~\bibnamefont
  {Yamamoto}}, \bibinfo {author} {\bibfnamefont {H.~M.}\ \bibnamefont
  {Shimizu}}, \bibinfo {author} {\bibfnamefont {M.}~\bibnamefont {Kitaguchi}},
  \bibinfo {author} {\bibfnamefont {K.}~\bibnamefont {Hirota}}, \bibinfo
  {author} {\bibfnamefont {T.}~\bibnamefont {Okudaira}}, \bibinfo {author}
  {\bibfnamefont {C.~C.}\ \bibnamefont {Haddock}}, \bibinfo {author}
  {\bibfnamefont {N.}~\bibnamefont {Oi}}, \bibinfo {author} {\bibfnamefont
  {I.}~\bibnamefont {Ito}}, \bibinfo {author} {\bibfnamefont {S.}~\bibnamefont
  {Endo}}, \bibinfo {author} {\bibfnamefont {S.}~\bibnamefont {Takada}},
  \bibinfo {author} {\bibfnamefont {J.}~\bibnamefont {Koga}}, \bibinfo {author}
  {\bibfnamefont {T.}~\bibnamefont {Yoshioka}}, \bibinfo {author}
  {\bibfnamefont {T.}~\bibnamefont {Ino}}, \bibinfo {author} {\bibfnamefont
  {K.}~\bibnamefont {Asahi}}, \bibinfo {author} {\bibfnamefont
  {T.}~\bibnamefont {Momose}}, \bibinfo {author} {\bibfnamefont
  {T.}~\bibnamefont {Iwata}}, \bibinfo {author} {\bibfnamefont
  {K.}~\bibnamefont {Sakai}}, \bibinfo {author} {\bibfnamefont
  {T.}~\bibnamefont {Oku}}, \bibinfo {author} {\bibfnamefont {A.}~\bibnamefont
  {Kimura}}, \bibinfo {author} {\bibfnamefont {M.}~\bibnamefont {Hino}},
  \bibinfo {author} {\bibfnamefont {T.}~\bibnamefont {Shima}},\ and\ \bibinfo
  {author} {\bibfnamefont {Y.}~\bibnamefont {Yamagata}},\ }\bibinfo {title}
  {Development of a neutron spin filter for a {T} violation search in compound
  nuclei},\ in\ \href {https://doi.org/10.7566/JPSCP.22.011018} {\emph
  {\bibinfo {booktitle} {Proceedings of the International Conference on Neutron
  Optics}}}\ (\bibinfo {year} {2017})\BibitemShut {NoStop}%
\bibitem [{\citenamefont {Beda}\ and\ \citenamefont {Skoy}(2007)}]{Beda2007}%
  \BibitemOpen
  \bibfield  {author} {\bibinfo {author} {\bibfnamefont {A.}~\bibnamefont
  {Beda}}\ and\ \bibinfo {author} {\bibfnamefont {V.}~\bibnamefont {Skoy}},\
  }\href {https://doi.org/10.1134/S1063779607060044} {\bibfield  {journal}
  {\bibinfo  {journal} {Physics of Particles and Nuclei}\ }\textbf {\bibinfo
  {volume} {38}},\ \bibinfo {pages} {775} (\bibinfo {year} {2007})}\BibitemShut
  {NoStop}%
\bibitem [{Bow(1997)}]{Bowman96V}%
  \BibitemOpen
  \bibinfo {title} {Numerous review articles appear in \textit{Parity and Time
  Reversal Violation in Compound States and Related Topics}}\ (\bibinfo
  {publisher} {edited by N. Auerbach and J. D. Bowman, World Scientific},\
  \bibinfo {year} {1997})\ \Eprint
  {https://arxiv.org/abs/https://www.worldscientific.com/doi/pdf/10.1142/3227}
  {https://www.worldscientific.com/doi/pdf/10.1142/3227} \BibitemShut {NoStop}%
\bibitem [{\citenamefont {Bowman}\ and\ \citenamefont
  {Gudkov}(2014)}]{PhysRevC.90.065503}%
  \BibitemOpen
  \bibfield  {author} {\bibinfo {author} {\bibfnamefont {J.~D.}\ \bibnamefont
  {Bowman}}\ and\ \bibinfo {author} {\bibfnamefont {V.}~\bibnamefont
  {Gudkov}},\ }\href {https://doi.org/10.1103/PhysRevC.90.065503} {\bibfield
  {journal} {\bibinfo  {journal} {Phys. Rev. C}\ }\textbf {\bibinfo {volume}
  {90}},\ \bibinfo {pages} {065503} (\bibinfo {year} {2014})}\BibitemShut
  {NoStop}%
\bibitem [{\citenamefont {Ryndin}(1968)}]{ryndin_1968}%
  \BibitemOpen
  \bibfield  {author} {\bibinfo {author} {\bibfnamefont {R.~M.}\ \bibnamefont
  {Ryndin}},\ }\bibinfo {title} {{$3^{\text{rd}}$} {P}roceedings of the {LNPI}
  {W}inter {S}chool (\textit{in {R}ussian)}}\ (\bibinfo {year} {Leningrad,
  USSR, 1968})\BibitemShut {NoStop}%
\bibitem [{\citenamefont {Bilenky}\ \emph {et~al.}(1964)\citenamefont
  {Bilenky}, \citenamefont {Lapidus},\ and\ \citenamefont
  {Ryndin}}]{Bilenky:1964pm}%
  \BibitemOpen
  \bibfield  {author} {\bibinfo {author} {\bibfnamefont {S.~M.}\ \bibnamefont
  {Bilenky}}, \bibinfo {author} {\bibfnamefont {L.~I.}\ \bibnamefont
  {Lapidus}},\ and\ \bibinfo {author} {\bibfnamefont {R.~M.}\ \bibnamefont
  {Ryndin}},\ }\href {https://doi.org/10.1070/PU1965v007n05ABEH003661}
  {\bibfield  {journal} {\bibinfo  {journal} {Usp. Fiz. Nauk}\ }\textbf
  {\bibinfo {volume} {84}},\ \bibinfo {pages} {243} (\bibinfo {year}
  {1964})}\BibitemShut {NoStop}%
\bibitem [{\citenamefont {Bilenky}\ \emph {et~al.}(1969)\citenamefont
  {Bilenky}, \citenamefont {Lapidus},\ and\ \citenamefont
  {Ryndin}}]{Bilenky:1969pm}%
  \BibitemOpen
  \bibfield  {author} {\bibinfo {author} {\bibfnamefont {S.~M.}\ \bibnamefont
  {Bilenky}}, \bibinfo {author} {\bibfnamefont {L.~I.}\ \bibnamefont
  {Lapidus}},\ and\ \bibinfo {author} {\bibfnamefont {R.~M.}\ \bibnamefont
  {Ryndin}},\ }\href@noop {} {\bibfield  {journal} {\bibinfo  {journal} {Phys.
  Usp.}\ }\textbf {\bibinfo {volume} {11}},\ \bibinfo {pages} {512} (\bibinfo
  {year} {1969})}\BibitemShut {NoStop}%
\bibitem [{\citenamefont {Kabir}(1988)}]{PhysRevD.37.1856}%
  \BibitemOpen
  \bibfield  {author} {\bibinfo {author} {\bibfnamefont {P.~K.}\ \bibnamefont
  {Kabir}},\ }\href {https://doi.org/10.1103/PhysRevD.37.1856} {\bibfield
  {journal} {\bibinfo  {journal} {Phys. Rev. D}\ }\textbf {\bibinfo {volume}
  {37}},\ \bibinfo {pages} {1856} (\bibinfo {year} {1988})}\BibitemShut
  {NoStop}%
\bibitem [{\citenamefont {Lamoreaux}\ and\ \citenamefont
  {Golub}(1994)}]{PhysRevD.50.5632}%
  \BibitemOpen
  \bibfield  {author} {\bibinfo {author} {\bibfnamefont {S.~K.}\ \bibnamefont
  {Lamoreaux}}\ and\ \bibinfo {author} {\bibfnamefont {R.}~\bibnamefont
  {Golub}},\ }\href {https://doi.org/10.1103/PhysRevD.50.5632} {\bibfield
  {journal} {\bibinfo  {journal} {Phys. Rev. D}\ }\textbf {\bibinfo {volume}
  {50}},\ \bibinfo {pages} {5632} (\bibinfo {year} {1994})}\BibitemShut
  {NoStop}%
\bibitem [{\citenamefont {Masuda}\ \emph {et~al.}(1992)\citenamefont {Masuda}
  \emph {et~al.}}]{masuda1992}%
  \BibitemOpen
  \bibfield  {author} {\bibinfo {author} {\bibfnamefont {Y.}~\bibnamefont
  {Masuda}} \emph {et~al.},\ }\bibinfo {title} {{P}roceedings of the {WEIN}
  `92}\ (\bibinfo {year} {edited by T. D. Vylov World Scientific, Singapore,
  1992})\BibitemShut {NoStop}%
\bibitem [{\citenamefont {Skoy}(1996)}]{PhysRevD.53.4070}%
  \BibitemOpen
  \bibfield  {author} {\bibinfo {author} {\bibfnamefont {V.~R.}\ \bibnamefont
  {Skoy}},\ }\href {https://doi.org/10.1103/PhysRevD.53.4070} {\bibfield
  {journal} {\bibinfo  {journal} {Phys. Rev. D}\ }\textbf {\bibinfo {volume}
  {53}},\ \bibinfo {pages} {4070} (\bibinfo {year} {1996})}\BibitemShut
  {NoStop}%
\bibitem [{\citenamefont {Lukashevich}\ \emph {et~al.}(2011)\citenamefont
  {Lukashevich}, \citenamefont {Aldushchenkov},\ and\ \citenamefont
  {Dallman}}]{PhysRevC.83.035501}%
  \BibitemOpen
  \bibfield  {author} {\bibinfo {author} {\bibfnamefont {V.~V.}\ \bibnamefont
  {Lukashevich}}, \bibinfo {author} {\bibfnamefont {A.~V.}\ \bibnamefont
  {Aldushchenkov}},\ and\ \bibinfo {author} {\bibfnamefont {D.}~\bibnamefont
  {Dallman}},\ }\href {https://doi.org/10.1103/PhysRevC.83.035501} {\bibfield
  {journal} {\bibinfo  {journal} {Phys. Rev. C}\ }\textbf {\bibinfo {volume}
  {83}},\ \bibinfo {pages} {035501} (\bibinfo {year} {2011})}\BibitemShut
  {NoStop}%
\bibitem [{\citenamefont {Masuda}(1998)}]{MASUDA1998479}%
  \BibitemOpen
  \bibfield  {author} {\bibinfo {author} {\bibfnamefont {Y.}~\bibnamefont
  {Masuda}},\ }\href
  {https://doi.org/https://doi.org/10.1016/S0375-9474(97)00726-4} {\bibfield
  {journal} {\bibinfo  {journal} {Nuclear Physics A}\ }\textbf {\bibinfo
  {volume} {629}},\ \bibinfo {pages} {479} (\bibinfo {year} {1998})},\ \bibinfo
  {note} {quark Lepton Nuclear Physics}\BibitemShut {NoStop}%
\bibitem [{\citenamefont {Masuda}(2000)}]{MASUDA2000632}%
  \BibitemOpen
  \bibfield  {author} {\bibinfo {author} {\bibfnamefont {Y.}~\bibnamefont
  {Masuda}},\ }\href
  {https://doi.org/https://doi.org/10.1016/S0168-9002(99)01053-0} {\bibfield
  {journal} {\bibinfo  {journal} {Nuclear Instruments and Methods in Physics
  Research Section A: Accelerators, Spectrometers, Detectors and Associated
  Equipment}\ }\textbf {\bibinfo {volume} {440}},\ \bibinfo {pages} {632}
  (\bibinfo {year} {2000})}\BibitemShut {NoStop}%
\bibitem [{\citenamefont {Kabir}(1989)}]{KABIR198963}%
  \BibitemOpen
  \bibfield  {author} {\bibinfo {author} {\bibfnamefont {P.}~\bibnamefont
  {Kabir}},\ }\href
  {https://doi.org/https://doi.org/10.1016/0168-9002(89)90249-0} {\bibfield
  {journal} {\bibinfo  {journal} {Nuclear Instruments and Methods in Physics
  Research Section A: Accelerators, Spectrometers, Detectors and Associated
  Equipment}\ }\textbf {\bibinfo {volume} {284}},\ \bibinfo {pages} {63}
  (\bibinfo {year} {1989})}\BibitemShut {NoStop}%
\bibitem [{\citenamefont {Baryshevsky}\ and\ \citenamefont
  {Podgoretsky}(1965)}]{Baryshevsky1965}%
  \BibitemOpen
  \bibfield  {author} {\bibinfo {author} {\bibfnamefont {V.}~\bibnamefont
  {Baryshevsky}}\ and\ \bibinfo {author} {\bibfnamefont {M.}~\bibnamefont
  {Podgoretsky}},\ }\href@noop {} {\bibfield  {journal} {\bibinfo  {journal}
  {Sov. Phys}\ }\textbf {\bibinfo {volume} {JETP 20}},\ \bibinfo {pages} {704}
  (\bibinfo {year} {1965})}\BibitemShut {NoStop}%
\bibitem [{\citenamefont {Abragam}\ and\ \citenamefont
  {Goldman}(1982)}]{Abragam1982}%
  \BibitemOpen
  \bibfield  {author} {\bibinfo {author} {\bibfnamefont {A.}~\bibnamefont
  {Abragam}}\ and\ \bibinfo {author} {\bibfnamefont {M.}~\bibnamefont
  {Goldman}},\ }\bibinfo {title} {Nuclear magnetism: Order and {D}isorder,
  {I}nternational {S}eries of {M}onographs on {P}hysics}\ (\bibinfo {year}
  {Oxford University Press, Oxford, UK, 1982})\BibitemShut {NoStop}%
\bibitem [{\citenamefont {Masuda}(1996)}]{MasudaBowman1996}%
  \BibitemOpen
  \bibfield  {author} {\bibinfo {author} {\bibfnamefont {Y.}~\bibnamefont
  {Masuda}},\ }\bibinfo {title} {{P}arity and {T}ime {R}eversal {V}iolation in
  {C}ompound {N}uclear {S}tates and {R}elated {T}opics}\ (\bibinfo {year}
  {edited by N. Auerbach and J. D. Bowman, World Scientific, Singapore, 1996})\
  p.~\bibinfo {pages} {83}\BibitemShut {NoStop}%
\bibitem [{\citenamefont {Serebrov}(1996)}]{Serebrov1996}%
  \BibitemOpen
  \bibfield  {author} {\bibinfo {author} {\bibfnamefont {A.~P.}\ \bibnamefont
  {Serebrov}},\ }\bibinfo {title} {{P}arity and {T}ime {R}eversal {V}iolation
  in {C}ompound {N}uclear {S}tates}\ (\bibinfo {year} {edited by N. Auerbach
  and J. D. Bowman, World Scientific, Singapore, 1996})\ p.\ \bibinfo {pages}
  {327}\BibitemShut {NoStop}%
\bibitem [{\citenamefont {Flambaum}(1993)}]{Flam93}%
  \BibitemOpen
  \bibfield  {author} {\bibinfo {author} {\bibfnamefont {V.~V.}\ \bibnamefont
  {Flambaum}},\ }\href {https://doi.org/10.1088/0031-8949/1993/t46/030}
  {\bibfield  {journal} {\bibinfo  {journal} {Physica Scripta}\ }\textbf
  {\bibinfo {volume} {T46}},\ \bibinfo {pages} {198} (\bibinfo {year}
  {1993})}\BibitemShut {NoStop}%
\bibitem [{\citenamefont {Flambaum}\ and\ \citenamefont
  {Vorov}(1993)}]{Flam93PRL}%
  \BibitemOpen
  \bibfield  {author} {\bibinfo {author} {\bibfnamefont {V.~V.}\ \bibnamefont
  {Flambaum}}\ and\ \bibinfo {author} {\bibfnamefont {O.~K.}\ \bibnamefont
  {Vorov}},\ }\href {https://doi.org/10.1103/PhysRevLett.70.4051} {\bibfield
  {journal} {\bibinfo  {journal} {Phys. Rev. Lett.}\ }\textbf {\bibinfo
  {volume} {70}},\ \bibinfo {pages} {4051} (\bibinfo {year}
  {1993})}\BibitemShut {NoStop}%
\bibitem [{\citenamefont {Flambaum}\ and\ \citenamefont
  {Izrailev}(1997{\natexlab{a}})}]{PhysRevE.56.5144}%
  \BibitemOpen
  \bibfield  {author} {\bibinfo {author} {\bibfnamefont {V.~V.}\ \bibnamefont
  {Flambaum}}\ and\ \bibinfo {author} {\bibfnamefont {F.~M.}\ \bibnamefont
  {Izrailev}},\ }\href {https://doi.org/10.1103/PhysRevE.56.5144} {\bibfield
  {journal} {\bibinfo  {journal} {Phys. Rev. E}\ }\textbf {\bibinfo {volume}
  {56}},\ \bibinfo {pages} {5144} (\bibinfo {year}
  {1997}{\natexlab{a}})}\BibitemShut {NoStop}%
\bibitem [{\citenamefont {Flambaum}\ \emph {et~al.}(2015)\citenamefont
  {Flambaum}, \citenamefont {Kozlov},\ and\ \citenamefont
  {Gribakin}}]{PhysRevA.91.052704}%
  \BibitemOpen
  \bibfield  {author} {\bibinfo {author} {\bibfnamefont {V.~V.}\ \bibnamefont
  {Flambaum}}, \bibinfo {author} {\bibfnamefont {M.~G.}\ \bibnamefont
  {Kozlov}},\ and\ \bibinfo {author} {\bibfnamefont {G.~F.}\ \bibnamefont
  {Gribakin}},\ }\href {https://doi.org/10.1103/PhysRevA.91.052704} {\bibfield
  {journal} {\bibinfo  {journal} {Phys. Rev. A}\ }\textbf {\bibinfo {volume}
  {91}},\ \bibinfo {pages} {052704} (\bibinfo {year} {2015})}\BibitemShut
  {NoStop}%
\bibitem [{\citenamefont {Flambaum}\ and\ \citenamefont
  {Vorov}(1994)}]{Flambaum1994V}%
  \BibitemOpen
  \bibfield  {author} {\bibinfo {author} {\bibfnamefont {V.~V.}\ \bibnamefont
  {Flambaum}}\ and\ \bibinfo {author} {\bibfnamefont {O.~K.}\ \bibnamefont
  {Vorov}},\ }\href {https://doi.org/10.1103/PhysRevC.49.1827} {\bibfield
  {journal} {\bibinfo  {journal} {Phys. Rev. C}\ }\textbf {\bibinfo {volume}
  {49}},\ \bibinfo {pages} {1827} (\bibinfo {year} {1994})}\BibitemShut
  {NoStop}%
\bibitem [{\citenamefont {Flambaum}\ and\ \citenamefont
  {Vorov}(1995{\natexlab{a}})}]{Flambaum1995a}%
  \BibitemOpen
  \bibfield  {author} {\bibinfo {author} {\bibfnamefont {V.~V.}\ \bibnamefont
  {Flambaum}}\ and\ \bibinfo {author} {\bibfnamefont {O.~K.}\ \bibnamefont
  {Vorov}},\ }\href {https://doi.org/10.1103/PhysRevC.51.1521} {\bibfield
  {journal} {\bibinfo  {journal} {Phys. Rev. C}\ }\textbf {\bibinfo {volume}
  {51}},\ \bibinfo {pages} {1521} (\bibinfo {year}
  {1995}{\natexlab{a}})}\BibitemShut {NoStop}%
\bibitem [{\citenamefont {Flambaum}\ and\ \citenamefont
  {Vorov}(1995{\natexlab{b}})}]{Flambaum1995b}%
  \BibitemOpen
  \bibfield  {author} {\bibinfo {author} {\bibfnamefont {V.~V.}\ \bibnamefont
  {Flambaum}}\ and\ \bibinfo {author} {\bibfnamefont {O.~K.}\ \bibnamefont
  {Vorov}},\ }\href {https://doi.org/10.1103/PhysRevC.51.2914} {\bibfield
  {journal} {\bibinfo  {journal} {Phys. Rev. C}\ }\textbf {\bibinfo {volume}
  {51}},\ \bibinfo {pages} {2914} (\bibinfo {year}
  {1995}{\natexlab{b}})}\BibitemShut {NoStop}%
\bibitem [{\citenamefont {Flambaum}\ and\ \citenamefont
  {Gribakin}(2000)}]{FlamGrib2000}%
  \BibitemOpen
  \bibfield  {author} {\bibinfo {author} {\bibfnamefont {V.}~\bibnamefont
  {Flambaum}}\ and\ \bibinfo {author} {\bibfnamefont {G.}~\bibnamefont
  {Gribakin}},\ }\href {https://doi.org/10.1080/13642810008205768} {\bibfield
  {journal} {\bibinfo  {journal} {Philosophical Magazine Part B}\ }\textbf
  {\bibinfo {volume} {80}},\ \bibinfo {pages} {2143} (\bibinfo {year}
  {2000})}\BibitemShut {NoStop}%
\bibitem [{\citenamefont {Flambaum}\ \emph {et~al.}(1994)\citenamefont
  {Flambaum}, \citenamefont {Gribakina}, \citenamefont {Gribakin},\ and\
  \citenamefont {Kozlov}}]{FlamGribakin1994}%
  \BibitemOpen
  \bibfield  {author} {\bibinfo {author} {\bibfnamefont {V.~V.}\ \bibnamefont
  {Flambaum}}, \bibinfo {author} {\bibfnamefont {A.~A.}\ \bibnamefont
  {Gribakina}}, \bibinfo {author} {\bibfnamefont {G.~F.}\ \bibnamefont
  {Gribakin}},\ and\ \bibinfo {author} {\bibfnamefont {M.~G.}\ \bibnamefont
  {Kozlov}},\ }\href {https://doi.org/10.1103/PhysRevA.50.267} {\bibfield
  {journal} {\bibinfo  {journal} {Phys. Rev. A}\ }\textbf {\bibinfo {volume}
  {50}},\ \bibinfo {pages} {267} (\bibinfo {year} {1994})}\BibitemShut
  {NoStop}%
\bibitem [{\citenamefont {Dzuba}\ \emph {et~al.}(2017)\citenamefont {Dzuba},
  \citenamefont {Berengut}, \citenamefont {Harabati},\ and\ \citenamefont
  {Flambaum}}]{DzubaBerengut2017}%
  \BibitemOpen
  \bibfield  {author} {\bibinfo {author} {\bibfnamefont {V.~A.}\ \bibnamefont
  {Dzuba}}, \bibinfo {author} {\bibfnamefont {J.~C.}\ \bibnamefont {Berengut}},
  \bibinfo {author} {\bibfnamefont {C.}~\bibnamefont {Harabati}},\ and\
  \bibinfo {author} {\bibfnamefont {V.~V.}\ \bibnamefont {Flambaum}},\ }\href
  {https://doi.org/10.1103/PhysRevA.95.012503} {\bibfield  {journal} {\bibinfo
  {journal} {Phys. Rev. A}\ }\textbf {\bibinfo {volume} {95}},\ \bibinfo
  {pages} {012503} (\bibinfo {year} {2017})}\BibitemShut {NoStop}%
\bibitem [{\citenamefont {Flambaum}\ \emph {et~al.}(1998)\citenamefont
  {Flambaum}, \citenamefont {Gribakina},\ and\ \citenamefont
  {Gribakin}}]{Flambaum1998StatisticsOE}%
  \BibitemOpen
  \bibfield  {author} {\bibinfo {author} {\bibfnamefont {V.}~\bibnamefont
  {Flambaum}}, \bibinfo {author} {\bibfnamefont {A.}~\bibnamefont
  {Gribakina}},\ and\ \bibinfo {author} {\bibfnamefont {G.}~\bibnamefont
  {Gribakin}},\ }\href@noop {} {\bibfield  {journal} {\bibinfo  {journal}
  {Physical Review A}\ }\textbf {\bibinfo {volume} {58}},\ \bibinfo {pages}
  {230} (\bibinfo {year} {1998})}\BibitemShut {NoStop}%
\bibitem [{\citenamefont {Flambaum}\ \emph {et~al.}(2002)\citenamefont
  {Flambaum}, \citenamefont {Gribakina}, \citenamefont {Gribakin},\ and\
  \citenamefont {Harabati}}]{FlamGribakin2002}%
  \BibitemOpen
  \bibfield  {author} {\bibinfo {author} {\bibfnamefont {V.~V.}\ \bibnamefont
  {Flambaum}}, \bibinfo {author} {\bibfnamefont {A.~A.}\ \bibnamefont
  {Gribakina}}, \bibinfo {author} {\bibfnamefont {G.~F.}\ \bibnamefont
  {Gribakin}},\ and\ \bibinfo {author} {\bibfnamefont {C.}~\bibnamefont
  {Harabati}},\ }\href {https://doi.org/10.1103/PhysRevA.66.012713} {\bibfield
  {journal} {\bibinfo  {journal} {Phys. Rev. A}\ }\textbf {\bibinfo {volume}
  {66}},\ \bibinfo {pages} {012713} (\bibinfo {year} {2002})}\BibitemShut
  {NoStop}%
\bibitem [{\citenamefont {Harabati}\ \emph {et~al.}(2017)\citenamefont
  {Harabati}, \citenamefont {Berengut}, \citenamefont {Flambaum},\ and\
  \citenamefont {Dzuba}}]{Harabati2017}%
  \BibitemOpen
  \bibfield  {author} {\bibinfo {author} {\bibfnamefont {C.}~\bibnamefont
  {Harabati}}, \bibinfo {author} {\bibfnamefont {J.~C.}\ \bibnamefont
  {Berengut}}, \bibinfo {author} {\bibfnamefont {V.~V.}\ \bibnamefont
  {Flambaum}},\ and\ \bibinfo {author} {\bibfnamefont {V.~A.}\ \bibnamefont
  {Dzuba}},\ }\href {https://doi.org/10.1088/1361-6455/aa6cd1} {\bibfield
  {journal} {\bibinfo  {journal} {Journal of Physics B: Atomic, Molecular and
  Optical Physics}\ }\textbf {\bibinfo {volume} {50}},\ \bibinfo {pages}
  {125004} (\bibinfo {year} {2017})}\BibitemShut {NoStop}%
\bibitem [{\citenamefont {Frankle}\ \emph {et~al.}(1992)\citenamefont
  {Frankle}, \citenamefont {Bowman}, \citenamefont {Bush}, \citenamefont
  {Delheij}, \citenamefont {Gould}, \citenamefont {Haase}, \citenamefont
  {Knudson}, \citenamefont {Mitchell}, \citenamefont {Penttil\"a},
  \citenamefont {Postma}, \citenamefont {Roberson}, \citenamefont {Seestrom},
  \citenamefont {Szymanski}, \citenamefont {Yoo}, \citenamefont {Yuan},\ and\
  \citenamefont {Zhu}}]{232Th}%
  \BibitemOpen
  \bibfield  {author} {\bibinfo {author} {\bibfnamefont {C.~M.}\ \bibnamefont
  {Frankle}}, \bibinfo {author} {\bibfnamefont {J.~D.}\ \bibnamefont {Bowman}},
  \bibinfo {author} {\bibfnamefont {J.~E.}\ \bibnamefont {Bush}}, \bibinfo
  {author} {\bibfnamefont {P.~P.~J.}\ \bibnamefont {Delheij}}, \bibinfo
  {author} {\bibfnamefont {C.~R.}\ \bibnamefont {Gould}}, \bibinfo {author}
  {\bibfnamefont {D.~G.}\ \bibnamefont {Haase}}, \bibinfo {author}
  {\bibfnamefont {J.~N.}\ \bibnamefont {Knudson}}, \bibinfo {author}
  {\bibfnamefont {G.~E.}\ \bibnamefont {Mitchell}}, \bibinfo {author}
  {\bibfnamefont {S.}~\bibnamefont {Penttil\"a}}, \bibinfo {author}
  {\bibfnamefont {H.}~\bibnamefont {Postma}}, \bibinfo {author} {\bibfnamefont
  {N.~R.}\ \bibnamefont {Roberson}}, \bibinfo {author} {\bibfnamefont {S.~J.}\
  \bibnamefont {Seestrom}}, \bibinfo {author} {\bibfnamefont {J.~J.}\
  \bibnamefont {Szymanski}}, \bibinfo {author} {\bibfnamefont {S.~H.}\
  \bibnamefont {Yoo}}, \bibinfo {author} {\bibfnamefont {V.~W.}\ \bibnamefont
  {Yuan}},\ and\ \bibinfo {author} {\bibfnamefont {X.}~\bibnamefont {Zhu}},\
  }\href {https://journals.aps.org/prc/abstract/10.1103/PhysRevC.46.778}
  {\bibfield  {journal} {\bibinfo  {journal} {Phys. Rev. C}\ }\textbf {\bibinfo
  {volume} {46}},\ \bibinfo {pages} {\\ 778} (\bibinfo {year}
  {1992})}\BibitemShut {NoStop}%
\bibitem [{\citenamefont {Stephenson}\ \emph {et~al.}(1998)\citenamefont
  {Stephenson}, \citenamefont {Bowman}, \citenamefont {Crawford}, \citenamefont
  {Delheij}, \citenamefont {Frankle}, \citenamefont {Iinuma}, \citenamefont
  {Knudson}, \citenamefont {Lowie}, \citenamefont {Masaike}, \citenamefont
  {Matsuda}, \citenamefont {Mitchell}, \citenamefont {Penttil\"a},
  \citenamefont {Postma}, \citenamefont {Roberson}, \citenamefont {Seestrom},
  \citenamefont {Sharapov}, \citenamefont {Yen},\ and\ \citenamefont
  {Yuan}}]{PhysRevC.58.1236}%
  \BibitemOpen
  \bibfield  {author} {\bibinfo {author} {\bibfnamefont {S.~L.}\ \bibnamefont
  {Stephenson}}, \bibinfo {author} {\bibfnamefont {J.~D.}\ \bibnamefont
  {Bowman}}, \bibinfo {author} {\bibfnamefont {B.~E.}\ \bibnamefont
  {Crawford}}, \bibinfo {author} {\bibfnamefont {P.~P.~J.}\ \bibnamefont
  {Delheij}}, \bibinfo {author} {\bibfnamefont {C.~M.}\ \bibnamefont
  {Frankle}}, \bibinfo {author} {\bibfnamefont {M.}~\bibnamefont {Iinuma}},
  \bibinfo {author} {\bibfnamefont {J.~N.}\ \bibnamefont {Knudson}}, \bibinfo
  {author} {\bibfnamefont {L.~Y.}\ \bibnamefont {Lowie}}, \bibinfo {author}
  {\bibfnamefont {A.}~\bibnamefont {Masaike}}, \bibinfo {author} {\bibfnamefont
  {Y.}~\bibnamefont {Matsuda}}, \bibinfo {author} {\bibfnamefont {G.~E.}\
  \bibnamefont {Mitchell}}, \bibinfo {author} {\bibfnamefont {S.~I.}\
  \bibnamefont {Penttil\"a}}, \bibinfo {author} {\bibfnamefont
  {H.}~\bibnamefont {Postma}}, \bibinfo {author} {\bibfnamefont {N.~R.}\
  \bibnamefont {Roberson}}, \bibinfo {author} {\bibfnamefont {S.~J.}\
  \bibnamefont {Seestrom}}, \bibinfo {author} {\bibfnamefont {E.~I.}\
  \bibnamefont {Sharapov}}, \bibinfo {author} {\bibfnamefont {Y.-F.}\
  \bibnamefont {Yen}},\ and\ \bibinfo {author} {\bibfnamefont {V.~W.}\
  \bibnamefont {Yuan}},\ }\href {https://doi.org/10.1103/PhysRevC.58.1236}
  {\bibfield  {journal} {\bibinfo  {journal} {Phys. Rev. C}\ }\textbf {\bibinfo
  {volume} {58}},\ \bibinfo {pages} {1236} (\bibinfo {year}
  {1998})}\BibitemShut {NoStop}%
\bibitem [{\citenamefont {{Desplanques}}\ and\ \citenamefont
  {{Noguera}}(1996)}]{Noguera1996}%
  \BibitemOpen
  \bibfield  {author} {\bibinfo {author} {\bibfnamefont {B.}~\bibnamefont
  {{Desplanques}}}\ and\ \bibinfo {author} {\bibfnamefont {S.}~\bibnamefont
  {{Noguera}}},\ }\href {https://doi.org/10.1016/0375-9474(95)00471-8}
  {\bibfield  {journal} {\bibinfo  {journal} {Nuclear Physics, A}\ }\textbf
  {\bibinfo {volume} {598}},\ \bibinfo {pages} {139} (\bibinfo {year}
  {1996})}\BibitemShut {NoStop}%
\bibitem [{\citenamefont {Blons}\ \emph {et~al.}(1984)\citenamefont {Blons},
  \citenamefont {Mazur}, \citenamefont {Paya}, \citenamefont {Ribrag},\ and\
  \citenamefont {Weigmann}}]{Blons:1984fqo}%
  \BibitemOpen
  \bibfield  {author} {\bibinfo {author} {\bibfnamefont {J.}~\bibnamefont
  {Blons}}, \bibinfo {author} {\bibfnamefont {C.}~\bibnamefont {Mazur}},
  \bibinfo {author} {\bibfnamefont {D.}~\bibnamefont {Paya}}, \bibinfo {author}
  {\bibfnamefont {M.}~\bibnamefont {Ribrag}},\ and\ \bibinfo {author}
  {\bibfnamefont {H.}~\bibnamefont {Weigmann}},\ }\href
  {https://doi.org/10.1016/0375-9474(84)90494-9} {\bibfield  {journal}
  {\bibinfo  {journal} {Nucl. Phys. A}\ }\textbf {\bibinfo {volume} {414}},\
  \bibinfo {pages} {1} (\bibinfo {year} {1984})}\BibitemShut {NoStop}%
\bibitem [{\citenamefont {Blons}(1989)}]{Blons1989}%
  \BibitemOpen
  \bibfield  {author} {\bibinfo {author} {\bibfnamefont {J.}~\bibnamefont
  {Blons}},\ }\href@noop {} {\bibfield  {journal} {\bibinfo  {journal} {Nucl.
  Phys. A}\ }\textbf {\bibinfo {volume} {502}},\ \bibinfo {pages} {121c}
  (\bibinfo {year} {1989})}\BibitemShut {NoStop}%
\bibitem [{\citenamefont {Flambaum}\ and\ \citenamefont
  {Sushkov}(1980)}]{FLAMBAUM1980277}%
  \BibitemOpen
  \bibfield  {author} {\bibinfo {author} {\bibfnamefont {V.}~\bibnamefont
  {Flambaum}}\ and\ \bibinfo {author} {\bibfnamefont {O.}~\bibnamefont
  {Sushkov}},\ }\href
  {https://doi.org/https://doi.org/10.1016/0370-2693(80)90877-1} {\bibfield
  {journal} {\bibinfo  {journal} {Physics Letters B}\ }\textbf {\bibinfo
  {volume} {94}},\ \bibinfo {pages} {277 } (\bibinfo {year}
  {1980})}\BibitemShut {NoStop}%
\bibitem [{\citenamefont {Sushkov}\ and\ \citenamefont
  {Flambaum}(1981{\natexlab{a}})}]{sushkovangular1981}%
  \BibitemOpen
  \bibfield  {author} {\bibinfo {author} {\bibfnamefont {O.}~\bibnamefont
  {Sushkov}}\ and\ \bibinfo {author} {\bibfnamefont {V.}~\bibnamefont
  {Flambaum}},\ }\href@noop {} {\bibfield  {journal} {\bibinfo  {journal} {Yad.
  Fiz. 33(3)}\ }\textbf {\bibinfo {volume} {629}},\ \bibinfo {pages} {1981.
  Sov. J. Nucl. Phys. 33, 329} (\bibinfo {year}
  {1981}{\natexlab{a}})}\BibitemShut {NoStop}%
\bibitem [{\citenamefont {Sushkov}\ and\ \citenamefont
  {Flambaum}(1981{\natexlab{b}})}]{sushkovnature1981}%
  \BibitemOpen
  \bibfield  {author} {\bibinfo {author} {\bibfnamefont {O.}~\bibnamefont
  {Sushkov}}\ and\ \bibinfo {author} {\bibfnamefont {V.}~\bibnamefont
  {Flambaum}},\ }\href@noop {} {\bibfield  {journal} {\bibinfo  {journal} {Yad.
  Fiz. 33(1)}\ }\textbf {\bibinfo {volume} {59}},\ \bibinfo {pages} {1981. Sov.
  J. Nucl. Phys. 33, 31} (\bibinfo {year} {1981}{\natexlab{b}})}\BibitemShut
  {NoStop}%
\bibitem [{\citenamefont {Flambaum}\ and\ \citenamefont
  {Zelevinsky}(1995)}]{Flambaum_1995}%
  \BibitemOpen
  \bibfield  {author} {\bibinfo {author} {\bibfnamefont {V.}~\bibnamefont
  {Flambaum}}\ and\ \bibinfo {author} {\bibfnamefont {V.}~\bibnamefont
  {Zelevinsky}},\ }\href {https://doi.org/10.1016/0370-2693(95)00325-f}
  {\bibfield  {journal} {\bibinfo  {journal} {Physics Letters B}\ }\textbf
  {\bibinfo {volume} {350}},\ \bibinfo {pages} {8–12} (\bibinfo {year}
  {1995})}\BibitemShut {NoStop}%
\bibitem [{\citenamefont {Auerbach}\ \emph {et~al.}(1995)\citenamefont
  {Auerbach}, \citenamefont {Bowman},\ and\ \citenamefont
  {Spevak}}]{PhysRevLett.74.2638}%
  \BibitemOpen
  \bibfield  {author} {\bibinfo {author} {\bibfnamefont {N.}~\bibnamefont
  {Auerbach}}, \bibinfo {author} {\bibfnamefont {J.~D.}\ \bibnamefont
  {Bowman}},\ and\ \bibinfo {author} {\bibfnamefont {V.}~\bibnamefont
  {Spevak}},\ }\href {https://doi.org/10.1103/PhysRevLett.74.2638} {\bibfield
  {journal} {\bibinfo  {journal} {Phys. Rev. Lett.}\ }\textbf {\bibinfo
  {volume} {74}},\ \bibinfo {pages} {2638} (\bibinfo {year}
  {1995})}\BibitemShut {NoStop}%
\bibitem [{\citenamefont {Spevak}\ and\ \citenamefont
  {Auerbach}(1995)}]{Spevak1995}%
  \BibitemOpen
  \bibfield  {author} {\bibinfo {author} {\bibfnamefont {V.}~\bibnamefont
  {Spevak}}\ and\ \bibinfo {author} {\bibfnamefont {N.}~\bibnamefont
  {Auerbach}},\ }\href
  {https://doi.org/https://doi.org/10.1016/0370-2693(95)01099-C} {\bibfield
  {journal} {\bibinfo  {journal} {Physics Letters B}\ }\textbf {\bibinfo
  {volume} {359}},\ \bibinfo {pages} {254} (\bibinfo {year}
  {1995})}\BibitemShut {NoStop}%
\bibitem [{\citenamefont {Butler}(2016)}]{Butler_2016}%
  \BibitemOpen
  \bibfield  {author} {\bibinfo {author} {\bibfnamefont {P.~A.}\ \bibnamefont
  {Butler}},\ }\href {https://doi.org/10.1088/0954-3899/43/7/073002} {\bibfield
   {journal} {\bibinfo  {journal} {Journal of Physics G: Nuclear and Particle
  Physics}\ }\textbf {\bibinfo {volume} {43}},\ \bibinfo {pages} {073002}
  (\bibinfo {year} {2016})}\BibitemShut {NoStop}%
\bibitem [{\citenamefont {Cao}\ \emph {et~al.}(2020)\citenamefont {Cao},
  \citenamefont {Agbemava}, \citenamefont {Afanasjev}, \citenamefont
  {Nazarewicz},\ and\ \citenamefont {Olsen}}]{Afanasjev}%
  \BibitemOpen
  \bibfield  {author} {\bibinfo {author} {\bibfnamefont {Y.}~\bibnamefont
  {Cao}}, \bibinfo {author} {\bibfnamefont {S.~E.}\ \bibnamefont {Agbemava}},
  \bibinfo {author} {\bibfnamefont {A.~V.}\ \bibnamefont {Afanasjev}}, \bibinfo
  {author} {\bibfnamefont {W.}~\bibnamefont {Nazarewicz}},\ and\ \bibinfo
  {author} {\bibfnamefont {E.}~\bibnamefont {Olsen}},\ }\href
  {https://doi.org/10.1103/PhysRevC.102.024311} {\bibfield  {journal} {\bibinfo
   {journal} {Phys. Rev. C}\ }\textbf {\bibinfo {volume} {102}},\ \bibinfo
  {pages} {024311} (\bibinfo {year} {2020})}\BibitemShut {NoStop}%
\bibitem [{\citenamefont {Butler}(2020)}]{Butler}%
  \BibitemOpen
  \bibfield  {author} {\bibinfo {author} {\bibfnamefont {P.}~\bibnamefont
  {Butler}},\ }\href {https://doi.org/10.1098/rspa.2020.0202} {\bibfield
  {journal} {\bibinfo  {journal} {Proceedings of the Royal Society A:
  Mathematical, Physical and Engineering Sciences}\ }\textbf {\bibinfo {volume}
  {476}},\ \bibinfo {pages} {20200202} (\bibinfo {year} {2020})}\BibitemShut
  {NoStop}%
\bibitem [{\citenamefont {Nazarewicz}(1994)}]{Nazarevich}%
  \BibitemOpen
  \bibfield  {author} {\bibinfo {author} {\bibfnamefont {W.}~\bibnamefont
  {Nazarewicz}},\ }\href
  {https://doi.org/https://doi.org/10.1016/0375-9474(94)90037-X} {\bibfield
  {journal} {\bibinfo  {journal} {Nuclear Physics A}\ }\textbf {\bibinfo
  {volume} {574}},\ \bibinfo {pages} {27 } (\bibinfo {year}
  {1994})}\BibitemShut {NoStop}%
\bibitem [{\citenamefont {Flambaum}\ and\ \citenamefont
  {Feldmeier}(2020)}]{FlambaumFeldmeier}%
  \BibitemOpen
  \bibfield  {author} {\bibinfo {author} {\bibfnamefont {V.~V.}\ \bibnamefont
  {Flambaum}}\ and\ \bibinfo {author} {\bibfnamefont {H.}~\bibnamefont
  {Feldmeier}},\ }\href {https://doi.org/10.1103/PhysRevC.101.015502}
  {\bibfield  {journal} {\bibinfo  {journal} {Phys. Rev. C}\ }\textbf {\bibinfo
  {volume} {101}},\ \bibinfo {pages} {015502} (\bibinfo {year}
  {2020})}\BibitemShut {NoStop}%
\bibitem [{nud()}]{nudat}%
  \BibitemOpen
  \href@noop {} {\bibinfo {title} {The {NUDAT} program for nuclear data}},\
  \bibinfo {howpublished} {\url{https://www.nndc.bnl.gov/nudat2/}}\BibitemShut
  {NoStop}%
\bibitem [{\citenamefont {Flambaum}\ and\ \citenamefont
  {Dzuba}(2020)}]{FlambaumDzuba}%
  \BibitemOpen
  \bibfield  {author} {\bibinfo {author} {\bibfnamefont {V.~V.}\ \bibnamefont
  {Flambaum}}\ and\ \bibinfo {author} {\bibfnamefont {V.~A.}\ \bibnamefont
  {Dzuba}},\ }\href {https://doi.org/10.1103/PhysRevA.101.042504} {\bibfield
  {journal} {\bibinfo  {journal} {Phys. Rev. A}\ }\textbf {\bibinfo {volume}
  {101}},\ \bibinfo {pages} {042504} (\bibinfo {year} {2020})}\BibitemShut
  {NoStop}%
\bibitem [{\citenamefont {Lane}\ and\ \citenamefont
  {Pendlebury}(1960)}]{LANE196039}%
  \BibitemOpen
  \bibfield  {author} {\bibinfo {author} {\bibfnamefont {A.}~\bibnamefont
  {Lane}}\ and\ \bibinfo {author} {\bibfnamefont {E.}~\bibnamefont
  {Pendlebury}},\ }\href
  {https://doi.org/https://doi.org/10.1016/0029-5582(60)90280-7} {\bibfield
  {journal} {\bibinfo  {journal} {Nuclear Physics}\ }\textbf {\bibinfo {volume}
  {15}},\ \bibinfo {pages} {39} (\bibinfo {year} {1960})}\BibitemShut {NoStop}%
\bibitem [{\citenamefont {Flambaum}(1992)}]{FlambaumPwave1992}%
  \BibitemOpen
  \bibfield  {author} {\bibinfo {author} {\bibfnamefont {V.~V.}\ \bibnamefont
  {Flambaum}},\ }\href {https://doi.org/10.1103/PhysRevC.45.437} {\bibfield
  {journal} {\bibinfo  {journal} {Phys. Rev. C}\ }\textbf {\bibinfo {volume}
  {45}},\ \bibinfo {pages} {437} (\bibinfo {year} {1992})}\BibitemShut
  {NoStop}%
\bibitem [{\citenamefont {Hussein}\ \emph {et~al.}(1995)\citenamefont
  {Hussein}, \citenamefont {Kerman},\ and\ \citenamefont {Lin}}]{Hussein1995}%
  \BibitemOpen
  \bibfield  {author} {\bibinfo {author} {\bibfnamefont {M.~S.}\ \bibnamefont
  {Hussein}}, \bibinfo {author} {\bibfnamefont {A.~K.}\ \bibnamefont
  {Kerman}},\ and\ \bibinfo {author} {\bibfnamefont {C.-Y.}\ \bibnamefont
  {Lin}},\ }\href@noop {} {\bibfield  {journal} {\bibinfo  {journal} {Z. Phys.
  A}\ }\textbf {\bibinfo {volume} {351}},\ \bibinfo {pages} {301} (\bibinfo
  {year} {1995})}\BibitemShut {NoStop}%
\bibitem [{\citenamefont {Feshbach}\ \emph {et~al.}(1996)\citenamefont
  {Feshbach}, \citenamefont {Hussein},\ and\ \citenamefont
  {Kerman}}]{Feshbach1996}%
  \BibitemOpen
  \bibfield  {author} {\bibinfo {author} {\bibfnamefont {H.}~\bibnamefont
  {Feshbach}}, \bibinfo {author} {\bibfnamefont {M.~S.}\ \bibnamefont
  {Hussein}},\ and\ \bibinfo {author} {\bibfnamefont {A.~K.}\ \bibnamefont
  {Kerman}},\ }\bibinfo {title} {{P}arity and {T}ime {R}eversal {V}iolation in
  {C}ompound {N}uclear {S}tates and {R}elated {T}opics}\ (\bibinfo {year}
  {edited by N. Auerbach and J. D. Bowman, World Scientific, Singapore, 1996})\
  p.\ \bibinfo {pages} {157}\BibitemShut {NoStop}%
\bibitem [{\citenamefont {Auerbach}(1992)}]{PhysRevC.45.R514}%
  \BibitemOpen
  \bibfield  {author} {\bibinfo {author} {\bibfnamefont {N.}~\bibnamefont
  {Auerbach}},\ }\href {https://doi.org/10.1103/PhysRevC.45.R514} {\bibfield
  {journal} {\bibinfo  {journal} {Phys. Rev. C}\ }\textbf {\bibinfo {volume}
  {45}},\ \bibinfo {pages} {R514} (\bibinfo {year} {1992})}\BibitemShut
  {NoStop}%
\bibitem [{\citenamefont {Auerbach}\ and\ \citenamefont
  {Bowman}(1992)}]{PhysRevC.46.2582}%
  \BibitemOpen
  \bibfield  {author} {\bibinfo {author} {\bibfnamefont {N.}~\bibnamefont
  {Auerbach}}\ and\ \bibinfo {author} {\bibfnamefont {J.~D.}\ \bibnamefont
  {Bowman}},\ }\href {https://doi.org/10.1103/PhysRevC.46.2582} {\bibfield
  {journal} {\bibinfo  {journal} {Phys. Rev. C}\ }\textbf {\bibinfo {volume}
  {46}},\ \bibinfo {pages} {2582} (\bibinfo {year} {1992})}\BibitemShut
  {NoStop}%
\bibitem [{\citenamefont {Auerbach}\ and\ \citenamefont
  {Spevak}(1994)}]{PhysRevC.50.1456}%
  \BibitemOpen
  \bibfield  {author} {\bibinfo {author} {\bibfnamefont {N.}~\bibnamefont
  {Auerbach}}\ and\ \bibinfo {author} {\bibfnamefont {V.}~\bibnamefont
  {Spevak}},\ }\href {https://doi.org/10.1103/PhysRevC.50.1456} {\bibfield
  {journal} {\bibinfo  {journal} {Phys. Rev. C}\ }\textbf {\bibinfo {volume}
  {50}},\ \bibinfo {pages} {1456} (\bibinfo {year} {1994})}\BibitemShut
  {NoStop}%
\bibitem [{\citenamefont {Bowman}\ \emph {et~al.}(1992)\citenamefont {Bowman},
  \citenamefont {Garvey}, \citenamefont {Gould}, \citenamefont {Hayes},\ and\
  \citenamefont {Johnson}}]{PhysRevLett.68.780}%
  \BibitemOpen
  \bibfield  {author} {\bibinfo {author} {\bibfnamefont {J.~D.}\ \bibnamefont
  {Bowman}}, \bibinfo {author} {\bibfnamefont {G.~T.}\ \bibnamefont {Garvey}},
  \bibinfo {author} {\bibfnamefont {C.~R.}\ \bibnamefont {Gould}}, \bibinfo
  {author} {\bibfnamefont {A.~C.}\ \bibnamefont {Hayes}},\ and\ \bibinfo
  {author} {\bibfnamefont {M.~B.}\ \bibnamefont {Johnson}},\ }\href
  {https://doi.org/10.1103/PhysRevLett.68.780} {\bibfield  {journal} {\bibinfo
  {journal} {Phys. Rev. Lett.}\ }\textbf {\bibinfo {volume} {68}},\ \bibinfo
  {pages} {780} (\bibinfo {year} {1992})}\BibitemShut {NoStop}%
\bibitem [{\citenamefont {Kerman}\ \emph {et~al.}(1963)\citenamefont {Kerman},
  \citenamefont {Rodberg},\ and\ \citenamefont {Young}}]{Kerman1963}%
  \BibitemOpen
  \bibfield  {author} {\bibinfo {author} {\bibfnamefont {A.~K.}\ \bibnamefont
  {Kerman}}, \bibinfo {author} {\bibfnamefont {L.~S.}\ \bibnamefont
  {Rodberg}},\ and\ \bibinfo {author} {\bibfnamefont {J.~E.}\ \bibnamefont
  {Young}},\ }\href {https://doi.org/10.1103/PhysRevLett.11.422} {\bibfield
  {journal} {\bibinfo  {journal} {Phys. Rev. Lett.}\ }\textbf {\bibinfo
  {volume} {11}},\ \bibinfo {pages} {422} (\bibinfo {year} {1963})}\BibitemShut
  {NoStop}%
\bibitem [{\citenamefont {Hussein}\ \emph {et~al.}(2016)\citenamefont
  {Hussein}, \citenamefont {Carlson},\ and\ \citenamefont
  {Kerman}}]{Hussein2016}%
  \BibitemOpen
  \bibfield  {author} {\bibinfo {author} {\bibfnamefont {M.~S.}\ \bibnamefont
  {Hussein}}, \bibinfo {author} {\bibfnamefont {B.~V.}\ \bibnamefont
  {Carlson}},\ and\ \bibinfo {author} {\bibfnamefont {A.~K.}\ \bibnamefont
  {Kerman}},\ }\href {https://doi.org/10.5506/APhysPolB.47.391} {\bibfield
  {journal} {\bibinfo  {journal} {Acta Physica Polonica. Series B}\ }\textbf
  {\bibinfo {volume} {47}},\ \bibinfo {pages} {391} (\bibinfo {year}
  {2016})}\BibitemShut {NoStop}%
\bibitem [{\citenamefont {Gudkov}\ and\ \citenamefont
  {Shimizu}(2018)}]{Gudkov2018}%
  \BibitemOpen
  \bibfield  {author} {\bibinfo {author} {\bibfnamefont {V.}~\bibnamefont
  {Gudkov}}\ and\ \bibinfo {author} {\bibfnamefont {H.~M.}\ \bibnamefont
  {Shimizu}},\ }\href {http://dx.doi.org/10.1103/PhysRevC.97.065502} {\bibfield
   {journal} {\bibinfo  {journal} {Phys. Rev. C}\ }\textbf {\bibinfo {volume}
  {97}} (\bibinfo {year} {2018})}\BibitemShut {NoStop}%
\bibitem [{\citenamefont {Flambaum}\ and\ \citenamefont
  {Gribakin}(1994)}]{FlamGrib94}%
  \BibitemOpen
  \bibfield  {author} {\bibinfo {author} {\bibfnamefont {V.~V.}\ \bibnamefont
  {Flambaum}}\ and\ \bibinfo {author} {\bibfnamefont {G.~F.}\ \bibnamefont
  {Gribakin}},\ }\href {https://doi.org/10.1103/PhysRevC.50.3122} {\bibfield
  {journal} {\bibinfo  {journal} {Phys. Rev. C}\ }\textbf {\bibinfo {volume}
  {50}},\ \bibinfo {pages} {3122} (\bibinfo {year} {1994})}\BibitemShut
  {NoStop}%
\bibitem [{\citenamefont {Bowman}\ \emph {et~al.}(1990)\citenamefont {Bowman},
  \citenamefont {Bowman}, \citenamefont {Bush}, \citenamefont {Delheij},
  \citenamefont {Frankle}, \citenamefont {Gould}, \citenamefont {Haase},
  \citenamefont {Knudson}, \citenamefont {Mitchell}, \citenamefont {Penttila},
  \citenamefont {Postma}, \citenamefont {Roberson}, \citenamefont {Seestrom},
  \citenamefont {Szymanski}, \citenamefont {Yuan},\ and\ \citenamefont
  {Zhu}}]{PhysRevLett.65.1192}%
  \BibitemOpen
  \bibfield  {author} {\bibinfo {author} {\bibfnamefont {J.~D.}\ \bibnamefont
  {Bowman}}, \bibinfo {author} {\bibfnamefont {C.~D.}\ \bibnamefont {Bowman}},
  \bibinfo {author} {\bibfnamefont {J.~E.}\ \bibnamefont {Bush}}, \bibinfo
  {author} {\bibfnamefont {P.~P.~J.}\ \bibnamefont {Delheij}}, \bibinfo
  {author} {\bibfnamefont {C.~M.}\ \bibnamefont {Frankle}}, \bibinfo {author}
  {\bibfnamefont {C.~R.}\ \bibnamefont {Gould}}, \bibinfo {author}
  {\bibfnamefont {D.~G.}\ \bibnamefont {Haase}}, \bibinfo {author}
  {\bibfnamefont {J.}~\bibnamefont {Knudson}}, \bibinfo {author} {\bibfnamefont
  {G.~E.}\ \bibnamefont {Mitchell}}, \bibinfo {author} {\bibfnamefont
  {S.}~\bibnamefont {Penttila}}, \bibinfo {author} {\bibfnamefont
  {H.}~\bibnamefont {Postma}}, \bibinfo {author} {\bibfnamefont {N.~R.}\
  \bibnamefont {Roberson}}, \bibinfo {author} {\bibfnamefont {S.~J.}\
  \bibnamefont {Seestrom}}, \bibinfo {author} {\bibfnamefont {J.~J.}\
  \bibnamefont {Szymanski}}, \bibinfo {author} {\bibfnamefont {V.~W.}\
  \bibnamefont {Yuan}},\ and\ \bibinfo {author} {\bibfnamefont
  {X.}~\bibnamefont {Zhu}},\ }\href
  {https://doi.org/10.1103/PhysRevLett.65.1192} {\bibfield  {journal} {\bibinfo
   {journal} {Phys. Rev. Lett.}\ }\textbf {\bibinfo {volume} {65}},\ \bibinfo
  {pages} {1192} (\bibinfo {year} {1990})}\BibitemShut {NoStop}%
\bibitem [{\citenamefont {Bohr}\ and\ \citenamefont
  {Mottelson}(1998)}]{NuclearStructure}%
  \BibitemOpen
  \bibfield  {author} {\bibinfo {author} {\bibfnamefont {A.}~\bibnamefont
  {Bohr}}\ and\ \bibinfo {author} {\bibfnamefont {B.~R.}\ \bibnamefont
  {Mottelson}},\ }\href {https://doi.org/10.1142/3530} {\emph {\bibinfo {title}
  {Nuclear Structure}}}\ (\bibinfo  {publisher} {World Scientific Publishing
  Company},\ \bibinfo {year} {1998})\BibitemShut {NoStop}%
\bibitem [{\citenamefont {Flambaum}\ \emph
  {et~al.}(1984{\natexlab{a}})\citenamefont {Flambaum}, \citenamefont
  {Khriplovich},\ and\ \citenamefont {Sushkov}}]{FLAMBAUM1984367}%
  \BibitemOpen
  \bibfield  {author} {\bibinfo {author} {\bibfnamefont {V.}~\bibnamefont
  {Flambaum}}, \bibinfo {author} {\bibfnamefont {I.}~\bibnamefont
  {Khriplovich}},\ and\ \bibinfo {author} {\bibfnamefont {O.}~\bibnamefont
  {Sushkov}},\ }\href
  {https://doi.org/https://doi.org/10.1016/0370-2693(84)90140-0} {\bibfield
  {journal} {\bibinfo  {journal} {\\ Phys. Lett. B}\ }\textbf {\bibinfo
  {volume} {146}},\ \bibinfo {pages} {367} (\bibinfo {year}
  {1984}{\natexlab{a}})}\BibitemShut {NoStop}%
\bibitem [{\citenamefont {Desplanques}\ \emph {et~al.}(1980)\citenamefont
  {Desplanques}, \citenamefont {Donoghue},\ and\ \citenamefont
  {Holstein}}]{DESPLANQUES1980449}%
  \BibitemOpen
  \bibfield  {author} {\bibinfo {author} {\bibfnamefont {B.}~\bibnamefont
  {Desplanques}}, \bibinfo {author} {\bibfnamefont {J.~F.}\ \bibnamefont
  {Donoghue}},\ and\ \bibinfo {author} {\bibfnamefont {B.~R.}\ \bibnamefont
  {Holstein}},\ }\href
  {https://doi.org/https://doi.org/10.1016/0003-4916(80)90217-1} {\bibfield
  {journal} {\bibinfo  {journal} {Annals of Physics}\ }\textbf {\bibinfo
  {volume} {124}},\ \bibinfo {pages} {449} (\bibinfo {year}
  {1980})}\BibitemShut {NoStop}%
\bibitem [{\citenamefont {Flambaum}\ and\ \citenamefont
  {Murray}(1997)}]{PhysRevC.56.1641}%
  \BibitemOpen
  \bibfield  {author} {\bibinfo {author} {\bibfnamefont {V.~V.}\ \bibnamefont
  {Flambaum}}\ and\ \bibinfo {author} {\bibfnamefont {D.~W.}\ \bibnamefont
  {Murray}},\ }\href {https://doi.org/10.1103/PhysRevC.56.1641} {\bibfield
  {journal} {\bibinfo  {journal} {Phys. Rev. C}\ }\textbf {\bibinfo {volume}
  {56}},\ \bibinfo {pages} {1641} (\bibinfo {year} {1997})}\BibitemShut
  {NoStop}%
\bibitem [{\citenamefont {Noguera}\ and\ \citenamefont
  {Desplanques}(1986)}]{NOGUERA1986189}%
  \BibitemOpen
  \bibfield  {author} {\bibinfo {author} {\bibfnamefont {S.}~\bibnamefont
  {Noguera}}\ and\ \bibinfo {author} {\bibfnamefont {B.}~\bibnamefont
  {Desplanques}},\ }\href
  {https://doi.org/https://doi.org/10.1016/0375-9474(86)90374-X} {\bibfield
  {journal} {\bibinfo  {journal} {Nuclear Physics A}\ }\textbf {\bibinfo
  {volume} {457}},\ \bibinfo {pages} {189} (\bibinfo {year}
  {1986})}\BibitemShut {NoStop}%
\bibitem [{\citenamefont {Dubovik}\ and\ \citenamefont
  {Zenkin}(1986)}]{DUBOVIK1986100}%
  \BibitemOpen
  \bibfield  {author} {\bibinfo {author} {\bibfnamefont {V.}~\bibnamefont
  {Dubovik}}\ and\ \bibinfo {author} {\bibfnamefont {S.}~\bibnamefont
  {Zenkin}},\ }\href
  {https://doi.org/https://doi.org/10.1016/0003-4916(86)90021-7} {\bibfield
  {journal} {\bibinfo  {journal} {Annals of Physics}\ }\textbf {\bibinfo
  {volume} {172}},\ \bibinfo {pages} {100} (\bibinfo {year}
  {1986})}\BibitemShut {NoStop}%
\bibitem [{\citenamefont {Feldman}\ \emph {et~al.}(1991)\citenamefont
  {Feldman}, \citenamefont {Crawford}, \citenamefont {Dubach},\ and\
  \citenamefont {Holstein}}]{PhysRevC.43.863}%
  \BibitemOpen
  \bibfield  {author} {\bibinfo {author} {\bibfnamefont {G.~B.}\ \bibnamefont
  {Feldman}}, \bibinfo {author} {\bibfnamefont {G.~A.}\ \bibnamefont
  {Crawford}}, \bibinfo {author} {\bibfnamefont {J.}~\bibnamefont {Dubach}},\
  and\ \bibinfo {author} {\bibfnamefont {B.~R.}\ \bibnamefont {Holstein}},\
  }\href {https://doi.org/10.1103/PhysRevC.43.863} {\bibfield  {journal}
  {\bibinfo  {journal} {Phys. Rev. C}\ }\textbf {\bibinfo {volume} {43}},\
  \bibinfo {pages} {863} (\bibinfo {year} {1991})}\BibitemShut {NoStop}%
\bibitem [{\citenamefont {Wasem}(2012)}]{Wasem2012}%
  \BibitemOpen
  \bibfield  {author} {\bibinfo {author} {\bibfnamefont {J.}~\bibnamefont
  {Wasem}},\ }\href {https://doi.org/10.1103/PhysRevC.85.022501} {\bibfield
  {journal} {\bibinfo  {journal} {Phys. Rev. C}\ }\textbf {\bibinfo {volume}
  {85}},\ \bibinfo {pages} {022501} (\bibinfo {year} {2012})}\BibitemShut
  {NoStop}%
\bibitem [{\citenamefont {Blyth}\ \emph {et~al.}(2018)\citenamefont {Blyth}
  \emph {et~al.}}]{PhysRevLett.121.242002}%
  \BibitemOpen
  \bibfield  {author} {\bibinfo {author} {\bibfnamefont {D.}~\bibnamefont
  {Blyth}} \emph {et~al.} (\bibinfo {collaboration} {NPDGamma Collaboration}),\
  }\href {https://doi.org/10.1103/PhysRevLett.121.242002} {\bibfield  {journal}
  {\bibinfo  {journal} {Phys. Rev. Lett.}\ }\textbf {\bibinfo {volume} {121}},\
  \bibinfo {pages} {242002} (\bibinfo {year} {2018})}\BibitemShut {NoStop}%
\bibitem [{\citenamefont {Haxton}\ and\ \citenamefont
  {Holstein}(2013)}]{Haxton2013}%
  \BibitemOpen
  \bibfield  {author} {\bibinfo {author} {\bibfnamefont {W.~C.}\ \bibnamefont
  {Haxton}}\ and\ \bibinfo {author} {\bibfnamefont {B.~R.}\ \bibnamefont
  {Holstein}},\ }\href
  {https://doi.org/https://doi.org/10.1016/j.ppnp.2013.03.009} {\bibfield
  {journal} {\bibinfo  {journal} {Progress in Particle and Nuclear Physics}\
  }\textbf {\bibinfo {volume} {71}},\ \bibinfo {pages} {185} (\bibinfo {year}
  {2013})},\ \bibinfo {note} {fundamental Symmetries in the Era of the
  LHC}\BibitemShut {NoStop}%
\bibitem [{\citenamefont {Flambaum}\ \emph
  {et~al.}(1984{\natexlab{b}})\citenamefont {Flambaum}, \citenamefont
  {Khriplovich},\ and\ \citenamefont {Sushkov}}]{Flambaum:154087}%
  \BibitemOpen
  \bibfield  {author} {\bibinfo {author} {\bibfnamefont {V.~V.}\ \bibnamefont
  {Flambaum}}, \bibinfo {author} {\bibfnamefont {I.~B.}\ \bibnamefont
  {Khriplovich}},\ and\ \bibinfo {author} {\bibfnamefont {O.~P.}\ \bibnamefont
  {Sushkov}},\ }\href {https://cds.cern.ch/record/154087} {\emph {\bibinfo
  {title} {{On the possibility to study P- and T-odd nuclear forces in atomic
  and molecular experiments}}}},\ \bibinfo {type} {Tech. Rep.}\ (\bibinfo
  {institution} {Akad. Nauk Novosibirsk. Inst. Yarn. Fiz.},\ \bibinfo {address}
  {Novosibirsk},\ \bibinfo {year} {1984})\BibitemShut {NoStop}%
\bibitem [{\citenamefont {Flambaum}\ \emph {et~al.}(1986)\citenamefont
  {Flambaum}, \citenamefont {Khriplovich},\ and\ \citenamefont
  {Sushkov}}]{FLAMBAUM1986750}%
  \BibitemOpen
  \bibfield  {author} {\bibinfo {author} {\bibfnamefont {V.}~\bibnamefont
  {Flambaum}}, \bibinfo {author} {\bibfnamefont {I.}~\bibnamefont
  {Khriplovich}},\ and\ \bibinfo {author} {\bibfnamefont {O.}~\bibnamefont
  {Sushkov}},\ }\href
  {https://doi.org/https://doi.org/10.1016/0375-9474(86)90331-3} {\bibfield
  {journal} {\bibinfo  {journal} {Nuclear Physics A}\ }\textbf {\bibinfo
  {volume} {449}},\ \bibinfo {pages} {750} (\bibinfo {year}
  {1986})}\BibitemShut {NoStop}%
\bibitem [{\citenamefont {Flambaum}\ \emph {et~al.}(2014)\citenamefont
  {Flambaum}, \citenamefont {DeMille},\ and\ \citenamefont
  {Kozlov}}]{PhysRevLett.113.103003}%
  \BibitemOpen
  \bibfield  {author} {\bibinfo {author} {\bibfnamefont {V.~V.}\ \bibnamefont
  {Flambaum}}, \bibinfo {author} {\bibfnamefont {D.}~\bibnamefont {DeMille}},\
  and\ \bibinfo {author} {\bibfnamefont {M.~G.}\ \bibnamefont {Kozlov}},\
  }\href {https://link.aps.org/doi/10.1103/PhysRevLett.113.103003} {\bibfield
  {journal} {\bibinfo  {journal} {Phys. Rev. Lett.}\ }\textbf {\bibinfo
  {volume} {113}},\ \bibinfo {pages} {103003} (\bibinfo {year}
  {2014})}\BibitemShut {NoStop}%
\bibitem [{\citenamefont {Haxton}\ and\ \citenamefont
  {Henley}(1983)}]{Haxton83}%
  \BibitemOpen
  \bibfield  {author} {\bibinfo {author} {\bibfnamefont {W.~C.}\ \bibnamefont
  {Haxton}}\ and\ \bibinfo {author} {\bibfnamefont {E.~M.}\ \bibnamefont
  {Henley}},\ }\href {https://doi.org/10.1103/PhysRevLett.51.1937} {\bibfield
  {journal} {\bibinfo  {journal} {Phys. Rev. Lett.}\ }\textbf {\bibinfo
  {volume} {51}},\ \bibinfo {pages} {1937} (\bibinfo {year}
  {1983})}\BibitemShut {NoStop}%
\bibitem [{\citenamefont {Khriplovich}\ and\ \citenamefont
  {Korkin}(2000)}]{Khriplovich2000}%
  \BibitemOpen
  \bibfield  {author} {\bibinfo {author} {\bibfnamefont {I.}~\bibnamefont
  {Khriplovich}}\ and\ \bibinfo {author} {\bibfnamefont {R.}~\bibnamefont
  {Korkin}},\ }\href
  {https://doi.org/https://doi.org/10.1016/S0375-9474(99)00403-0} {\bibfield
  {journal} {\bibinfo  {journal} {Nuclear Physics A}\ }\textbf {\bibinfo
  {volume} {665}},\ \bibinfo {pages} {365} (\bibinfo {year}
  {2000})}\BibitemShut {NoStop}%
\bibitem [{\citenamefont {Dmitriev}\ and\ \citenamefont
  {Sen’kov}(2003)}]{Dmitriev03}%
  \BibitemOpen
  \bibfield  {author} {\bibinfo {author} {\bibfnamefont {V.~F.}\ \bibnamefont
  {Dmitriev}}\ and\ \bibinfo {author} {\bibfnamefont {R.~A.}\ \bibnamefont
  {Sen’kov}},\ }\href {https://doi.org/10.1134/1.1619505} {\bibfield
  {journal} {\bibinfo  {journal} {Physics of Atomic Nuclei}\ }\textbf {\bibinfo
  {volume} {66}},\ \bibinfo {pages} {1940–1945} (\bibinfo {year}
  {2003})}\BibitemShut {NoStop}%
\bibitem [{\citenamefont {Yamanaka}\ \emph {et~al.}(2017)\citenamefont
  {Yamanaka}, \citenamefont {Sahoo}, \citenamefont {Yoshinaga}, \citenamefont
  {Sato}, \citenamefont {Asahi},\ and\ \citenamefont {Das}}]{Yamanaka2017}%
  \BibitemOpen
  \bibfield  {author} {\bibinfo {author} {\bibfnamefont {N.}~\bibnamefont
  {Yamanaka}}, \bibinfo {author} {\bibfnamefont {B.}~\bibnamefont {Sahoo}},
  \bibinfo {author} {\bibfnamefont {N.}~\bibnamefont {Yoshinaga}}, \bibinfo
  {author} {\bibfnamefont {T.}~\bibnamefont {Sato}}, \bibinfo {author}
  {\bibfnamefont {K.}~\bibnamefont {Asahi}},\ and\ \bibinfo {author}
  {\bibfnamefont {B.}~\bibnamefont {Das}},\ }\href
  {https://link.springer.com/article/10.1140/epja/i2017-12237-2} {\bibfield
  {journal} {\bibinfo  {journal} {The European Physical Journal A}\ }\textbf
  {\bibinfo {volume} {53}},\ \bibinfo {pages} {1} (\bibinfo {year}
  {2017})}\BibitemShut {NoStop}%
\bibitem [{\citenamefont {de~Vries}\ \emph {et~al.}(2015)\citenamefont
  {de~Vries}, \citenamefont {Mereghetti},\ and\ \citenamefont
  {Walker-Loud}}]{Vries2015}%
  \BibitemOpen
  \bibfield  {author} {\bibinfo {author} {\bibfnamefont {J.}~\bibnamefont
  {de~Vries}}, \bibinfo {author} {\bibfnamefont {E.}~\bibnamefont
  {Mereghetti}},\ and\ \bibinfo {author} {\bibfnamefont {A.}~\bibnamefont
  {Walker-Loud}},\ }\href {https://doi.org/10.1103/PhysRevC.92.045201}
  {\bibfield  {journal} {\bibinfo  {journal} {Phys. Rev. C}\ }\textbf {\bibinfo
  {volume} {92}},\ \bibinfo {pages} {045201} (\bibinfo {year}
  {2015})}\BibitemShut {NoStop}%
\bibitem [{\citenamefont {Pospelov}\ and\ \citenamefont
  {Ritz}(2005)}]{POSPELOV2005119}%
  \BibitemOpen
  \bibfield  {author} {\bibinfo {author} {\bibfnamefont {M.}~\bibnamefont
  {Pospelov}}\ and\ \bibinfo {author} {\bibfnamefont {A.}~\bibnamefont
  {Ritz}},\ }\href {https://doi.org/https://doi.org/10.1016/j.aop.2005.04.002}
  {\bibfield  {journal} {\bibinfo  {journal} {Annals of Physics}\ }\textbf
  {\bibinfo {volume} {318}},\ \bibinfo {pages} {119} (\bibinfo {year}
  {2005})},\ \bibinfo {note} {special Issue}\BibitemShut {NoStop}%
\bibitem [{\citenamefont {Palos}(2018)}]{Palos2018}%
  \BibitemOpen
  \bibfield  {author} {\bibinfo {author} {\bibfnamefont {L.~B.}\ \bibnamefont
  {Palos}},\ }\href
  {https://conferences.lbl.gov/event/137/session/18/contribution/139/material/slides/0.pdf}
  {\bibinfo {title} {{(NOPTREX Collaboration), talk at the Thirteenth
  Conference on the Intersections of Particle and Nuclear Physics
  (unpublished)}}} (\bibinfo {year} {2018})\BibitemShut {NoStop}%
\bibitem [{\citenamefont {Moody}\ and\ \citenamefont {Wilczek}(1984)}]{Moody}%
  \BibitemOpen
  \bibfield  {author} {\bibinfo {author} {\bibfnamefont {J.~E.}\ \bibnamefont
  {Moody}}\ and\ \bibinfo {author} {\bibfnamefont {F.}~\bibnamefont
  {Wilczek}},\ }\href {https://doi.org/10.1103/PhysRevD.30.130} {\bibfield
  {journal} {\bibinfo  {journal} {Phys. Rev. D}\ }\textbf {\bibinfo {volume}
  {30}},\ \bibinfo {pages} {130} (\bibinfo {year} {1984})}\BibitemShut
  {NoStop}%
\bibitem [{\citenamefont {{Marsh}}(2016)}]{Marsh}%
  \BibitemOpen
  \bibfield  {author} {\bibinfo {author} {\bibfnamefont {D.~J.~E.}\
  \bibnamefont {{Marsh}}},\ }\href
  {https://doi.org/10.1016/j.physrep.2016.06.005} {\bibfield  {journal}
  {\bibinfo  {journal} {Physics Reports}\ }\textbf {\bibinfo {volume} {643}},\
  \bibinfo {pages} {1} (\bibinfo {year} {2016})},\ \Eprint
  {https://arxiv.org/abs/1510.07633} {arXiv:1510.07633 [astro-ph.CO]}
  \BibitemShut {NoStop}%
\bibitem [{\citenamefont {Graham}\ \emph {et~al.}(2015)\citenamefont {Graham},
  \citenamefont {Kaplan},\ and\ \citenamefont {Rajendran}}]{Graham}%
  \BibitemOpen
  \bibfield  {author} {\bibinfo {author} {\bibfnamefont {P.~W.}\ \bibnamefont
  {Graham}}, \bibinfo {author} {\bibfnamefont {D.~E.}\ \bibnamefont {Kaplan}},\
  and\ \bibinfo {author} {\bibfnamefont {S.}~\bibnamefont {Rajendran}},\ }\href
  {https://doi.org/10.1103/PhysRevLett.115.221801} {\bibfield  {journal}
  {\bibinfo  {journal} {Phys. Rev. Lett.}\ }\textbf {\bibinfo {volume} {115}},\
  \bibinfo {pages} {221801} (\bibinfo {year} {2015})}\BibitemShut {NoStop}%
\bibitem [{\citenamefont {Gupta}\ \emph {et~al.}(2016)\citenamefont {Gupta},
  \citenamefont {Komargodski}, \citenamefont {Perez} \emph {et~al.}}]{Gupta}%
  \BibitemOpen
  \bibfield  {author} {\bibinfo {author} {\bibfnamefont {R.~S.}\ \bibnamefont
  {Gupta}}, \bibinfo {author} {\bibfnamefont {Z.}~\bibnamefont {Komargodski}},
  \bibinfo {author} {\bibfnamefont {G.}~\bibnamefont {Perez}}, \emph {et~al.},\
  }\href@noop {} {\bibfield  {journal} {\bibinfo  {journal} {J. High Energ.
  Phys. 02}\ ,\ \bibinfo {pages} {166}} (\bibinfo {year} {2016})}\BibitemShut
  {NoStop}%
\bibitem [{\citenamefont {Flacke}\ \emph {et~al.}(2017)\citenamefont {Flacke},
  \citenamefont {Frugiuele}, \citenamefont {Fuchs}, \citenamefont {Gupta},\
  and\ \citenamefont {Perez}}]{Flacke}%
  \BibitemOpen
  \bibfield  {author} {\bibinfo {author} {\bibfnamefont {T.}~\bibnamefont
  {Flacke}}, \bibinfo {author} {\bibfnamefont {C.}~\bibnamefont {Frugiuele}},
  \bibinfo {author} {\bibfnamefont {E.}~\bibnamefont {Fuchs}}, \bibinfo
  {author} {\bibfnamefont {R.~S.}\ \bibnamefont {Gupta}},\ and\ \bibinfo
  {author} {\bibfnamefont {G.}~\bibnamefont {Perez}},\ }\href
  {https://doi.org/10.1007/jhep06(2017)050} {\bibfield  {journal} {\bibinfo
  {journal} {Journal of High Energy Physics}\ }\textbf {\bibinfo {volume}
  {6}},\ \bibinfo {pages} {1} (\bibinfo {year} {2017})}\BibitemShut {NoStop}%
\bibitem [{\citenamefont {Stadnik}\ \emph {et~al.}(2018)\citenamefont
  {Stadnik}, \citenamefont {Dzuba},\ and\ \citenamefont
  {Flambaum}}]{Stadnik2017}%
  \BibitemOpen
  \bibfield  {author} {\bibinfo {author} {\bibfnamefont {Y.~V.}\ \bibnamefont
  {Stadnik}}, \bibinfo {author} {\bibfnamefont {V.~A.}\ \bibnamefont {Dzuba}},\
  and\ \bibinfo {author} {\bibfnamefont {V.~V.}\ \bibnamefont {Flambaum}},\
  }\href {https://doi.org/10.1103/PhysRevLett.120.013202} {\bibfield  {journal}
  {\bibinfo  {journal} {Phys. Rev. Lett.}\ }\textbf {\bibinfo {volume} {120}},\
  \bibinfo {pages} {013202} (\bibinfo {year} {2018})}\BibitemShut {NoStop}%
\bibitem [{\citenamefont {Graner}\ \emph {et~al.}(2016)\citenamefont {Graner},
  \citenamefont {Chen}, \citenamefont {Lindahl},\ and\ \citenamefont
  {Heckel}}]{Graner2016}%
  \BibitemOpen
  \bibfield  {author} {\bibinfo {author} {\bibfnamefont {B.}~\bibnamefont
  {Graner}}, \bibinfo {author} {\bibfnamefont {Y.}~\bibnamefont {Chen}},
  \bibinfo {author} {\bibfnamefont {E.~G.}\ \bibnamefont {Lindahl}},\ and\
  \bibinfo {author} {\bibfnamefont {B.~R.}\ \bibnamefont {Heckel}},\ }\href
  {https://doi.org/10.1103/PhysRevLett.116.161601} {\bibfield  {journal}
  {\bibinfo  {journal} {Phys. Rev. Lett.}\ }\textbf {\bibinfo {volume} {116}},\
  \bibinfo {pages} {161601} (\bibinfo {year} {2016})}\BibitemShut {NoStop}%
\bibitem [{\citenamefont {Swallows}\ \emph {et~al.}(2013)\citenamefont
  {Swallows}, \citenamefont {Loftus}, \citenamefont {Griffith}, \citenamefont
  {Heckel}, \citenamefont {Fortson},\ and\ \citenamefont
  {Romalis}}]{Swallows2013}%
  \BibitemOpen
  \bibfield  {author} {\bibinfo {author} {\bibfnamefont {M.~D.}\ \bibnamefont
  {Swallows}}, \bibinfo {author} {\bibfnamefont {T.~H.}\ \bibnamefont
  {Loftus}}, \bibinfo {author} {\bibfnamefont {W.~C.}\ \bibnamefont
  {Griffith}}, \bibinfo {author} {\bibfnamefont {B.~R.}\ \bibnamefont
  {Heckel}}, \bibinfo {author} {\bibfnamefont {E.~N.}\ \bibnamefont
  {Fortson}},\ and\ \bibinfo {author} {\bibfnamefont {M.~V.}\ \bibnamefont
  {Romalis}},\ }\href {https://doi.org/10.1103/PhysRevA.87.012102} {\bibfield
  {journal} {\bibinfo  {journal} {Phys. Rev. A}\ }\textbf {\bibinfo {volume}
  {87}},\ \bibinfo {pages} {012102} (\bibinfo {year} {2013})}\BibitemShut
  {NoStop}%
\bibitem [{\citenamefont {Mantry}\ \emph {et~al.}(2014)\citenamefont {Mantry},
  \citenamefont {Pitschmann},\ and\ \citenamefont {Ramsey-Musolf}}]{musolf}%
  \BibitemOpen
  \bibfield  {author} {\bibinfo {author} {\bibfnamefont {S.}~\bibnamefont
  {Mantry}}, \bibinfo {author} {\bibfnamefont {M.}~\bibnamefont {Pitschmann}},\
  and\ \bibinfo {author} {\bibfnamefont {M.~J.}\ \bibnamefont
  {Ramsey-Musolf}},\ }\bibfield  {journal} {\bibinfo  {journal} {Physical
  Review D}\ }\textbf {\bibinfo {volume} {90}},\ \href
  {https://doi.org/10.1103/physrevd.90.054016} {10.1103/physrevd.90.054016}
  (\bibinfo {year} {2014})\BibitemShut {NoStop}%
\bibitem [{\citenamefont {Yuan}\ \emph {et~al.}(1991)\citenamefont {Yuan},
  \citenamefont {Bowman}, \citenamefont {Bowman}, \citenamefont {Bush},
  \citenamefont {Delheij}, \citenamefont {Frankle}, \citenamefont {Gould},
  \citenamefont {Haase}, \citenamefont {Knudson}, \citenamefont {Mitchell},
  \citenamefont {Penttil\"a}, \citenamefont {Postma}, \citenamefont {Roberson},
  \citenamefont {Seestrom}, \citenamefont {Szymanski},\ and\ \citenamefont
  {Zhu}}]{PhysRevC.44.2187}%
  \BibitemOpen
  \bibfield  {author} {\bibinfo {author} {\bibfnamefont {V.~W.}\ \bibnamefont
  {Yuan}}, \bibinfo {author} {\bibfnamefont {C.~D.}\ \bibnamefont {Bowman}},
  \bibinfo {author} {\bibfnamefont {J.~D.}\ \bibnamefont {Bowman}}, \bibinfo
  {author} {\bibfnamefont {J.~E.}\ \bibnamefont {Bush}}, \bibinfo {author}
  {\bibfnamefont {P.~P.~J.}\ \bibnamefont {Delheij}}, \bibinfo {author}
  {\bibfnamefont {C.~M.}\ \bibnamefont {Frankle}}, \bibinfo {author}
  {\bibfnamefont {C.~R.}\ \bibnamefont {Gould}}, \bibinfo {author}
  {\bibfnamefont {D.~G.}\ \bibnamefont {Haase}}, \bibinfo {author}
  {\bibfnamefont {J.~N.}\ \bibnamefont {Knudson}}, \bibinfo {author}
  {\bibfnamefont {G.~E.}\ \bibnamefont {Mitchell}}, \bibinfo {author}
  {\bibfnamefont {S.}~\bibnamefont {Penttil\"a}}, \bibinfo {author}
  {\bibfnamefont {H.}~\bibnamefont {Postma}}, \bibinfo {author} {\bibfnamefont
  {N.~R.}\ \bibnamefont {Roberson}}, \bibinfo {author} {\bibfnamefont {S.~J.}\
  \bibnamefont {Seestrom}}, \bibinfo {author} {\bibfnamefont {J.~J.}\
  \bibnamefont {Szymanski}},\ and\ \bibinfo {author} {\bibfnamefont
  {X.}~\bibnamefont {Zhu}},\ }\href {https://doi.org/10.1103/PhysRevC.44.2187}
  {\bibfield  {journal} {\bibinfo  {journal} {Phys. Rev. C}\ }\textbf {\bibinfo
  {volume} {44}},\ \bibinfo {pages} {2187} (\bibinfo {year}
  {1991})}\BibitemShut {NoStop}%
\bibitem [{\citenamefont {Sharapov}\ \emph {et~al.}(2000)\citenamefont
  {Sharapov}, \citenamefont {Bowman}, \citenamefont {Crawford}, \citenamefont
  {Delheij}, \citenamefont {Frankle}, \citenamefont {Iinuma}, \citenamefont
  {Knudson}, \citenamefont {Lowie}, \citenamefont {Lynch}, \citenamefont
  {Masaike}, \citenamefont {Matsuda}, \citenamefont {Mitchell}, \citenamefont
  {Penttil\"a}, \citenamefont {Postma}, \citenamefont {Roberson}, \citenamefont
  {Seestrom}, \citenamefont {Stephenson}, \citenamefont {Yen},\ and\
  \citenamefont {Yuan}}]{PhysRevCLastExperiment}%
  \BibitemOpen
  \bibfield  {author} {\bibinfo {author} {\bibfnamefont {E.~I.}\ \bibnamefont
  {Sharapov}}, \bibinfo {author} {\bibfnamefont {J.~D.}\ \bibnamefont
  {Bowman}}, \bibinfo {author} {\bibfnamefont {B.~E.}\ \bibnamefont
  {Crawford}}, \bibinfo {author} {\bibfnamefont {P.~P.~J.}\ \bibnamefont
  {Delheij}}, \bibinfo {author} {\bibfnamefont {C.~M.}\ \bibnamefont
  {Frankle}}, \bibinfo {author} {\bibfnamefont {M.}~\bibnamefont {Iinuma}},
  \bibinfo {author} {\bibfnamefont {J.~N.}\ \bibnamefont {Knudson}}, \bibinfo
  {author} {\bibfnamefont {L.~Y.}\ \bibnamefont {Lowie}}, \bibinfo {author}
  {\bibfnamefont {J.~E.}\ \bibnamefont {Lynch}}, \bibinfo {author}
  {\bibfnamefont {A.}~\bibnamefont {Masaike}}, \bibinfo {author} {\bibfnamefont
  {Y.}~\bibnamefont {Matsuda}}, \bibinfo {author} {\bibfnamefont {G.~E.}\
  \bibnamefont {Mitchell}}, \bibinfo {author} {\bibfnamefont {S.~I.}\
  \bibnamefont {Penttil\"a}}, \bibinfo {author} {\bibfnamefont
  {H.}~\bibnamefont {Postma}}, \bibinfo {author} {\bibfnamefont {N.~R.}\
  \bibnamefont {Roberson}}, \bibinfo {author} {\bibfnamefont {S.~J.}\
  \bibnamefont {Seestrom}}, \bibinfo {author} {\bibfnamefont {S.~L.}\
  \bibnamefont {Stephenson}}, \bibinfo {author} {\bibfnamefont {Y.-F.}\
  \bibnamefont {Yen}},\ and\ \bibinfo {author} {\bibfnamefont {V.~W.}\
  \bibnamefont {Yuan}},\ }\href {https://doi.org/10.1103/PhysRevC.61.025501}
  {\bibfield  {journal} {\bibinfo  {journal} {Phys. Rev. C}\ }\textbf {\bibinfo
  {volume} {61}},\ \bibinfo {pages} {025501} (\bibinfo {year}
  {2000})}\BibitemShut {NoStop}%
\bibitem [{\citenamefont {Auerbach}\ \emph {et~al.}(1996)\citenamefont
  {Auerbach}, \citenamefont {Flambaum},\ and\ \citenamefont
  {Spevak}}]{PhysRevLett.76.4316}%
  \BibitemOpen
  \bibfield  {author} {\bibinfo {author} {\bibfnamefont {N.}~\bibnamefont
  {Auerbach}}, \bibinfo {author} {\bibfnamefont {V.~V.}\ \bibnamefont
  {Flambaum}},\ and\ \bibinfo {author} {\bibfnamefont {V.}~\bibnamefont
  {Spevak}},\ }\href {https://doi.org/10.1103/PhysRevLett.76.4316} {\bibfield
  {journal} {\bibinfo  {journal} {Phys. Rev. Lett.}\ }\textbf {\bibinfo
  {volume} {76}},\ \bibinfo {pages} {4316} (\bibinfo {year}
  {1996})}\BibitemShut {NoStop}%
\bibitem [{\citenamefont {Spevak}\ \emph {et~al.}(1997)\citenamefont {Spevak},
  \citenamefont {Auerbach},\ and\ \citenamefont {Flambaum}}]{PhysRevC.56.1357}%
  \BibitemOpen
  \bibfield  {author} {\bibinfo {author} {\bibfnamefont {V.}~\bibnamefont
  {Spevak}}, \bibinfo {author} {\bibfnamefont {N.}~\bibnamefont {Auerbach}},\
  and\ \bibinfo {author} {\bibfnamefont {V.~V.}\ \bibnamefont {Flambaum}},\
  }\href {https://doi.org/10.1103/PhysRevC.56.1357} {\bibfield  {journal}
  {\bibinfo  {journal} {Phys. Rev. C}\ }\textbf {\bibinfo {volume} {56}},\
  \bibinfo {pages} {1357} (\bibinfo {year} {1997})}\BibitemShut {NoStop}%
\bibitem [{\citenamefont {Kadmenskii}\ \emph {et~al.}(1982)\citenamefont
  {Kadmenskii}, \citenamefont {Markushev},\ and\ \citenamefont
  {Furman}}]{Kadmenskii1982}%
  \BibitemOpen
  \bibfield  {author} {\bibinfo {author} {\bibfnamefont {S.~G.}\ \bibnamefont
  {Kadmenskii}}, \bibinfo {author} {\bibfnamefont {V.~P.}\ \bibnamefont
  {Markushev}},\ and\ \bibinfo {author} {\bibfnamefont {V.~I.}\ \bibnamefont
  {Furman}},\ }\href@noop {} {\bibfield  {journal} {\bibinfo  {journal} {Soviet
  Journal of Nuclear Physics 35(2)}\ ,\ \bibinfo {pages} {166}} (\bibinfo
  {year} {1982})}\BibitemShut {NoStop}%
\bibitem [{\citenamefont {Kadmenskii}\ \emph {et~al.}(1984)\citenamefont
  {Kadmenskii}, \citenamefont {Markushev}, \citenamefont {Popov},\ and\
  \citenamefont {Furman}}]{Kadmenskii1984}%
  \BibitemOpen
  \bibfield  {author} {\bibinfo {author} {\bibfnamefont {S.~G.}\ \bibnamefont
  {Kadmenskii}}, \bibinfo {author} {\bibfnamefont {V.~P.}\ \bibnamefont
  {Markushev}}, \bibinfo {author} {\bibfnamefont {Y.~P.}\ \bibnamefont
  {Popov}},\ and\ \bibinfo {author} {\bibfnamefont {V.~I.}\ \bibnamefont
  {Furman}},\ }\href@noop {} {\bibfield  {journal} {\bibinfo  {journal} {Soviet
  Journal of Nuclear Physics 39(1)}\ ,\ \bibinfo {pages} {4}} (\bibinfo {year}
  {1984})}\BibitemShut {NoStop}%
\bibitem [{\citenamefont {Chishti}\ \emph {et~al.}(2020)\citenamefont
  {Chishti}, \citenamefont {O{\textquoteright}Donnell}, \citenamefont
  {Battaglia}, \citenamefont {Bowry}, \citenamefont {Jaroszynski},
  \citenamefont {Singh}, \citenamefont {Scheck}, \citenamefont {Spagnoletti},\
  and\ \citenamefont {Smith}}]{Thorium228}%
  \BibitemOpen
  \bibfield  {author} {\bibinfo {author} {\bibfnamefont {M.}~\bibnamefont
  {Chishti}}, \bibinfo {author} {\bibfnamefont {D.}~\bibnamefont
  {O{\textquoteright}Donnell}}, \bibinfo {author} {\bibfnamefont
  {G.}~\bibnamefont {Battaglia}}, \bibinfo {author} {\bibfnamefont
  {M.}~\bibnamefont {Bowry}}, \bibinfo {author} {\bibfnamefont
  {D.}~\bibnamefont {Jaroszynski}}, \bibinfo {author} {\bibfnamefont
  {B.}~\bibnamefont {Singh}}, \bibinfo {author} {\bibfnamefont
  {M.}~\bibnamefont {Scheck}}, \bibinfo {author} {\bibfnamefont
  {P.}~\bibnamefont {Spagnoletti}},\ and\ \bibinfo {author} {\bibfnamefont
  {J.}~\bibnamefont {Smith}},\ }\href
  {https://doi.org/10.1038/s41567-020-0899-4} {\bibfield  {journal} {\bibinfo
  {journal} {Nature Physics}\ }\textbf {\bibinfo {volume} {16}},\ \bibinfo
  {pages} {853} (\bibinfo {year} {2020})}\BibitemShut {NoStop}%
\bibitem [{\citenamefont {Flambaum}\ \emph {et~al.}(1996)\citenamefont
  {Flambaum}, \citenamefont {Gribakin},\ and\ \citenamefont
  {Izrailev}}]{Izrailev1996}%
  \BibitemOpen
  \bibfield  {author} {\bibinfo {author} {\bibfnamefont {V.~V.}\ \bibnamefont
  {Flambaum}}, \bibinfo {author} {\bibfnamefont {G.~F.}\ \bibnamefont
  {Gribakin}},\ and\ \bibinfo {author} {\bibfnamefont {F.~M.}\ \bibnamefont
  {Izrailev}},\ }\href {https://doi.org/10.1103/PhysRevE.53.5729} {\bibfield
  {journal} {\bibinfo  {journal} {Phys. Rev. E}\ }\textbf {\bibinfo {volume}
  {53}},\ \bibinfo {pages} {5729} (\bibinfo {year} {1996})}\BibitemShut
  {NoStop}%
\bibitem [{\citenamefont {Horoi}\ \emph {et~al.}(1995)\citenamefont {Horoi},
  \citenamefont {Zelevinsky},\ and\ \citenamefont {Brown}}]{Horoi1995}%
  \BibitemOpen
  \bibfield  {author} {\bibinfo {author} {\bibfnamefont {M.}~\bibnamefont
  {Horoi}}, \bibinfo {author} {\bibfnamefont {V.}~\bibnamefont {Zelevinsky}},\
  and\ \bibinfo {author} {\bibfnamefont {B.~A.}\ \bibnamefont {Brown}},\ }\href
  {https://doi.org/10.1103/PhysRevLett.74.5194} {\bibfield  {journal} {\bibinfo
   {journal} {Phys. Rev. Lett.}\ }\textbf {\bibinfo {volume} {74}},\ \bibinfo
  {pages} {5194} (\bibinfo {year} {1995})}\BibitemShut {NoStop}%
\bibitem [{\citenamefont {Zelevinsky}\ \emph {et~al.}(1995)\citenamefont
  {Zelevinsky}, \citenamefont {Horoi},\ and\ \citenamefont {{A.
  Brown}}}]{Zelevinsky1995}%
  \BibitemOpen
  \bibfield  {author} {\bibinfo {author} {\bibfnamefont {V.}~\bibnamefont
  {Zelevinsky}}, \bibinfo {author} {\bibfnamefont {M.}~\bibnamefont {Horoi}},\
  and\ \bibinfo {author} {\bibfnamefont {B.}~\bibnamefont {{A. Brown}}},\
  }\href {https://doi.org/https://doi.org/10.1016/0370-2693(95)00324-E}
  {\bibfield  {journal} {\bibinfo  {journal} {Physics Letters B}\ }\textbf
  {\bibinfo {volume} {350}},\ \bibinfo {pages} {141} (\bibinfo {year}
  {1995})}\BibitemShut {NoStop}%
\bibitem [{\citenamefont {Zaretsky}\ and\ \citenamefont
  {Sirotkin}(1983)}]{Zaratsky}%
  \BibitemOpen
  \bibfield  {author} {\bibinfo {author} {\bibfnamefont {D.}~\bibnamefont
  {Zaretsky}}\ and\ \bibinfo {author} {\bibfnamefont {V.}~\bibnamefont
  {Sirotkin}},\ }\href@noop {} {\bibfield  {journal} {\bibinfo  {journal} {Yad.
  Fiz. 37(1)}\ }\textbf {\bibinfo {volume} {45}},\ \bibinfo {pages} {1983. Sov.
  J. Nucl. Phys. 37, 361} (\bibinfo {year} {1983})}\BibitemShut {NoStop}%
\bibitem [{\citenamefont {Koonin}\ \emph {et~al.}(1992)\citenamefont {Koonin},
  \citenamefont {Johnson},\ and\ \citenamefont {Vogel}}]{Koonin1992}%
  \BibitemOpen
  \bibfield  {author} {\bibinfo {author} {\bibfnamefont {S.~E.}\ \bibnamefont
  {Koonin}}, \bibinfo {author} {\bibfnamefont {C.~W.}\ \bibnamefont
  {Johnson}},\ and\ \bibinfo {author} {\bibfnamefont {P.}~\bibnamefont
  {Vogel}},\ }\href {https://doi.org/10.1103/PhysRevLett.69.1163} {\bibfield
  {journal} {\bibinfo  {journal} {Phys. Rev. Lett.}\ }\textbf {\bibinfo
  {volume} {69}},\ \bibinfo {pages} {1163} (\bibinfo {year}
  {1992})}\BibitemShut {NoStop}%
\bibitem [{\citenamefont {Carlson}\ and\ \citenamefont
  {Hussein}(1993)}]{Carlson1993}%
  \BibitemOpen
  \bibfield  {author} {\bibinfo {author} {\bibfnamefont {B.~V.}\ \bibnamefont
  {Carlson}}\ and\ \bibinfo {author} {\bibfnamefont {M.~S.}\ \bibnamefont
  {Hussein}},\ }\href {https://doi.org/10.1103/PhysRevC.47.376} {\bibfield
  {journal} {\bibinfo  {journal} {Phys. Rev. C}\ }\textbf {\bibinfo {volume}
  {47}},\ \bibinfo {pages} {376} (\bibinfo {year} {1993})}\BibitemShut
  {NoStop}%
\bibitem [{\citenamefont {Lewenkopf}\ and\ \citenamefont
  {Weidenm\"uller}(1992)}]{Lewenkopf1992}%
  \BibitemOpen
  \bibfield  {author} {\bibinfo {author} {\bibfnamefont {C.~H.}\ \bibnamefont
  {Lewenkopf}}\ and\ \bibinfo {author} {\bibfnamefont {H.~A.}\ \bibnamefont
  {Weidenm\"uller}},\ }\href {https://doi.org/10.1103/PhysRevC.46.2601}
  {\bibfield  {journal} {\bibinfo  {journal} {Phys. Rev. C}\ }\textbf {\bibinfo
  {volume} {46}},\ \bibinfo {pages} {2601} (\bibinfo {year}
  {1992})}\BibitemShut {NoStop}%
\bibitem [{\citenamefont {Flambaum}\ and\ \citenamefont
  {Izrailev}(1997{\natexlab{b}})}]{Izrailev1997}%
  \BibitemOpen
  \bibfield  {author} {\bibinfo {author} {\bibfnamefont {V.~V.}\ \bibnamefont
  {Flambaum}}\ and\ \bibinfo {author} {\bibfnamefont {F.~M.}\ \bibnamefont
  {Izrailev}},\ }\href {https://doi.org/10.1103/PhysRevE.55.R13} {\bibfield
  {journal} {\bibinfo  {journal} {Phys. Rev. E}\ }\textbf {\bibinfo {volume}
  {55}},\ \bibinfo {pages} {R13} (\bibinfo {year}
  {1997}{\natexlab{b}})}\BibitemShut {NoStop}%
\end{thebibliography}%

\end{document}